\documentclass[12pt, a4paper]{scrbook}

\usepackage[utf8]{inputenc}
\usepackage[english,brazil]{babel}
\usepackage{indentfirst}
\usepackage{graphicx, fancyhdr}
\usepackage{amsmath,amssymb,amsfonts,amsthm}
\usepackage{setspace}
\usepackage{booktabs}
\usepackage{braket}
\usepackage{nicefrac}
\usepackage{bbm}                    
\usepackage{subfigure}              
\usepackage{fancyhdr}               
\usepackage[nottoc]{tocbibind}
\usepackage{hyperref}
\hypersetup{
citebordercolor=0 .75 1,
linktocpage=true,
backref=true
}

\newcommand{\eq}[1]{(\ref{#1})}     
\fancypagestyle{plain}{
\fancyhead{}                        
 }
\renewcommand{\sin}{\operatorname{sen}}
\renewcommand{\csc}{\operatorname{cosec}}

\theoremstyle{definition}
\newtheorem{example}{Exemplo}[chapter]

\begin{document}


\begin{titlepage}


\thispagestyle{empty}

\newcommand{\lyxline}[1][1pt]{
  \par\noindent
    \rule[.5ex]{\linewidth}{#1}\par}
\vspace*{-2cm}

\parskip = .4\baselineskip

\begin{center}
UNIVERSIDADE DE SÃO PAULO \\
INSTITUTO DE FÍSICA
\par\end{center}

\vspace{3cm}

\begin{doublespace}
\begin{center}
\textbf{\huge Um estudo sobre a Supersimetria}\textbf{\Large{} }\\
\vspace{0.2cm}
\textbf{\huge no contexto da Mecânica Quântica}\textbf{\Large{} }\\

\par\end{center}{\Large \par}
\end{doublespace}

\vspace{0.8cm}

\vspace{1.0cm}

\begin{center}
\textbf{\Large Fabricio Marques do Carmo}
\par\end{center}{\Large \par}

\vspace{2.4cm}


\begin{flushright}

\begin{minipage}[c][1\totalheight][t]{10.5cm}%
\textbf{Orientador: Prof. Dr. Adilson José da Silva}

\vspace{0.4cm}

\vspace{0.2cm}

Dissertação apresentada ao Instituto de Física da \linebreak Universidade
de São Paulo para a obtenção do título de Mestre em Ciências.

\vspace{0.2cm}


\vspace{0.8cm}

\end{minipage}
\par\end{flushright}

\vspace{2.2cm}

\noindent \textbf{\small Banca Examinadora: }\\
Prof. Dr. Adilson José da Silva (USP) \\
Prof. Dr. Alex Gomes Dias (UFABC) \\
Prof. Dr. Emerson José Veloso de Passos (USP)
\vspace{1.4cm}


\begin{center}
São Paulo\\
2011
\par
\end{center}

\end{titlepage}

                                    %
\newpage                            %
\baselineskip 8.75 mm               %
\thispagestyle{empty}               %
\vspace*{-2.0cm}                    %
\hspace{-3.0cm}
\includegraphics[scale=0.98]{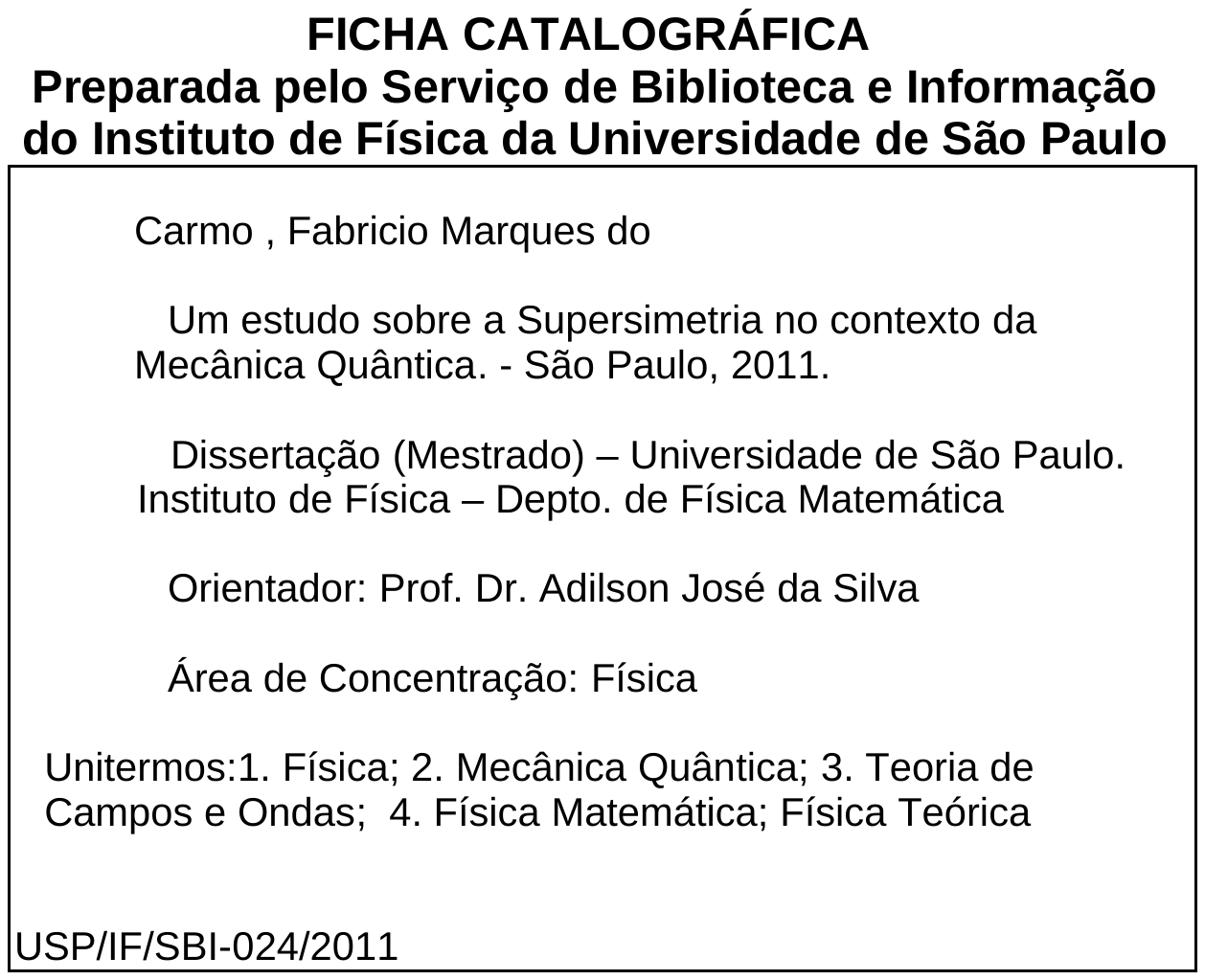}
                                    %
                                    %
                                    %

\newpage
\thispagestyle{empty}

\vspace*{17.0 cm}
\begin{flushright}
 \parbox{5.4in}{
        \begin{center}
                \begin{flushleft}
                        \textit{\hspace{8eM}
Ao meu pai Edson com quem tive minhas primeiras
discussões filosóficas profundas sobre o Universo, assunto esse que eu viria a conhecer 
mais tarde com o nome de Física. Eu gostaria que, de algum modo, ele pudesse me ver agora.
                        }
                \end{flushleft}
        \end{center}
        }
                           
        \vspace{1.0 cm}
\end{flushright}

\newpage
\thispagestyle{empty}
\frontmatter


\newpage
\thispagestyle{empty}

\vspace*{17.0 cm}
\begin{flushright}
 \parbox{5.4in}{
        \begin{center}
                \begin{flushleft}       
                        \textit{\hspace{8eM}
À minha querida esposa Luciana, amor da minha vida, que encontrei uma certa vez, há muito 
tempo, na final de um torneio de damas.
                        }
                \end{flushleft}
        \end{center}
        }
                           
        \vspace{1.0 cm}
\end{flushright}

\newpage
\thispagestyle{empty}
\frontmatter


\vspace*{8cm}

\hspace{3cm}\parbox[c]{20cm}{ {\it
Não importa onde você parou...\\
em que momento da vida você cansou...\\
o que importa é que sempre é possível e necessário
``Recomeçar''.\\
Recomeçar é dar uma nova chance a si mesmo...\\
é renovar as esperanças na vida e o mais importante...\\
acreditar em você de novo...\\
Sofreu muito nesse período? Foi aprendizado.\\
Chorou muito? Foi limpeza da alma.\\
Ficou com raiva das pessoas? Foi para perdoá-las um dia.\\
Tem tanta gente esperando apenas um sorriso seu para ``chegar''
perto de você.\\
Recomeçar...\\
hoje é um bom dia para começar novos desafios.\\
Onde você quer chegar?\\
Ir alto... Sonhe alto...\\
queira o melhor do melhor...\\
pensando assim trazemos para nós aquilo que desejamos...\\
Se pensarmos pequeno coisas pequenas teremos...\\
Já se desejarmos fortemente o melhor e principalmente lutarmos pelo melhor,\\
o melhor vai se instalar em nossa vida.\\
Porque sou do tamanho daquilo que vejo, e não do tamanho da minha altura.
}}
\vspace{0.7cm}

\hspace{4cm}{\footnotesize Carlos Drummond de Andrade,  ``\textit{Recomeçar}''}

\thispagestyle{empty}


\newpage
\thispagestyle{empty}
\frontmatter

\bigskip
\bigskip
\begin{center}
{\Large{\bf Agradecimentos}}
\end{center}
\vspace{1cm}

\begin{quote}

Ao Professor Adilson José da Silva, orientador, cuja fantástica receptividade, disposição,
habilidade didática e força bruta para resolver problemas, se mostraram ímpares desde o princípio;
\medskip

Aos meus amigos do Instituto de Física, Claudio Padilha, Felipe Villaverde, Ronaldo
Batista, Henrique Xavier, Roberto Maluf, Enrique Alberto \mbox{Gallegos} Collado, Kazuo Teramoto e 
Osvaldo Negrini que estiveram sempre dispostos a ajudar de forma relevante na resolução de 
problemas, em geral, totalmente não relacionados com seja lá o que for que eles faziam;
\medskip

Aos meus outros amigos e professores favoritos não citados acima, pois a lista de nomes é, 
felizmente, extensa demais para ser aqui colocada;
\medskip

Ao meu pai Edson, à minha mãe Ely e ao meu irmão Fabio, que me ensinaram a ser quem eu sou;
\medskip

À minha querida esposa Luciana, pelo extremamente relevante e enorme (com ênfase em enorme) amor 
que fez e faz toda a diferença em minha vida, que me faz feliz, e;
\medskip

Ao CNPq (Conselho Nacional de Desenvolvimento Científico e Tecnológico), pelo apoio
concedido durante o período de realização deste trabalho.

\end{quote}

\newpage
\hspace{1cm}
\newpage
                                    %

\centerline{\large \textbf{RESUMO}}

\vspace{3.0cm}

\begin{quote}
Neste trabalho apresentamos uma introdução à Supersimetria no contexto da Mecânica Quântica em $1$
dimensão espacial. Para isso desenvolvemos o conceito de fatorização de hamiltonianos por meio
do exemplo do oscilador harmônico simples, introduzimos o oscilador supersimétrico e, em
seguida, generalizamos esses conceitos para introduzir os fundamentos da Mecânica Quântica
Supersimétrica. Discutimos também a respeito de ferramentas úteis na resolução de problemas de
Mecânica Quântica que estão intrinsecamente relacionadas à Supersimetria como a hierarquia de
hamiltonianos e a invariância de forma. Apresentamos dois métodos de aproximação que serão
particularmente úteis; o bem conhecido Método Variacional e também a Teoria de Perturbações
Logarítmica, esta última intimamente relacionada com os conceitos de superpotenciais e
hierarquia de hamiltonianos. Por fim, apresentamos problemas associados a superpotenciais que
são monômios em potências pares da coordenada $x$ multiplicados pela função sinal
$\varepsilon(x)$, o que aparentemente constitui uma classe inédita de problemas na Mecânica
Quântica Supersimétrica.
\end{quote}

\newpage

\begin{otherlanguage}{english}
\centerline{\large \textbf{ABSTRACT}} 

\vspace{3.0cm}

\begin{quote}
In this work we present an introduction to Supersymmetry in the context of $1$-dimensional 
Quantum Mechanics. For that purpose we develop the concept of hamiltonians factorization
using the simple harmonic oscillator as an example, we introduce the supersymmetric oscilator and,
next, we generalize these concepts to introduce the fundamentals of Supersymmetric Quantum
Mechanics. We also discuss useful tools to solve problems in Quantum Mechanics which are
intrinsecally related to Supersymmetry as hierarchy of hamiltonians and shape invariance. We
present two approximation methods which will be specially useful: the well known Variational
Method and the Logarithmic Perturbation Theory, the latter being closely related to the
concepts of superpotentials and hierarchy of hamiltonians. Finally, we present problems related to
superpotentials which are monomials in even powers of the $x$ coordinate multiplied by the sign 
function $\varepsilon(x)$, which seems to be a new class of problems in Supersymmetric Quantum 
Mechanics.
\end{quote}
\end{otherlanguage}
\newpage

                                    %
\renewcommand{\contentsname}{Índice}
\tableofcontents                    
\mainmatter                         %
                                    %

\chapter{Introdução}
\label{chapter:intro}

\section{O que é Supersimetria?}

A Supersimetria (SUSI) é uma simetria que relaciona estados bosônicos e fermiônicos. Ela foi
originalmente desenvolvida no contexto da Teoria Quântica de Campos tendo sido proposta em 
1966 por Miyazawa \cite{Miyazawa} como uma simetria entre mésons e bárions. Foi redescoberta 
no início dos anos 70 e em 1974 foi trazida à atenção da comunidade científica 
por meio dos trabalhos de Wess e Zumino \cite{WessZumino1}. Esses trabalhos  
apresentaram o que ficou conhecido como \emph{Modelo de Wess-Zumino}, o primeiro modelo de 
teoria quântica de campos em $(3+1)$ dimensões supersimétrico.

A SUSI surgiu como um caminho para estender o Grupo de Poincaré de modo a incluir simetrias 
internas. Antes isso não era permitido devido ao teorema \emph{no-go} de Coleman-Mandula
\cite{ColemanMandula}, que impunha uma restrição sobre essa possibilidade afirmando que a 
álgebra de Lie mais geral das simetrias da matriz $S$ consistente com uma teoria quântica de 
campos relativística deveria conter, além dos geradores do Grupo de Poincaré (aqui denotados 
$P_{\mu}$ para translações no espaço-tempo e $J_{\mu \nu}$ para rotações no espaço e 
\emph{boosts}), no máximo um número finito de operadores escalares $B_{\mu}$ pertencentes à 
álgebra de Lie de um grupo compacto. Esse obstáculo foi finalmente contornado devido a um 
resultado de Haag, Sohnius e Lopuszanski \cite{Haag} 
que mostrou que, para incluir as simetrias internas, a álgebra de Lie do grupo de Poincaré 
deveria ser estendida para uma álgebra de Lie graduada de modo a incluir além de 
comutadores também \emph{anti-}comutadores \cite{WessBagger}. Essa álgebra estendida é então 
chamada \emph{álgebra SUSI} ou simplesmente \emph{super-álgebra}.

A relação estabelecida pela SUSI entre estados bosônicos e fermiônicos se dá por meio das
transformações SUSI, cujos geradores denotamos $Q$ e $\bar{Q}$. A parte da
super-álgebra que é incluída no Grupo de Poincaré estabelecendo relações entre os geradores 
SUSI e os geradores das translações no espaço-tempo é da forma \footnote{Aqui, uma vez que este
trabalho se refere à Mecânica Quântica, não nos preocuparemos em detalhar assuntos como 
\emph{tipos de índice}, \emph{spinores de Majorana}, \emph{transformações SUSI}, etc. Para uma 
introdução didática do assunto ver, por exemplo, \cite{Ryder} ou \cite{Balin}.}:

\begin{equation}
  \begin{aligned}
  &  \{Q_{\alpha},\bar{Q}_{\dot{\beta}}\} = 2 \sigma ^{\mu}_{\phantom{\mu} \alpha
    \dot{\beta}}P_{\mu}\\
  &  \{Q_{\alpha},Q_{\beta}\} = \{\bar{Q}_{\dot{\alpha}},\bar{Q}_{\dot{\beta}}\} = 0\\
  &  \left[P_{\mu},Q_{\alpha} \right] = \left[P_{\mu},\bar{Q}_{\dot{\alpha}} \right] = 0 
  \label{SuperAlgebraQFT}
  \end{aligned}
\end{equation}

Embora \eq{SuperAlgebraQFT} seja apenas \emph{parte} da super-álgebra, neste trabalho nos 
referiremos a conjuntos de equações desse tipo simplesmente como \emph{super-álgebra}.

A SUSI prevê que, para cada partícula (bóson ou férmion), deve haver uma outra 
partícula correspondente com a mesma massa e diferindo de $\frac{1}{2}$ no valor do spin
(férmion ou bóson). Essas partículas são então chamadas de 
\emph{parceiras supersimétricas}. O parceiro supersimétrico do fóton, que é um bóson de spin 
$1$ e massa de repouso nula, seria então um férmion de spin $\frac{1}{2}$ com massa de 
repouso nula ao qual chamamos \emph{fotino}. O parceiro supersimétrico do elétron, que é um 
férmion de spin $\frac{1}{2}$ e massa $m_e$, seria, por sua vez, um bóson de spin \emph{zero}
com massa $m_e$ ao qual chamamos \mbox{\emph{s-elétron}} (\emph{scalar-electron}). Como os 
parceiros supersimétricos das partículas do modelo padrão, contrariando a expectativa, jamais 
foram observados, então, se a SUSI for uma simetria da natureza, ela deva ter 
sido \emph{quebrada} em algum momento. E daí vem o interesse em estudar a quebra da SUSI a
medida que a temperatura do Universo diminuiu.

\section{A Supersimetria no Contexto da Mecânica Quântica}

No início da década de 80, Witten \cite{Witten1} \cite{Witten2} propôs como um caminho para o
entendimento da quebra da SUSI a construção de um
esquema em Mecânica Quântica onde estavam presentes os principais ingredientes da SUSI. Esse
esquema é o que chamamos de \emph{Mecânica Quântica Supersimétrica} (MQ SUSI). O interesse 
em estudar MQ SUSI é, desde então, justificado pela simplicidade dessa formulação da
SUSI em comparação com a original. Sendo mais simples esperamos que seu estudo forneça pistas
relevantes para o melhor entendimento da SUSI na Teoria Quântica de Campos, em especial no que 
se refere à quebra da SUSI. 

A super-álgebra da MQ SUSI, devidamente adaptada para o tipo de sistema que representa, tem 
a mesma forma da super-álgebra dada em \eq{SuperAlgebraQFT} podendo ser escrita, por 
exemplo, como: 

\begin{equation}
  \begin{aligned}
    &  \{Q ,Q^{\dagger} \} = 2 H \\
    &  \{Q ,Q \} = \{Q^{\dagger} ,Q^{\dagger} \} = 0\\
    &  \left[H,Q \right] = \left[H,Q^{\dagger} \right] = 0 
  \label{SuperAlgebraMQSUSI}
  \end{aligned}
\end{equation}

Em comparação com a SUSI original, dizemos que a MQ SUSI relaciona estados respectivamente
chamados ``bosônicos'' e ``fermiônicos''. Essa nomenclatura não tem, entretanto, 
nada a ver com o spin de partículas sendo inteiro ou semi-inteiro (o que caracterizaria a 
definição de bósons e férmions) e em MQ SUSI representa, conforme veremos, apenas uma analogia. 

A MQ SUSI é frequentemente chamada de \emph{modelo de brinquedo} uma
vez que, em sua concepção original, o objetivo foi fornecer ferramentas para o entendimento de
algo mais profundo. Apesar disso podemos destacar inúmeras possibilidades de empregar a MQ SUSI
como uma ferramenta na resolução de problemas da própria Mecânica Quântica. Um
primeiro exemplo que podemos citar consiste na aplicação da MQ SUSI ao problema do oscilador
harmônico simples (OHS). Uma
generalização do método algébrico de resolução do OHS (aquele que utiliza operadores de criação 
e destruição $a^{\dagger}$ e $a$) permite tratar outros tipos de problema e está diretamente
ligada aos conceitos da MQ SUSI. Uma outra aplicação interessante consiste em empregar
os métodos da MQ SUSI na resolução da equação radial do problema do átomo de hidrogênio, o que é
uma opção muito mais simples do que a usual.

Nesse trabalho apresentaremos um estudo da Supersimetria no contexto da Mecânica Quântica em
$1$ dimensão espacial. Primeiramente apresentaremos os fundamentos da MQ SUSI, introduzindo-a
por meio da fatorização de hamiltonianos em analogia com o método algébrico de resolução do OHS.
Também serão apresentados métodos para a resolução de problemas em Mecânica Quântica utilizando 
conceitos da MQ SUSI como hierarquia de hamiltonianos e invariância de forma. Alguns métodos de
aproximação que serão úteis em seguida são apresentados. Por fim, apresentamos alguns problemas
para os quais utilizamos os métodos da MQ SUSI e os métodos de aproximação
previamente expostos. Em especial, nos concentramos na resolução de uma classe de problemas
ainda não explorada pela literatura acerca de potenciais do tipo 
$W(x) = g \varepsilon(x) x^{2n}$, isto é, na forma de monômios em potências pares de $x$
multiplicados pela função sinal $ \varepsilon(x)$.
                                    %

\chapter{Mecânica Quântica Supersimétrica}
\label{chapter:mqsusi}

\section{Fatorização de Hamiltonianos}

\subsection{Fatorização do Hamiltoniano do OHS}\label{subsection:fatorizacaoOH}
A fatorização do hamiltoniano do oscilador harmônico simples (OHS) consiste em tentar 
escrevê-lo como um produto de dois operadores de primeira ordem. 

%

O hamiltoniano do OHS, tendo escolhido por simplicidade $\hbar=2m=\omega=1$, é dado por:

\begin{equation}
\label{hamiltOHS}
H = p^2 + x^2
\end{equation}
%

Sabendo que em Mecânica Quântica $[x,p] \neq 0$, definimos:

\begin{equation}
\begin{aligned}
\label{escada}
a^{\dagger} = \frac{1}{\sqrt{2}}(x-ip)\\
a = \frac{1}{\sqrt{2}}(x+ip)
\end{aligned}
\end{equation}
e notamos que
\begin{equation*}
 a^{\dagger}a + a a^{\dagger} = p^2 + x^2 = H
\end{equation*}
onde os termos contendo $[x,p]$ são cancelados.

Sendo $[x,p]=i$ (pois estamos usando $\hbar = 1$) e tendo definido $a^{\dagger}$ e $a$ 
em função de $x$ e $p$ em \eq{escada},
encontramos que $[a,a^{\dagger}]=1$, o que permite escrever:

\begin{equation}
  H = a^{\dagger}a + a a^{\dagger} = a^{\dagger}a + \frac{1}{2} = a a^{\dagger} - \frac{1}{2}
\label{hamiltOHS2}
\end{equation}
ou seja, podemos redefinir o hamiltoniano a menos de um fator constante, de modo que ele seja
simplesmente:

\begin{equation}
\label{hamiltOHredef}
H =  a^{\dagger}a
\end{equation}
(ou, de forma análoga, podemos também redefini-lo de de modo que ele seja $H = a a^{\dagger}$).

Com isso temos o hamiltoniano do OHS fatorizado, isto é, escrito como um produto dos operadores
$a^{\dagger}$ e $a$.

\subsection{Fatorização de um Hamiltoniano Geral}\label{subsection:fatorizacaoGeral}
Inspirados pela fatorização do hamiltoniano do OHS, vamos tentar estender o método para um
hamiltoniano mais geral. Para isso definimos:

\begin{equation}
\begin{aligned}
\label{superescada}
A^{\dagger} =  \left( W(x)-ip \right)  \\
A = \left( W(x) + ip \right)
\end{aligned}
\end{equation}
onde $W(x)$ é uma função de $x$.

Vamos agora construir um hamiltoniano na mesma forma do hamiltoniano OHS dado em
\eq{hamiltOHredef}, ou seja:

\begin{equation}
\label{hamiltfatorizado1}
H_{-} =  A^{\dagger}A
\end{equation}

Também podemos, e de fato vamos, construir um hamiltoniano $H_{+} = AA^{\dagger}$.

Substituindo \eq{superescada} em \eq{hamiltfatorizado1}, usando que $p =
-i \frac{d}{dx}$ e sendo $W'(x) = \frac{d}{dx}W(x)$, temos:

\begin{equation}
\begin{aligned}
\label{hamiltfatorizadoexplicito1}
H_{-} & =  A^{\dagger}A  
        =  \left( W(x)-ip \right) \left( W(x) + ip \right)\\ 
      & = p^2 + \left( W(x)^2 + i [W(x),p] \right)
        = p^2 + \left( W(x)^2 - W'(x) \right)
\end{aligned}
\end{equation}
e de forma idêntica para $H_{+} = AA^{\dagger}$, de modo que:

\begin{equation*}
H_{\mp} = p^2 + \left( W(x)^2 \mp W'(x) \right)
\end{equation*}
o que permite definir os potenciais $V_{\mp}(x)$ dos hamiltonianos $H_{\mp}$ como:

\begin{equation}
\label{riccati}
V_{\mp}(x) =  W(x)^2 \mp W'(x)
\end{equation}
que é conhecida como equação de Riccati. 

Assim, vemos que para fatorizar um
hamiltoniano geral devemos, uma vez que conhecemos seu potencial $V(x)$, resolver a equação
de Riccati \eq{riccati} para determinar a função $W(x)$. Se isso for possível, podemos então,
conforme \eq{superescada} construir os operadores $A^{\dagger}$ e $A$ e com isso obter o
hamiltoniano $H = A^{\dagger}A$ (e também $H = AA^{\dagger}$) na forma fatorizada.


\section{Osciladores Harmônicos e Supersimetria}

\subsection{Osciladores Bosônico e Fermiônico}

Vamos introduzir a seguir o conceito de osciladores bosônico e fermiônico. Esses conceitos serão
utilizados em seguida para construir um exemplo didático do funcionamento e aplicabilidade do 
formalismo da SUSI à Mecânica Quântica.

O oscilador bosônico que vamos definir agora nada mais é do que o próprio oscilador harmônico
simples da Mecânica Quântica. Em
termos de operadores de criação e destruição, o hamiltoniano do sistema é dado por (ver
\eq{hamiltOHS2}):

\begin{equation}
\label{hamiltOHbose}
H_B = \frac{1}{2}\left( a^{\dagger}a + a a^{\dagger}  \right) 
\end{equation}
onde $a^{\dagger}$ e $a$ são,respectivamente, os operadores de criação e 
destruição. Esses operadores satisfazem as seguintes relações de comutação:

\begin{equation}
\begin{aligned}
\label{comutOHbose}
& [ a,a^{\dagger} ] = 1 \\
& [ a,a ] = [ a^{\dagger},a^{\dagger} ] = 0
\end{aligned}
\end{equation}
e o espectro de $H_B$ é formado pelos vetores de estado $\ket{n}$ de tal
forma que:

\begin{equation}
  a \ket{0} = 0 
  \qquad \text{e} \qquad 
  \frac{ (a^{\dagger})^n}{\sqrt{n!}} \ket{0} = \ket{n}
  \label{espectroOHbose}
\end{equation}
tendo $n$ a possibilidade de assumir os valores $0,1,2,\ldots$.

Em contraste com a expressão \eq{hamiltOHbose}, do hamiltoniano do oscilador bosônico, que é
simétrica pela troca das posições de $a$ e $a^{\dagger}$, definimos o oscilador fermiônico por
meio de seu hamiltoniano:

\begin{equation}
\label{hamiltOHfermi}
H_F = \frac{1}{2}\left( b^{\dagger}b - b b^{\dagger}  \right)
\end{equation}
de modo que esta seja antissimétrica pela troca das posições de $b$ e $b^{\dagger}$. Aqui 
$b$ e $b^{\dagger}$ são, respectivamente, operadores de criação e destruição fermiônicos. Esses
operadores são definidos como elementos grassmanianos, satisfazendo as seguintes relações de
\emph{anti-}comutação:

\begin{equation}
\begin{aligned}
\label{acomutOHfermi}
& \lbrace b,b^{\dagger} \rbrace = 1 \\
& \lbrace b,b \rbrace = \lbrace b^{\dagger},b^{\dagger} \rbrace=0
\end{aligned}
\end{equation}

O espectro de $H_F$ é definido em \eq{espectroOHfermi} de forma semelhante a que foi feita para
o espectro de $H_B$ em \eq{espectroOHbose}. Aqui porém, a nilpotência dos operadores $b$ e
$b^{\dagger}$, manifestada na segunda linha de \eq{acomutOHfermi}, restringe os possíveis
valores de $n$, de modo que:

\begin{equation}
  b \ket{0} = 0 \qquad \text{e} \qquad b^{\dagger} \ket{0} = \ket{1}
  \label{espectroOHfermi}
\end{equation}
e com isso o espectro de $H_F$ é formado por apenas dois estados, $\ket{m}$ com $m=0,1.$.

Dadas as relações de comutação \eq{comutOHbose} e anticomutação \eq{acomutOHfermi},
podemos utilizá-las para reescrever os hamiltonianos \eq{hamiltOHbose} e \eq{hamiltOHfermi}
como:

\begin{equation}
\begin{aligned}
\label{hamiltsOHbf}
& H_B = \left( N + \frac{1}{2}  \right) \\
& H_F = \left( M - \frac{1}{2}  \right)
\end{aligned}
\end{equation}
onde $N = a^{\dagger}a$ e $M=b^{\dagger}b$ são operadores número, isto é, são operadores tais
que, conforme \eq{espectroOHbose} e \eq{espectroOHfermi}, as seguintes equações de autovalores 
e autoestados são satisfeitas:

\begin{equation}
\begin{aligned}
\label{opnumero}
& N \ket{n} = n \ket{n}\\
& M \ket{m} = m \ket{m}
\end{aligned}
\end{equation}
onde podemos ter $n = 0,1,2, \ldots$ e $m = 0,1.$.

Utilizando a definição dos operadores número acima e as relações de anti-comutação 
\eq{acomutOHfermi}, vemos que, para o operador número fermiônico, vale:

\begin{equation*}
\begin{aligned}
M^2 & = b^{\dagger}bb^{\dagger}b = b^{\dagger} (1 -b^{\dagger}b)b = b^{\dagger}b = M \\
\Rightarrow
& \quad M (M - 1)=0
\end{aligned}
\end{equation*}
de modo que os autovalores desse operador só podem assumir os valores $0$ ou $1$, o que é uma
outra forma de ver a restrição sobre os valores de $m$.

Sendo válidas as equações de autovalores e autovetores \eq{opnumero}, considerando a forma dos
hamiltonianos $H_B$ e $H_F$ conforme \eq{hamiltsOHbf} e sabendo quais são os valores que 
$n$ e $m$ podem assumir, temos que os autovalores desses hamiltonianos são:

\begin{equation}
  \begin{aligned}
    & E^B_n = \left( n + \frac{1}{2} \right) \; , \qquad \qquad \qquad \qquad n=0,1,2,\ldots \\
    & E^F_m = \left( m - \frac{1}{2} \right) \; , \qquad \qquad \qquad \qquad m=0,1.
    \label{autovalOHbf}
  \end{aligned}
\end{equation}

Cabe notar aqui que denominar esses osciladores ``bosônico'' e ``fermiônico'' é uma analogia
associada, respectivamente, à simetria e antissimetria dos hamiltonianos correspondentes. Além
disso, a restrição sobre os possíveis valores de $m$ é um análogo do princípio da exclusão de
Pauli.


\subsection{Oscilador Supersimétrico}
Vamos definir um hamiltoniano que é a soma dos osciladores bosônico e fermiônico:

\begin{equation}
\label{hamiltOHsusi}
H = H_B + H_F = \left( a^{\dagger}a + b^{\dagger}b  \right)
\end{equation}
onde o lado direito foi obtido substituindo as equações \eq{hamiltsOHbf}.

Tendo em vista a definição dos operadores número conforme \eq{opnumero}, temos que os
autoestados do hamiltoniano \eq{hamiltOHsusi} são:

\begin{equation}
\ket{n, m} = \ket{n} \otimes \ket{m}
\end{equation}
e a equação de Schrödinger para esse hamiltoniano fica:

\begin{equation}
\label{schrOHsusi}
H \ket{n, m} = \left( n + m \right) \ket{n, m}
\end{equation}
onde a forma dos autovalores foi determinada a partir de \eq{autovalOHbf}, somando $E^B_n$ e
$E^F_m$.

O estado fundamental desse sistema é $\ket{0, 0}$ com energia $E_0 = 0$. Todos os outros estados
são duplamente degenerados uma vez que, para $n > 0$, podemos ter $m = 0$ ou $m = 1$, de 
modo que os estados $\ket{n, 0}$ e $\ket{n - 1, 1}$, conforme \eq{schrOHsusi}, tem a
mesma energia.

A seguir, definimos os operadores:

\begin{equation}
\begin{aligned}
\label{supercargasOHsusi}
Q = a^{\dagger}b\\
Q^{\dagger} = b^{\dagger}a
\end{aligned}
\end{equation}
que misturam operadores de criação e destruição bosônicos e fermiônicos.

Usando as relações de comutação \eq{comutOHbose} e anti-comutação \eq{acomutOHfermi} juntamente
com a definição do hamiltoniano \eq{hamiltOHsusi} e com a definição dos operadores $Q$ e 
$Q^{\dagger}$, mostramos que:

\begin{align}
[Q,H] = [Q^{\dagger},H]=0 \label{comutQH} \\
\lbrace Q,Q \rbrace = \lbrace Q^{\dagger},Q^{\dagger} \rbrace = 0 \label{acomutQQ}\\
\lbrace Q,Q^{\dagger} \rbrace = H \label{acomutQQdag}
\end{align}

Essas relações de comutação e anti-comutação definem a nossa super-álgebra, 
sendo $Q$ e $Q^{\dagger}$ os geradores das transformações SUSI. A ação desses
operadores sobre os autoestados do hamiltoniano $H$ pode ser avaliada por meio das seguintes 
relações:

\begin{align*}
[Q, H_B] & = [a^{\dagger}b,a^{\dagger}a + \tfrac{1}{2}] = [a^{\dagger}b,a^{\dagger}a]
           = a^{\dagger}[a^{\dagger},a]b = - a^{\dagger}b = -Q \\
[Q, H_F] & = [a^{\dagger}b,b^{\dagger}b - \tfrac{1}{2}] = [a^{\dagger}b,b^{\dagger}b] 
           = a^{\dagger} (b b^{\dagger}) b
           = a^{\dagger} \lbrace b^{\dagger},b \rbrace b = a^{\dagger}b = Q
\end{align*}
onde \eq{comutOHbose} e \eq{acomutOHfermi} foram usadas. De forma idêntica temos:

\begin{align*}
[Q^{\dagger}, H_B] & = [b^{\dagger}a,a^{\dagger}a + \tfrac{1}{2}] = [b^{\dagger}a,a^{\dagger}a]
           = b^{\dagger}[a,a^{\dagger}]a = b^{\dagger}a = Q^{\dagger} \\
[Q^{\dagger}, H_F] & = [b^{\dagger}a,b^{\dagger}b - \tfrac{1}{2}] = [b^{\dagger}a,b^{\dagger}b] 
           = - b^{\dagger} (b b^{\dagger}) a
           = - b^{\dagger} \lbrace b, b^{\dagger} \rbrace a = - b^{\dagger}a = - Q^{\dagger}
\end{align*}

Com isso, vamos agora avaliar a ação dos operadores $H_B$ e $H_F$ sobre um vetor de estado 
$Q \ket{n,m}$:

\begin{equation}
\begin{aligned}
\label{sobeOHbose}
H_B Q \ket{n,m} & = (Q H_B - [Q, H_B]) \ket{n,m}
                      = Q \left( n + \frac{1}{2} \right) \ket{n,m} + Q \ket{n,m} \\
                    & = \left( n + 1 + \frac{1}{2} \right) Q \ket{n,m}
\end{aligned}
\end{equation}
\begin{equation}
\begin{aligned}
\label{desceOHfermi}
H_F Q \ket{n,m} & = (Q H_F - [Q, H_F]) \ket{n,m} 
                      = Q \left( m - \frac{1}{2} \right) \ket{n,m} - Q \ket{n,m} \\
                    & = \left( m - 1 - \frac{1}{2} \right) Q \ket{n,m}
\end{aligned}
\end{equation}
ou seja, para $m \neq 0$, $Q \ket{n,m}$ é autoestado de $H_B$ e $H_F$ com autovalores 
$\left( n + 1 + \frac{1}{2} \right)$ e $\left( m - 1 - \frac{1}{2}\right)$, 
respectivamente. Para $m = 0$, o operador $Q$ aniquila o vetor de estado 
$\ket{n, 0}$, ou seja, $Q \ket{n, 0} = 0$.

O mesmo desenvolvimento feito em \eq{sobeOHbose} e \eq{desceOHfermi} pode ser
feito para um vetor de estado $Q^{\dagger} \ket{n,m}$, levando à concluir que, 
para $n \neq 0$ e $m \neq 1$, esse estado é autoestado de $H_B$ e $H_F$ com autovalores 
$n - 1 + \frac{1}{2}$ e $m + 1 - \frac{1}{2}$, respectivamente. Para $n = 0$ ou $m = 1$,
por outro lado, temos que $Q^{\dagger} \ket{n, m} = 0$.

Em outras palavras, o operador $Q$ atua nos autoestados de $H$ levando o número quântico 
fermiônico de $m=1$ a $m=0$ e o número quântico bosônico de $n$ a $n + 1$. O operador 
$Q^{\dagger}$, por sua vez, leva $m=0$ em $m=1$ e $n$ em $n - 1$. Com isso vemos que,
uma vez que vale \eq{comutQH} e levando em conta a
normalização, temos, de acordo com a equação de Schrödinger \eq{schrOHsusi}:

\begin{equation}
\label{degOHsusi}
\left.
\begin{aligned}
Q \ket{n,m}           & = \delta_{m,1}\sqrt{n+1}\ket{n+1,m-1} \\
Q^{\dagger} \ket{n,m} & = \delta_{m,0} \sqrt{n} \ket{n-1,m+1}
\end{aligned}
\right \}
\qquad
E_{nm} = \left( n + m \right)
\end{equation}
ou seja, exceto pelo autoestado $\ket{0, 0}$, de energia $E_0 = 0$, todos os outros 
autoestados de $H$ satisfazem \eq{degOHsusi}. Esses estados $Q \ket{n,m}$ 
e $Q^{\dagger} \ket{n,m}$ são degenerados, tendo ambos energia 
$E_{n,m} = \left( n + m \right)$. Isso é a manifestação da \emph{supersimetria} desse sistema
que é então denominado oscilador \emph{supersimétrico}.


\section{Mecânica Quântica Supersimétrica}

\subsection{Generalização do Oscilador Supersimétrico}

A Mecânica Quântica Supersimétrica (MQ SUSI) pode ser entendida como uma generalização do
modelo do oscilador supersimétrico no mesmo sentido em que a fatorização de hamiltonianos em
termos dos operadores $A$ e $A^{\dagger}$ feita na seção \ref{subsection:fatorizacaoGeral}
é uma generalização da fatorização do hamiltoniano do OHS.

O oscilador supersimétrico constitui, na verdade, o sistema supersimétrico mais simples da 
Mecânica Quântica. Na definição dele, empregamos, conforme o que foi feito na seção 
\ref{subsection:fatorizacaoOH}, a forma fatorizada de hamiltonianos de osciladores harmônicos, 
o oscilador bosônico e o oscilador fermiônico. Podemos então procurar generalizar esse método 
para sistemas de Mecânica Quântica com hamiltonianos fatorizáveis, isto é, hamiltonianos que
tenham potenciais $V(x)$ tais que a equação de Riccati \eq{riccati} admita solução $W(x)$ 
(ver seção \ref{subsection:fatorizacaoGeral}). Se houver essa possibilidade, essa função 
$W(x)$ assim determinada é denominada \emph{superpotencial}, podendo ser entendida de algum 
modo como um objeto mais fundamental em uma teoria MQ SUSI do que o próprio potencial.

Vamos então definir um hamiltoniano $H$ de tal forma que ele satisfaça a álgebra definida por
\eq{comutQH} e \eq{acomutQQdag}. Com essa finalidade definimos os geradores $Q$ e 
$Q^{\dagger}$ de forma semelhante a \eq{supercargasOHsusi}, mas agora como:

\begin{equation}
\begin{aligned}
\label{supercargasMQsusi}
Q = A^{\dagger} \theta \\
Q^{\dagger} = \theta^{\dagger}A
\end{aligned}
\end{equation}
com os operadores $A$ e $A^{\dagger}$ no lugar de $a$ e $a^{\dagger}$ e sendo $\theta$ e
$\theta^{\dagger}$ elementos grassmanianos satisfazendo as relações de anti-comutação 
\eq{acomutOHfermi}. Uma possível realização dessas relações de anti-comutação ocorre para
$\theta$ e $\theta^{\dagger}$ dados por:

\begin{equation}
\begin{aligned}
\label{grassMQsusi}
\theta = \sigma_{+} = \frac{1}{2} \left(\sigma_1 + i \sigma_2 \right) 
       = \begin{pmatrix}0 & 1 \\ 0 & 0\end{pmatrix}  \\
\theta^{\dagger} = \sigma_{-} = \frac{1}{2} \left( \sigma_1 - i \sigma_2 \right)
                 = \begin{pmatrix} 0 & 0 \\ 1 & 0 \end{pmatrix}
\end{aligned}
\end{equation}
onde $\sigma_1 = \left( \begin{smallmatrix}0 & 1 \\ 1 & 0\end{smallmatrix} \right)$ e
$\sigma_2 = \left( \begin{smallmatrix}0 & -i \\ i & 0\end{smallmatrix} \right)$ são matrizes de 
Pauli.

Com isso temos, a partir de \eq{acomutQQdag}, uma matriz $H$ \footnote{Essa matriz 
$H = \left( \begin{smallmatrix} H_{-} & 0 \\ 0 & H_{+} \end{smallmatrix} \right)$ 
é algumas vezes chamada de \emph{super-hamiltoniano}.} 
dada por:

\begin{equation}
\label{superH}
H = \lbrace Q,Q^{\dagger} \rbrace =
\begin{pmatrix} 
A^{\dagger}A  & 0           \\
0             & AA^{\dagger}
\end{pmatrix} =
\begin{pmatrix} 
H_{-}  & 0    \\
0      & H_{+}
\end{pmatrix}
\end{equation}
sendo também satisfeitas, como se pode facilmente verificar, as relações 
\eq{comutQH} e \eq{acomutQQ}, ou seja:

\begin{align*}
[Q,H] = [Q^{\dagger},H] = 0 \\
\lbrace Q,Q \rbrace = \lbrace Q^{\dagger},Q^{\dagger} \rbrace = 0
\end{align*}

Podemos ainda escrever $H$ explicitamente em termos do superpotencial como:

\begin{equation}
\label{superHW}
H = \left( p^2 + W(x)^2 \right) \mathbbm{1} - W'(x) \sigma_3
\end{equation}
onde $\mathbbm{1}$ é a matriz identidade $2 \times 2$ e 
$\sigma_3 = \left( \begin{smallmatrix} 1 & 0 \\ 0 & -1 \end{smallmatrix} \right)$ é uma matriz de
Pauli.

Com isso, $H$ definida em \eq{superH} e os geradores $Q$ e $Q^{\dagger}$ definidos 
em \eq{supercargasMQsusi} satisfazem as relações de comutação e anti-comutação \eq{comutQH},
\eq{acomutQQ} e \eq{acomutQQdag} definindo, portanto, uma super-álgebra.

\subsection{Autovalores e Autoestados de $H$}\label{subsection:autovalH}

Os candidatos a autoestados de $H$ são matrizes 
$\Psi =\left( \begin{smallmatrix} \psi^{-}_n \\ \psi^{+}_m \end{smallmatrix} \right)$, onde 
$\psi^{-}_n$ e $\psi^{+}_m$ são, respectivamente, os autoestados dos hamiltonianos $H_{-}$ e 
$H_{+}$ com autovalores $E^{-}_n$ e $E^{+}_m$. Assim temos:

\begin{align*}
H \Psi = 
\begin{pmatrix} 
H_{-} & 0      \\
0     & H_{+}
\end{pmatrix}
\begin{pmatrix} \psi^{-}_n \\ \psi^{+}_m  \end{pmatrix} = 
\begin{pmatrix} E^{-}_n \psi^{-} \\ E^{+}_m \psi^{+}_m \end{pmatrix}
\end{align*}
o que permite concluir que, para $E^{-}_n = E^{+}_m$, $\Psi$ será autoestado 
de $H$.

Como, conforme \eq{comutQH}, os geradores $Q$ e $Q^{\dagger}$ comutam com $H$, sabemos que se
$\Psi$ é um autoestado de $H$, então $Q \Psi$ e $Q^{\dagger} \Psi$ também são. Assim, 
examinando a ação de $H$ sobre $Q \Psi$ e $Q^{\dagger} \Psi$, temos:

\begin{align*}
H Q \Psi &= \begin{pmatrix} H_{-} & 0 \\ 0 & H_{+} \end{pmatrix} 
                \begin{pmatrix} 0 & A^{\dagger} \\ 0 & 0\end{pmatrix}
                \begin{pmatrix} \psi^{-}_n \\ \psi^{+}_m   \end{pmatrix} \\
         &= \begin{pmatrix} H_{-} A^{\dagger} \psi^{+}_m \\ 0   \end{pmatrix}
         = \begin{pmatrix} A^{\dagger} A A^{\dagger} \psi^{+}_m \\ 0   \end{pmatrix}
         = \begin{pmatrix} A^{\dagger} H_{+} \psi^{+}_m \\ 0   \end{pmatrix}
         = E^{+}_m Q \Psi
\end{align*}
e
\begin{align*}
H Q^{\dagger} \Psi &= \begin{pmatrix} H_{-} & 0 \\ 0 & H_{+} \end{pmatrix} 
                \begin{pmatrix} 0 & 0 \\ A & 0\end{pmatrix}
                \begin{pmatrix} \psi^{-}_n \\ \psi^{+}_m   \end{pmatrix} \\
         &= \begin{pmatrix} 0 \\ H_{+} A \psi^{-}_n \end{pmatrix}
         = \begin{pmatrix} 0 \\ A A^{\dagger} A \psi^{-}_n   \end{pmatrix}
         = \begin{pmatrix} 0 \\ A H_{-} \psi^{-}_n   \end{pmatrix}
         = E^{-}_n Q^{\dagger} \Psi
\end{align*}
ou seja, $Q \Psi$ e $Q^{\dagger} \Psi$ são, respectivamente, autoestados degenerados de $H$ 
com autovalores $E^{+}_m = E^{-}_n$.

Precisamos ainda ter um estado fundamental $\Psi_0$ de tal forma que, digamos, 
$Q^{\dagger} \Psi_0 = 0$. Para isso escolhemos o hamiltoniano $H_{-} = A^{\dagger}A$ de tal 
forma que o estado fundamental dele tenha energia $E^{-}_0 = 0$. Assim:

\begin{equation*}
H_{-} \psi^{-}_0 = A^{\dagger}A \psi^{-}_0 = 0
\end{equation*}

Uma forma de conseguir isso é impondo que $A \psi^{-}_0 = 0$ (isso é o análogo de agir com o
operador de destruição $a$ sobre o estado fundamental $\ket{0}$ de um OHS). Dessa forma, temos:

\begin{align*}
A \psi^{-}_0 (x) = \left( W(x) + i p  \right) \psi^{-}_0 (x) 
             = \left(W(x) + \frac{d}{dx} \right) \psi^{-}_0 (x) = 0 \\
\Rightarrow \quad \frac{d}{dx} \psi^{-}_0 (x) = - W(x) \psi^{-}_0 (x) 
\end{align*}
e, portanto,

\begin{equation}
\label{estfundMQsusi}
\psi^{-}_0 (x) = \mathcal{N} \exp{\left( - \int ^{x} W(y) dy \right)}
\end{equation}
é o estado fundamental de $H_{-}$. Isso é verdade desde 
que esse estado seja normalizável, ou seja, desde que se possa encontrar uma constante 
$\mathcal{N}$ de modo que $\braket{\psi^{-}_0 | \psi^{-}_0} = 1$. Se não houver essa possibilidade dizemos
que ocorre \emph{quebra da supersimetria} ou que temos \emph{SUSI quebrada}.

Havendo essa possibilidade porém e sendo o estado fundamental aniquilado por $Q^{\dagger}$ (ou $Q$), 
dizemos que o sistema \emph{preserva SUSI} e o estado fundamental de $H$ é dado por:

\begin{equation}
\label{estfundMQsusimatriz}
\Psi_0 = \begin{pmatrix} \psi^{-}_0 \\ 0 \end{pmatrix}
\end{equation}

Os demais autoestados de $H$, com $n=1,2,3, \ldots$, são:

\begin{equation}
\label{estadosMQsusimatriz}
Q^{\dagger} \Psi_n \qquad \text{e} \qquad Q \Psi_n
\end{equation}
onde:

\begin{equation}
\label{estadosMQsusimatriz2}
\Psi_n = \begin{pmatrix} \psi^{-}_n \\ \psi^{+}_{n-1} \end{pmatrix}
\end{equation}

Nessa situação temos um estado fundamental $\Psi_0$ com energia $E^{-}_0 = 0$ e todos os outros 
estados $Q^{\dagger} \Psi_n$ e $Q \Psi_n$ degenerados com energia 
$E^{-}_{n} = E^{+}_{n-1} > 0$. 

Daqui em diante, a menos que mencionemos explicitamente a quebra de SUSI, consideraremos sempre 
a situação de SUSI preservada.

\subsection{Parceiros Supersimétricos}\label{subsection:parceirossusi}

Nesse ponto podemos abrir mão da notação matricial em favor da simplicidade. Isso é feito
notando que os componentes não nulos dos autoestados $Q^{\dagger} \Psi_n$ e $Q \Psi_n$ de $H$ 
são, respectivamente, autoestados de $H_{-}$ e $H_{+}$, ou seja,

\begin{equation*}
\Psi_0 = \begin{pmatrix} \psi^{-}_0 \\ 0 \end{pmatrix} \qquad \text{,} \qquad
Q^{\dagger} \Psi_n = \begin{pmatrix} 0 \\ A \psi^{-}_n \end{pmatrix} \qquad \text{e} \qquad
Q \Psi_n =  \begin{pmatrix} A^{\dagger} \psi^{+}_{n-1} \\ 0 \end{pmatrix} 
\end{equation*}
são autoestados de $H$ com energias, respectivamente, $E^{-}_0 = 0$ e 
$E^{-}_n = E^{+}_{n-1}$, $n \geq 1$. Com isso, uma vez que temos: 

\begin{align*}
& H_{-} \psi^{-}_0 = 0 \\
& H_{+} A \psi^{-}_n = A A^{\dagger} A \psi^{-}_n 
                     = A H_{-} \psi^{-}_n 
                     = E^{-}_n A \psi^{-}_n \\
& H_{-} A^{\dagger} \psi^{+}_{n-1} = A^{\dagger} A A^{\dagger} \psi^{+}_{n-1} 
                                   = A^{\dagger} H_{+} \psi^{+}_{n-1} 
                                   = E^{+}_{n-1} A^{\dagger} \psi^{+}_{n-1}
\end{align*}
os componentes não nulos $A \psi^{-}_n$ e $A^{\dagger} \psi^{+}_{n-1}$ são, respectivamente, 
autoestados de $H_{+}$ e $H_{-}$ com energia $E^{-}_n = E^{+}_{n-1}$ e o estado fundamental 
da teoria passa a ser simplesmente o autoestado $\psi^{-}_0$ de $H_{-}$.

Nesse contexto, os hamiltonianos $H_{\mp}$ que compõem a matriz $H$ tem, conforme a equação de 
Riccati \eq{riccati}, potenciais:

\begin{equation*}
V_{\mp} = W(x)^2 \mp W'(x)
\end{equation*}
que são então chamados \emph{potenciais parceiros supersimétricos}.

Utilizando essa linguagem, dizemos que temos dois hamiltonianos parceiros, $H_{-}$ e $H_{+}$, 
sendo que um deles (conforme nossa escolha, $H_{-}$) tem um estado fundamental $\psi^{-}_0$ 
de energia $E^{-}_0 = 0$ e os demais estados com energias $E^{-}_n>0$, $n=1,2,3,\ldots$.
Enquanto isso o outro hamiltoniano, $H_{+}$, tem níveis de energia 
$E^{+}_{n-1}$, $n=1,2,3,\ldots$ de tal forma que $E^{+}_{n-1} = E^{-}_n$. Além disso, vemos 
que os autoestados de $H_{-}$ e $H_{+}$ se relacionam de modo que 
$\psi^{+}_{n-1} \propto A \psi^{-}_n$ e $\psi^{-}_n \propto A^{\dagger} \psi^{+}_{n-1}$. Levando
em conta a normalização e considerando $n = 1,2,3,\ldots$, podemos sumarizar isso como:

\begin{align}
& E^{-}_0 = 0 \label{naodegE0}\\
& E^{+}_{n-1} = E^{-}_n \label{degE}\\
& \psi^{+}_{n-1} = \left( E^{-}_n \right)^{-\nicefrac{1}{2}} A \psi^{-}_n \label{conectaestados1}\\
& \psi^{-}_n = \left( E^{+}_{n-1} \right)^{-\nicefrac{1}{2}} A^{\dagger} \psi^{+}_{n-1}
\label{conectaestados2}
\end{align}

\begin{figure}[h!]
\begin{center}
\includegraphics[width=0.75\textwidth]{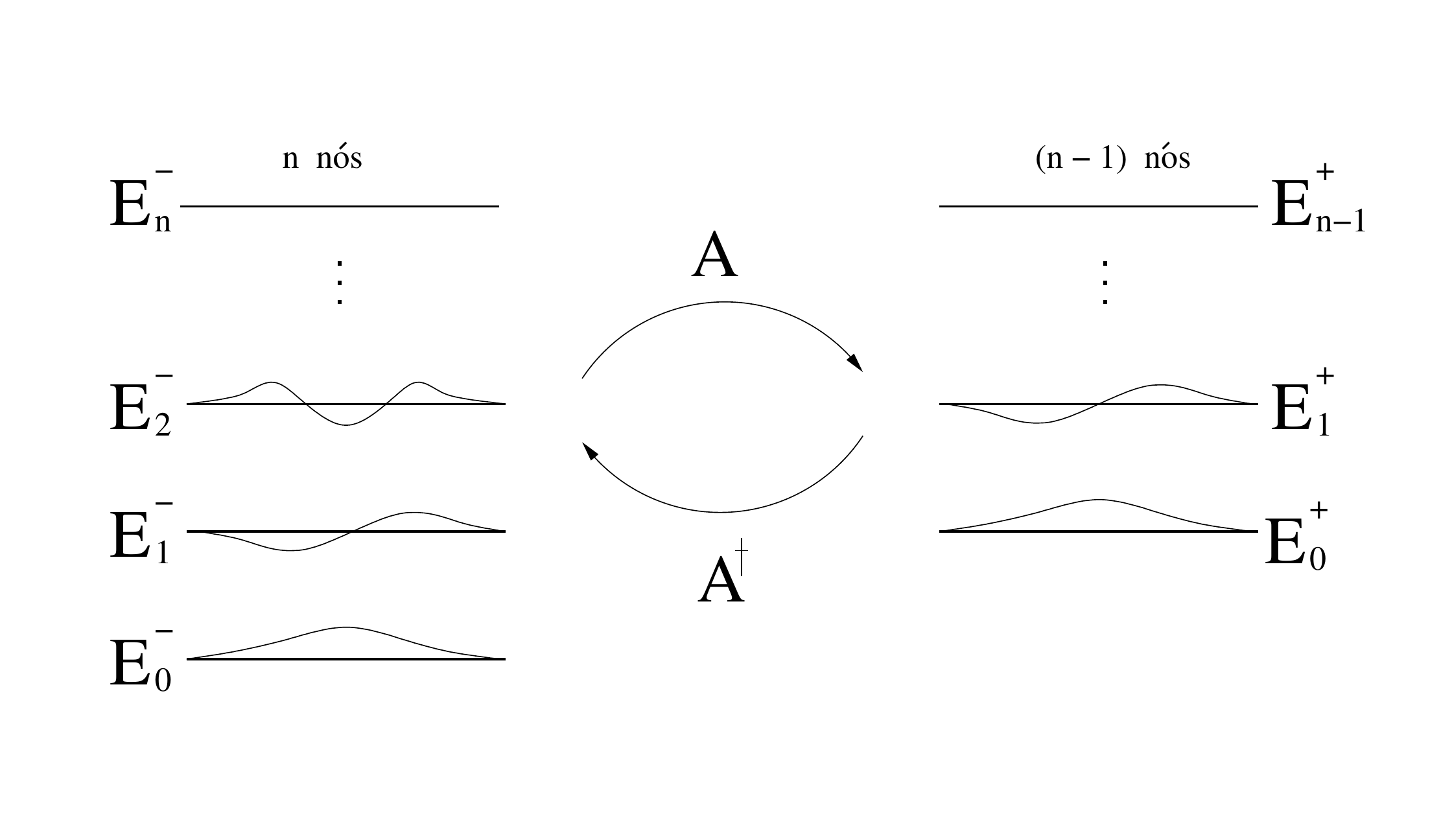}
\end{center}
\caption{\label{fig:acaoAAdag} Ação dos operadores $A$ e $A^{\dagger}$ sobre as autofunções
associadas a um nível de energia $E^{-}_n = E^{+}_{n-1}$.}
\end{figure}

Isso permite não somente relacionar níveis de energia desses dois hamiltonianos, mas também
construir as autofunções correspondentes. Os operadores $A$ e $A^{\dagger}$ permitem construir as
autofunções de $H_{+}$ a partir das autofunções de $H_{-}$ e vice-versa sendo que, ao fazer isso 
(conforme ilustra a figura \ref{fig:acaoAAdag}), esses operadores modificam a 
forma das funções originais destruindo ou criando nós.

\begin{example}[Poço Quadrado Infinito]
  \label{example:pocoinf}
Seguindo \cite{Cooper}, vamos considerar o exemplo do Poço Quadrado Infinito, ou seja, 
do sistema de Mecânica Quântica definido pelo potencial:

\begin{equation}
  V(x) = \left\{ \begin{array}{cl}
    0 & , \quad 0 \leq x \leq L \\
    \infty & , \quad x < 0 \;\; \text{ou} \;\; x>L
  \end{array}
    \right.
  \label{pocoinf}
\end{equation}

As soluções da equação de Schrödinger com o potencial \eq{pocoinf} na região $0 \leq x \leq L$ são:

\begin{align}
  & \psi_n(x) = \left( \frac{2}{L} \right)^{\nicefrac{1}{2}}\sin{\left( \frac{(n+1)\pi x}{L} \right)}
  \label{funcondapocoinf}\\
  & E_n = \frac{\pi ^2}{L^2}(n+1)^2
  \label{energiaspocoinf}
\end{align}
onde podemos ter $n=0,1,2,\ldots$.

Definindo um hamiltoniano $H_{-}$ de poço infinito deslocado de tal modo que esse novo hamiltoniano
tenha energia do estado fundamental zero, ou seja, de tal modo que o potencial de $H_{-}$ seja 
$V_{-}(x) = V(x) - E_0$, teremos como soluções:

\begin{align}
  & \psi^{-}_n(x) = \left( \frac{2}{L} \right)^{\nicefrac{1}{2}}\sin{\left( \frac{(n+1)\pi x}{L} \right)}
  \label{funcondapocoinf1}\\
  & E^{-}_n = \frac{\pi ^2}{L^2}n(n+2)
  \label{energiaspocoinf1}
\end{align}

Sendo o estado fundamental de $H_{-}$ dado por \eq{estfundMQsusi}, podemos derivar
aquela expressão com respeito a $x$ e obter o superpotencial:

\begin{equation}
  W(x) = - \frac{ \psi^{- \prime}_0(x)}{ \psi^{-}_0(x)} = -
  \frac{\pi}{L}\cot{\left(\frac{\pi x}{L}\right)}
  \label{superpotpocoinf}
\end{equation}

A partir da equação de Riccati \eq{riccati} vemos que $V_{-}(x) = W(x)^2 - W'(x)$ e definimos
seu parceiro supersimétrico como $V_{+}(x) = W(x)^2 + W'(x)$. Na região $0 \leq x \leq L$ esses
potenciais são:

\begin{equation}
  \begin{aligned}
   & V_{-}(x) = - \frac{\pi^2}{L^2}\\
   & V_{+}(x) = - \frac{\pi^2}{L^2} + 2 \frac{\pi^2}{L^2} \csc^2{\left( \frac{\pi x}{L} \right)}
  \end{aligned}
  \label{parceirossusipocoinf}
\end{equation}

Utilizando as relações \eq{degE} e \eq{conectaestados1} respectivamente para as soluções 
\eq{energiaspocoinf1} e \eq{funcondapocoinf1} correspondentes ao hamiltoniano $H_{-}$,
encontramos as soluções correspondentes ao hamiltoniano $H_{+}$ que são:

\begin{align}
  & \psi^{+}_{n-1}(x) = \left( \frac{2}{n(n+2) L} \right)^{\nicefrac{1}{2}}
  \left\{ -\cot{\left( \frac{\pi x}{L} \right)} \sin{\frac{(n+1)\pi x}{L}} 
  + (n+1)\cos{\left( \frac{(n+1)\pi x}{L} \right)} \right\}  
  \label{funcondapocoinf2}\\
  & E^{+}_{n-1} = \frac{\pi ^2}{L^2}n(n+2)
  \label{energiaspocoinf2}
\end{align}
onde podemos ter $n=1,2,3,\ldots$.

Esse exemplo ilustra uma possível utilidade da MQ SUSI ao relacionar os estados de sistemas
consideravelmente distintos. Se conhecemos as soluções de um desses sistemas, então podemos
obter facilmente as soluções do outro.
\end{example}

\subsection{Quebra da Supersimetria}\label{subsection:quebra}

Conforme vimos nas seções anteriores, quando a função de onda do estado fundamental de um
hamiltoniano $H_{-}$ pode ser determinada por \eq{estfundMQsusi}, então esse hamiltoniano
$H_{-}$ pode ser fatorizado, isto é, pode ser escrito na forma:

\begin{equation}
  H_{-} = A^{\dagger}A = p^2 + W^2(x) - W'(x)
  \label{HmenosQuebra}
\end{equation}
onde $W(x)$ é o superpotencial e teremos um hamiltoniano parceiro SUSI
$H_{+}$ da forma:

\begin{equation}
  H_{+} = A A^{\dagger} = p^2 + W^2(x) + W'(x)
  \label{HmaisQuebra}
\end{equation}

Vamos considerar agora um problema em que nos é dado inicialmente um superpotencial $W(x)$ e
queremos saber se os hamiltonianos $H_{-}$ e $H_{+}$ definidos conforme \eq{HmenosQuebra} e
\eq{HmaisQuebra} são parceiros SUSI, ou seja, queremos saber se esses hamiltonianos obedecem às
propriedades descritas na seção \ref{subsection:parceirossusi} acima. Para que a SUSI se
manifeste no nosso sistema devemos ter um estado fundamental normalizável de energia zero que
deve satisfazer a equação de Schrödinger ou para $H_{-}$ ou para $H_{+}$. Em outras palavras 
devemos ter ou:

\begin{equation}
  A \psi^{-}_0(x) = 0 \quad \Rightarrow \quad \psi^{-}_0(x) 
                  = \mathcal{N} \exp{\left( - \int^{x} W(y) dy \right)}
  \label{caso1estfundexiste}
\end{equation}
ou
\begin{equation}
  A^{\dagger} \psi^{+}_0(x) = 0 \quad \Rightarrow \quad \psi^{+}_0(x) 
                            = \mathcal{N} \exp{\left( + \int^{x} W(y) dy \right)}
  \label{caso2estfundexiste}
\end{equation}
de tal forma que, para o superpotencial $W(x)$ dado, ou \eq{caso1estfundexiste} ou
\eq{caso2estfundexiste} será o estado fundamental normalizável de energia zero.

Sendo \eq{caso1estfundexiste} o estado fundamental procurado, então a SUSI se manifesta conforme
descrito nas seções acima com $H_{-}$ sendo o hamiltoniano para o qual temos um estado 
fundamental de energia zero. Se, ao invés de \eq{caso1estfundexiste}, \eq{caso2estfundexiste} for
esse estado fundamental podemos simplesmente fazer uma redefinição 
$W(x) \rightarrow \tilde{W}(x) = - W(x)$ e, com isso teremos novamente 
\eq{caso1estfundexiste} sendo o estado fundamental procurado. Desse modo, 
desde que \eq{caso1estfundexiste} ou \eq{caso2estfundexiste} sejam o estado 
fundamental normalizável de energia zero, podemos escolher $H_{-}$ como sendo o hamiltoniano
para o qual esse estado de energia zero é o estado fundamental e a SUSI se manifestará nesse sistema. 

Se, por outro lado, não existirem soluções normalizáveis ou na forma \eq{caso1estfundexiste} 
ou na forma \eq{caso2estfundexiste}, então não poderemos definir $H_{-}$ com um estado
fundamental de energia zero e, nesse caso, dizemos que a SUSI é \emph{quebrada}. Se a SUSI for
quebrada as expressões \eq{naodegE0} e \eq{degE} que relacionam os níveis de energia 
de $H_{-}$ e $H_{+}$, bem como \eq{conectaestados1} e \eq{conectaestados2} que relacionam os
estados não mais valerão e no lugar delas teremos:

\begin{align}
& E^{+}_{n} = E^{-}_n  \, , \qquad n=0,1,2,\ldots  \label{degEQuebra}\\
& \psi^{+}_{n} = \left( E^{-}_n \right)^{-\nicefrac{1}{2}} A \psi^{-}_n \label{conectaestados1Quebra}\\
& \psi^{-}_n = \left( E^{+}_{n} \right)^{-\nicefrac{1}{2}} A^{\dagger} \psi^{+}_{n}
\label{conectaestados2Quebra}
\end{align}
onde todos os níveis de energia de $H_{-}$ e $H_{+}$, inclusive os estados fundamentais estarão
emparelhados e os operadores $A$ e $A^{\dagger}$ não mais modificam o número de nós nas funções
de onda.

\begin{figure}[h!]
  \begin{center}
    \includegraphics[width=0.70\textwidth]{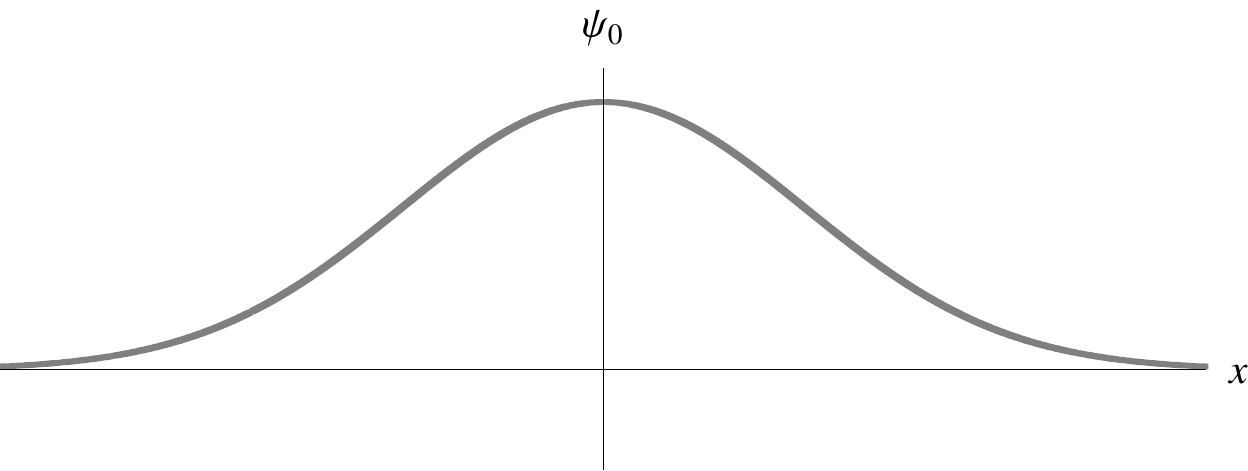}
  \end{center}
  \caption{Forma geral de uma função de onda $\psi_0(x)$ normalizável sem nós.}
  \label{fig:estfundnormnonodes}
\end{figure}

De acordo com \eq{caso1estfundexiste} e \eq{caso2estfundexiste}, o superpotencial pode ser
escrito em termos das funções de onda do estado fundamental de energia zero 
(ou de $H_{-}$ ou de $H_{+}$) como:

\begin{equation}
  W(x) = \mp \frac{ \psi_0^{\mp \prime}(x)}{ \psi_0^{\mp}(x)}
  \label{condicaosuperpotQuebra}
\end{equation}
onde o sinal ``-'' corresponde a escolha de $H_{-}$ como sendo o hamiltoniano com o estado
fundamental de energia zero.

Se a função de onda que aparece em \eq{condicaosuperpotQuebra} for a função de onda do estado 
fundamental de energia zero normalizável e sem nós, por exemplo, da forma da figura
\ref{fig:estfundnormnonodes}, então de acordo com \eq{condicaosuperpotQuebra}, com o sinal de
$\psi_0(x)$ sendo sempre o mesmo e com o sinal de $\psi_0 '(x)$ mudando conforme o sinal de
$x$ muda, vemos que os superpotenciais para os quais a SUSI não é quebrada devem ter sinais
opostos conforme $x$ é positivo ou negativo. De modo mais geral a função de onda do estado
fundamental pode ter outras formas diferentes daquela da figura \ref{fig:estfundnormnonodes},
porém, para que ocorra a manifestação da SUSI ainda é preciso que o superpotencial $W(x)$ tenha
sinais opostos para $x \rightarrow +\infty$ e $x \rightarrow -\infty$, ou em outras palavras,
$W(x)$ deve ter um número ímpar de zeros.

Em particular para a escolha de $H_{-}$ como o hamiltoniano que tem o estado fundamental
normalizável de energia zero, deveremos ter, conforme \eq{condicaosuperpotQuebra}, $W(x)<0$ para
$x \rightarrow -\infty$ e $W(x)>0$ para $x \rightarrow +\infty$.

\subsection{Hierarquia de Hamiltonianos}\label{subsection:hierarquia}

Consideremos um conjunto de hamiltonianos $H_n$, $n=1,2,3,\ldots$ com estados fundamentais 
$\psi^n_0$ de energias $E^n_0$ não necessariamente nulas. Podemos, a partir dos $H_n$, definir 
novos hamiltonianos $\tilde{H}_n = H_n - E^n_0$. Esses novos hamiltonianos $\tilde{H}_n$ tem 
energias dos estados fundamentais nulas e seus autoestados são os mesmos dos hamiltonianos 
$H_n$. Agora vamos supor que possamos escrever esses novos hamiltonianos em formas fatorizadas 
$\tilde{H}_n = A_n^{\dagger} A_n$, ou seja, vamos supor que seja possível encontrar
superpotenciais $W_n (x)$ de modo que os potenciais dos hamiltonianos $\tilde{H}_n$ sejam dados 
por:

\begin{equation}
\label{riccatiVn}
\tilde{V}_n(x) = V_n(x) - E^n_0 = W_n(x)^2 - W_n'(x)
\end{equation}
(que é simplesmente a equação de Riccati, \eq{riccati}). Além disso vamos supor que vale a
equação \eq{estfundMQsusi}, ou seja, vamos supor que possamos escrever os estados fundamentais
$\psi^n_0$ como:

\begin{equation}
\label{estfundMQsusin}
\psi^n_0 (x) = {\cal N}_n \exp{\left( - \int ^{x} W_n(y) dy \right)}
\end{equation}

Tendo determinado os superpotenciais $W_n(x)$, podemos definir, conforme \eq{superescada}, 
os operadores $A_n^{\dagger}$ e $A_n$ como:

\begin{equation}
\begin{aligned}
\label{superescadan}
A_n^{\dagger} =  \left( W_n(x)-ip \right)  \\
A_n = \left( W_n(x) + ip \right)
\end{aligned}
\end{equation}
e com isso, os hamiltonianos originais $H_n$ podem ser escritos como:

\begin{equation}
\label{Hn}
H_n = A_n^{\dagger} A_n + E^n_0
\end{equation}

Consideremos agora que $\tilde{H}_2$ é o parceiro supersimétrico de $\tilde{H}_1$, 
que $\tilde{H}_3$ é o parceiro supersimétrico de $\tilde{H}_2$ e assim por diante, ou seja, 
consideremnos que $\tilde{H}_{n+1}$ é o parceiro SUSI de $\tilde{H}_n$. Sendo assim, se
$\tilde{H}_n = A_n^{\dagger} A_n$, então teremos $\tilde{H}_{n+1} = A_n A_n^{\dagger}$
Alternativamente, podemos dizer que, se $H_n$ é dado por \eq{Hn}, então podemos escrever 
$H_{n+1}$ como:

\begin{equation}
\label{Hnmais1}
H_{n+1} = A_n A_n^{\dagger} + E^n_0
\end{equation}

\begin{figure}[h!]
\begin{center}
\includegraphics[width=0.80\textwidth]{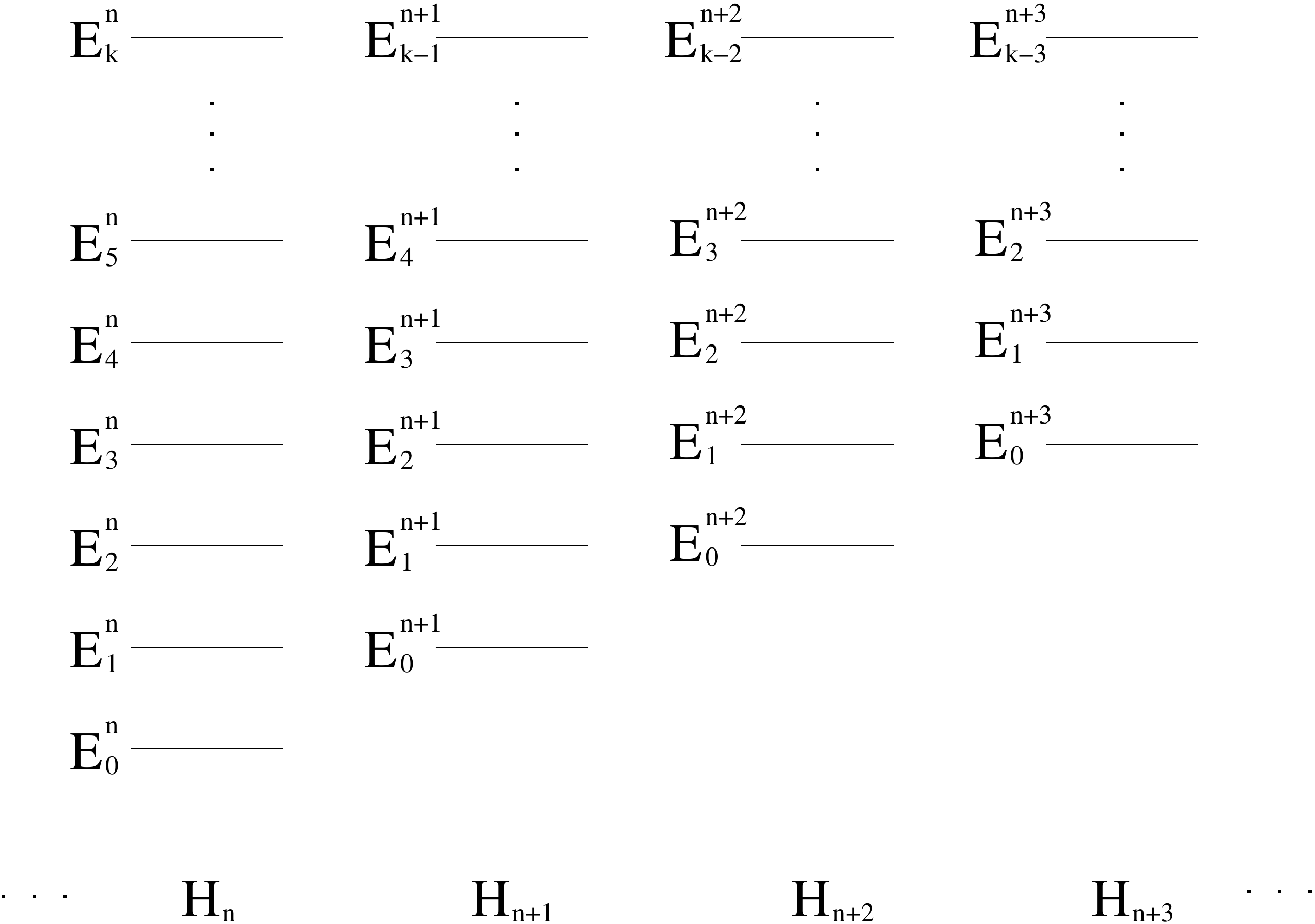}
\end{center}
\caption{\label{fig:hierarquia} Hierarquia de hamiltonianos.}
\end{figure}

Desse modo, sendo $H_n$ e $H_{n+1}$ dados, respectivamente, por \eq{Hn} e \eq{Hnmais1},
sabemos, conforme o que vimos nas seções \ref{subsection:autovalH} e
\ref{subsection:parceirossusi}, que o estado fundamental de $H_n$ terá energia $E^n_0$ e todos
os outros estados terão energias $E^n_k = E^{n+1}_{k-1}$, $k=1,2,3,\ldots$. Na nossa notação 
$E^{b}_{a}$ é a energia do nível $a$ do hamiltoniano $H_{b}$. A figura \ref{fig:hierarquia} 
exemplifica isso para uma sequência de hamiltonianos.

A essa sequência de hamiltonianos, definida dessa forma, damos o nome de 
\emph{hierarquia de hamiltonianos}. 

De acordo com a equação \eq{degE} (e conforme ilustrado na figura \ref{fig:hierarquia}),
sabemos que os níveis de energia se relacionam simplesmente como:

\begin{equation}
\label{degEn}
E^{n+j}_{k-j} = E^n_k 
\end{equation}
onde $j=\pm 1,\pm 2, \pm 3, \ldots$ e $(k-j) \geq 0$. 

\vbox{
Além disso, por meio das equações \eq{conectaestados1} e \eq{conectaestados2}, determinamos a
relação entre os autoestados correspondentes a cada nível de energia como:

\begin{align}
\label{conectaestadosn1}
\psi^{n+j}_{k-j} & = 
\left( 
\prod_{l=1}^{j} \left( E^{n+j-l}_{k-j+l} - E^{n+j-l}_{0} \right)^{-\nicefrac{1}{2}}  
A_{n+j-l}
\right) 
\psi^n_k
\\
\label{conectaestadosn2}
\psi^n_k & = 
\left(
\prod_{l=1}^{j} \left( E^{n+l}_{k+j-l} - E^{n+l}_{0} \right)^{-\nicefrac{1}{2}}
A^{\dagger}_{n+l-1}
\right)
\psi^{n+j}_{k-j}
\end{align}
onde $j>0$ e $(k-j) \geq 0$.
}

A hierarquia de hamiltonianos pode ser construída recursivamente a partir de um hamiltoniano 
$H_n = p^2 + V_n(x)$, supondo que este possa ser escrito na forma \eq{Hn}. Nesse caso o
superpotencial $W_n(x)$ é determinado pela equação de Riccati \eq{riccatiVn}. Uma vez que
tenhamos determinado $W_n(x)$, podemos utilizar \eq{Hnmais1} para determinar $H_{n+1}$, cujo 
potencial é $V_{n+1}$. Conhecendo $V_{n+1}$ podemos utilizar novamente a equação de Riccati 
\eq{riccatiVn} para determinar $W_{n+1}$, e assim por diante. Seguindo esse procedimento,
podemos obter os potenciais (e consequentemente os próprios hamiltonianos) de $j$ posições à
direita na hierarquia como:

\begin{equation}
\label{VcomW}
V_{n+j} = V_n + 2 \frac{d}{dx} \sum_{l=1}^{j} W_{n+l-1}
\end{equation}

Alternativamente, se pudermos determinar o estado fundamental $\psi^n_0$ de $H_n$, podemos
utilizar \eq{estfundMQsusin} para escrever o superpotencial $W_n$ como:

\begin{equation*}
W_n = - \frac{d}{dx} \ln{\left( \psi^n_0 \right)}
\end{equation*}
que substituído na equação \eq{VcomW} resulta em:

\begin{equation}
\label{Vcompsi0}
V_{n+j} = V_n - 2 \frac{d^2}{dx^2}\ln{\left( \prod_{l=1}^{j} \psi^{n+l-1}_0 \right)}
\end{equation}

Essa construção pode ser utilizada para, uma vez conhecidos os $k$ primeiros estados 
de um hamiltoniano $H_n$, determinar os $(k-j)$ primeiros estados de outros $j$ hamiltonianos
$H_{n+j}$ ($j<n$) à direita de $H_n$ na hierarquia de hamiltonianos. Também podemos determinar
$k$ estados excitados acima do nível $(j-1)$ dos hamiltonianos $H_{n-j}$ à esquerda nessa 
mesma sequência. Podemos ainda utilizar a hierarquia de hamiltonianos para, uma vez 
conhecidos os estados fundamentais de $k$ hamiltonianos na sequência, determinar os $k$ 
primeiros estados do primeiro hamiltoniano. 

Para exemplificar isso podemos imaginar, por exemplo, o caso do OHS, cujo superpotencial 
associado é $W(x) \propto x$. Se considerarmos uma hierarquia formada pelos hamiltonianos de
osciladores deslocados, ou seja, de tal forma que o $n$-ésimo hamiltoniano à direita na
sequência seja $H_n = H_{OHS} + n E^{OHS}_0$, então podemos usar o procedimento descrito acima
para construir tantos estados do OHS quantos forem os hamiltonianos na sequência. Isso nada mais
é do que o conhecido método algébrico para obtenção das soluções do OHS o que, como vemos, 
corresponde a uma aplicação a um caso particular do procedimento mais geral que foi descrito 
nesta seção.

\section{Invariância de Forma}\label{section:shapeinvariance}

\subsection{Potenciais Invariantes de Forma}\label{subsection:PIF}

Vamos considerar um par de potenciais $V_n$ e $V_{n+1}$ podendo depender, além da
coordenada $x$, de conjuntos de parâmetros $a_n$ e $a_{n+1}$, respectivamente. Se $a_n$ for uma 
função dos parâmetros $a_{n+1}$, ou seja, se $a_n \equiv a_n (a_{n+1})$, e se, além disso, for 
possível escrever os potenciais $V_n$ e $V_{n+1}$ respeitando uma relação do tipo:

\begin{equation}
  V_{n+1} (x;a_{n+1}) = V_n (x;a_n (a_{n+1})) + R(a_{n+1})
  \label{shapeinvariance}
\end{equation}
onde $R(a_{n+1})$ é independente de $x$, então dizemos que $V_n$ e $V_{n+1}$ são 
\emph{potenciais invariantes de forma} (PIF).

A invariância de forma, juntamente com a hierarquia de hamiltonianos (que foi discutida na 
seção \ref{subsection:hierarquia}), desempenha um papel importante na determinação de soluções
em MQ SUSI. Como exemplo de invariância de forma podemos pensar, novamente, no OHS. Conforme
comentamos no último parágrafo da seção \ref{subsection:hierarquia}, podemos considerar uma
hierarquia de hamiltonianos da forma $H_n = H_{\text{OHS}} + n E^{\text{OHS}}_0$. Para essa 
hierarquia vemos que, tomando qualquer par de hamiltonianos consecutivos na sequência, os potenciais 
correspondentes são, além de parceiros SUSI, invariantes de forma. Tomando um par de
hamiltonianos não consecutivos, embora estes já não sejam parceiros SUSI, ainda assim serão
invariantes de forma. 

O interesse em estudar a invariância de forma é encontrar um método de resolução de problemas 
em Mecânica Quântica análogo ao método algébrico do OHS que se estenda a sistemas cujos potenciais 
sejam invariantes de forma. 


\subsection{Invariância de Forma e a Hierarquia de Hamiltonianos}
\label{susection:shapeinvHierarquia}

Consideremos novamente o par de potenciais parceiros SUSI e PIF $V_n$ e $V_{n+1}$ da seção
\ref{subsection:PIF} acima, respeitando a condição de invariância de forma \eq{shapeinvariance}. 
Esses potenciais correspondem, respectivamente, a hamiltonianos
parceiros $H_n$ e $H_{n+1}$, sendo $H_n$ tal que:

\begin{equation}
  E_0^n (a_n ) = 0 
  \qquad \text{e} \qquad 
  \psi_0^n (x;a_n) = \mathcal{N} \exp{\left[ -\int^x W(y;a_n) dy \right]}
  \label{energiazeroHn}
\end{equation}

Podemos utilizar a condição \eq{shapeinvariance} recursivamente para construir uma sequência de
potenciais da seguinte maneira:

\begin{align*}
& V_n (x;a_n) \\
& V_{n+1} (x;a_{n+1}) =  V_n (x;a_n(a_{n+1})) + R_n (a_{n+1}) \\
& V_{n+2} (x;a_{n+2}) =  V_{n+1} (x;a_{n+1}(a_{n+2})) + R_{n+1} (a_{n+2}) \\
&     \qquad \qquad  =  V_n (x;a_n(a_{n+1}(a_{n+2}))) + R_n (a_{n+1}(a_{n+2})) + R_{n+1} (a_{n+2}) \\
&  \vdots                    \\
& V_{n+j} (x;a_{n+j}) =  V_n (x;a_n(a_{n+1}(\cdots(a_{n+j-1}(a_{n+j})) \cdots )) ) 
             + R_n (a_{n+1}(a_{n+2}(\cdots(a_{n+j-1}(a_{n+j}))\cdots ))) \\
&     \qquad \qquad   + R_{n+1}(a_{n+2}(a_{n+3}(\cdots(a_{n+j-1}(a_{n+j}))\cdots ))) 
                                                       + \cdots + R_{n+j-1}(a_{n+j}) \\
&  \vdots                  
\end{align*}

Nessa sequência, os elementos $(n+j)$ e $(n+j+1)$ (onde $j \geq 1$) são dados, 
respectivamente, por:

\begin{align}
& V_{n+j} (x;a_{n+j}) = V_{n} (x;a_n) + \sum_{k=n}^{n+j-1} R_{k} (a_{k+1}) \label{termo1SIP} \\
& V_{n+j+1} (x;a_{n+j+1}) = V_{n} (x;a_n) + \sum_{k=n}^{n+j} R_{k} (a_{k+1}) 
              = V_{n+1} (x;a_{n+1}) + \sum_{k=n}^{n+j-1} R_{k+1} (a_{k+2}) \label{termo2SIP}
\end{align}
(aqui a dependência dos parâmetros em relação a outros parâmetros foi omitida).

Se, ao utilizarmos a condição de invariância de forma \eq{shapeinvariance} recursivamente 
para construir essa sequência, considerarmos que:

\begin{equation*}
  a_p (a_q) = f (a_q) 
  \qquad \text{e} \qquad 
  R_p (a_q) = R (a_q) 
  \quad , \qquad \qquad \forall p, q
\end{equation*}
então os hamiltonianos correspondentes aos potenciais \eq{termo1SIP} e \eq{termo2SIP} 
serão, respectivamente:

\begin{align}
  H_{n+j}   &= H_n + \sum_{k=n}^{n+j-1} R (a_{k+1}) \label{termo1hierarquiaSIP}    \\
  H_{n+j+1} &= H_{n+1} + \sum_{k=n}^{n+j-1} R (a_{k+1}) \label{termo2hierarquiaSIP}
\end{align}
e, como $H_n$ e $H_{n+1}$ são parceiros SUSI, então $H_{n+j}$ e $H_{n+j+1}$ também são,
ou seja, a sequência de hamiltonianos assim construída forma uma 
\emph{hierarquia de hamiltonianos}. Saber disso permite, utilizando os conhecimentos
apresentados na seção \ref{subsection:hierarquia}, encontrar os valores das energias 
dos diversos níveis, bem como as autofunções correspondentes, de um sistema que 
apresenta invariância de forma.

A partir de \eq{termo1hierarquiaSIP} e sabendo que $E_0^n = 0$, encontramos que a energia do
estado fundamental de $H_{n+j}$ é dada por:

\begin{equation}
  E_0^{n+j} = \sum_{k=n}^{n+j-1} R (a_{k+1}) 
  \label{energiaestfundtermo1SIP}
\end{equation}
e, conforme o que foi visto na seção \ref{subsection:hierarquia}, os valores das 
energias do hamiltoniano $H_n$ se relacionam com os valores das energias dos estados 
fundamentais dos hamiltonianos $H_{n+j}$ à direita na hierarquia de modo que
$E_j^n = E_0^{n+j}$ ($j \geq 1$). Além disso sabemos que $E_0^n = 0$. 
Assim, utlizando \eq{degEn} e substituindo \eq{energiaestfundtermo1SIP}, temos:

\begin{equation}
  E_0^n (a_n) = 0 \qquad \text{e} \qquad E_j^n (a_n) =  \sum_{k=n}^{n+j-1} R (a_{k+1}) 
  \label{espectroSIP}
\end{equation}
que são os valores de todos os níveis de energia do hamiltoniano $H_n$, sendo estes dependentes
dos parâmetros $a_n$.

Os estados de $H_n$ também podem ser determinados, a partir de $\psi_0^n$ (dado em 
\eq{energiazeroHn}), por meio da hierarquia de hamiltonianos. Para fazer isso devemos notar que,
conforme \eq{termo1hierarquiaSIP},
o hamiltoniano $H_{n+j}$ difere de $H_n$ simplismente por uma soma de termos constantes. Essa
soma é, conforme \eq{energiaestfundtermo1SIP}, a energia $E_0^{n+j}$ do estado fundamental de
$H_{n+j}$. Assim, a partir da equação de Schrödinger com o hamiltoniano $H_{n+j}$ e com seu estado
fundamental, temos:

\begin{equation}
  \begin{aligned}
    H_{n+j} \psi_0^{n+j}  = \left( H_n + E_0^{n+j} \right) \psi_0^{n+j} 
                          = E_0^{n+j} \psi_0^{n+j} 
    \quad & \Rightarrow \quad
    H_n \psi_0^{n+j} = 0 \\
&    \Rightarrow \quad 
    \psi_0^{n+j} = \psi_0^n
    \label{relacaoestadosfundamentais}
  \end{aligned}
\end{equation}

Agora, utilizando \eq{conectaestadosn2}, que relaciona estados de uma hierarquia de 
hamiltonianos, com $k=j$ e, de acordo com \eq{relacaoestadosfundamentais}, 
substituindo $\psi_0^{n+j}$ por $\psi_0^n$, podemos escrever:

\begin{equation}
  \psi_j^n = \left( \prod_{l=1}^{j} \left( E_{2j - l}^{n+l} - E_0^{n+l} \right)^{- \nicefrac{1}{2}} 
  A^{\dagger}_{n+l-1} \right) \psi_0^n
  \label{estadosSIP}
\end{equation}
que permite determinar, a partir do estado fundamental $\psi_0^n$, todos os outros estados por
meio da aplicação de operadores $A^{\dagger}$. No caso particular do OHS esses operadores serão
justamente os operadores de criação (ou operadores escada) $a^{\dagger}$ apresentados na seção
\ref{subsection:fatorizacaoOH}, expressão \eq{escada}.

\begin{example}[Problema Radial do Átomo de Hidrogênio]
  \label{example:hidrogenio}
Um exemplo interessante da aplicabilidade dos conceitos de invariância de forma e hierarquia de
hamiltonianos consiste em sua aplicação na resolução da equação radial do problema do átomo de
hidrogênio. Essa possibilidade é estudada, por exemplo, em \cite{Valance}.

Em Mecânica Quântica resolver o problema do átomo de Hidrogênio consiste em resolver a equação 
de Schrödinger em $3$ dimensões em coordenadas esféricas com um potencial coulombiano 
$V_c(r) = -\frac{e^2}{r}$. Para fazer isso empregamos o método de se\-pa\-ra\-ção de variáveis 
supondo soluções da forma $\frac{u(r)}{r} Y(\theta , \varphi)$. A partir disso surgem duas 
equações diferenciais. Uma delas, a qual chamamos \emph{equação angular}, é a equação 
diferencial parcial em $\theta$ e $\varphi$ cujas soluções são os harmônicos esféricos 
$Y_{lm}(\theta , \varphi)$. A outra, a qual chamamos \emph{equação radial}, é uma equação 
diferencial ordinária que tem a mesma forma de uma equação de Schrödinger em $1$ dimensão 
(a coordenada $r$), mas com potencial:

\begin{equation}
  V_r(r) = -\frac{e^2}{r} + \frac{l(l+1)}{r^2}
  \label{potradial}
\end{equation}

Podemos construir uma sequência de hamiltonianos com potenciais invariantes de forma semelhantes
ao potencial $V_r(r)$ dado em \eq{potradial}. Essa sequência invariante de forma é constituida
de potenciais que dependem de um parâmetro $l$ de tal modo que ir de um elemento da sequência
para outro elemento consecutivo consiste em realizar uma translação do tipo $l \rightarrow l+1$.
Escolhendo o potencial do primeiro hamiltoniano da sequência de tal modo que, para $l=0$, a
energia do estado fundamental desse hamiltoniano seja zero, temos:

\begin{equation}
  V_1 (r;l) = -\frac{e^2}{r} + \frac{l(l+1)}{r^2} + \frac{e^4}{4(l+1)^2}
  \label{potradial2}
\end{equation}

Os demais potenciais da sequência devem estar relacionados com o primeiro por meio da soma de
fatores $R(l)$. Se esses fatores forem da forma:

\begin{equation}
  R(l) = \frac{e^4}{4l^2} - \frac{e^4}{4(l+1)^2}
  \label{fatorR}
\end{equation}
então a sequência será:

\begin{align*}
  & V_1 (r;l) = -\frac{e^2}{r} + \frac{l(l+1)}{r^2} + \frac{e^4}{4(l+1)^2} 
  + \frac{e^4}{4(l+1)^2} - \frac{e^4}{4(l+1)^2}
 \\
& V_2 (r;l+1) =  V_1 (r;l+1)) + R(l+1)  \\
&   \qquad \qquad  =  -\frac{e^2}{r} + \frac{(l+1)( (l+1)+1)}{r^2} 
  + \frac{e^4}{4( (l+1)+1)^2} + \frac{e^4}{4(l+1)^2} - \frac{e^4}{4( (l+1)+1)^2}
 \\
& V_3 (r;l+2) =  V_1 (r;l+2) + R(l+1) + R(l+2) \\ 
&     \qquad \qquad  =  -\frac{e^2}{r} + \frac{(l+2)( (l+2)+1)}{r^2} 
  + \frac{e^4}{4( (l+2)+1)^2} + \frac{e^4}{4(l+1)^2} - \frac{e^4}{4( (l+2)+1)^2}
 \\
&  \vdots                    \\
& V_{j+1} (r;l+j) =  V_1 (r;l+j) + \sum_{k=1}^{j} R(l+k) \\ 
&    \qquad \qquad =  -\frac{e^2}{r} + \frac{(l+j)( (l+j)+1)}{r^2} 
  + \frac{e^4}{4( (l+j)+1)^2} + \frac{e^4}{4(l+1)^2} - \frac{e^4}{4( (l+j)+1)^2}
 \\
&  \vdots                  
\end{align*}

Para formar uma hierarquia de hamiltonianos a partir dessa sequência de potenciais, 
devemos determinar superpotenciais que fatorizem cada um dos hamiltonianos 
correspondentes. Esses superpotenciais são:

\begin{equation}
  W_j(r) = \frac{e^2}{2(l+j)} - \frac{(l+j)}{r}
  \label{superpothidrogen}
\end{equation}
de modo que a sequência de potenciais invariantes de forma pode ser redefinida como:

\begin{align*}
  & \tilde{V}_1 =  V_1 - \left[ \frac{e^4}{4(l+1)^2} - \frac{e^4}{4(l+1)^2} \right] 
                = W_1^2 - W'_1
\qquad \qquad \qquad \qquad \\
 & \tilde{V}_2 =  V_2 - \left[ \frac{e^4}{4(l+1)^2} - \frac{e^4}{4( (l+1)+1)^2} \right] 
                = W_2^2 - W'_2
 \qquad \qquad \qquad \qquad \\
&  \vdots                    \\
& \tilde{V}_{j+1} =  V_{j+1} - \left[ \frac{e^4}{4(l+1)^2} - \frac{e^4}{4( (l+j)+1)^2} \right] 
                  = W_{j+1}^2 - W'_{j+1}
 \qquad \qquad \qquad \qquad \\
&  \vdots                  
\end{align*}
que são potenciais dois a dois parceiros SUSI, isto é, são potenciais que correspondem aos
hamiltonianos de uma hierarquia.

De acordo com \eq{energiaestfundtermo1SIP}, a energia do estado fundamental de um 
hamiltoniano na posição $(j+1)$, $j=0,1,2,\ldots$, da sequência é dada por:

\begin{equation}
  E^{j+1}_0 = \sum_{k=1}^{j} R(l+k) 
            = \frac{e^4}{4(l+1)^2} - \frac{e^4}{4( (l+j)+1)^2}
  \label{energiasestfundtermo1SIPhidrogen}
\end{equation}
e, levando em consideração o emparelhamento de níveis de energia da hierarquia, de acordo
com \eq{espectroSIP}, temos:

\begin{equation}
  E^1_j = \frac{e^4}{4(l+1)^2} - \frac{e^4}{4( (l+j)+1)^2}
  \label{energiashierarquiahidrogen}
\end{equation}

O estado fundamental do hamiltoniano $(j+1)$, conforme \eq{energiazeroHn} e passando a
explicitar a dependência em $l$, é dado por:

\begin{equation}
  \psi_{0,l}^{j+1} = \mathcal{N}_{j+1,l} \exp{\left(  - \int^r dr W_{j+1,l}(r) \right)}
  \label{estadosfundhierarquiahidrogendef}
\end{equation}
o que, definindo de acordo com nosso sistema de unidades 
$a \equiv \frac{2}{e^2}$, resulta em:

\begin{equation}
  \psi_{0,l}^{j+1}(r) = \mathcal{N}_{j+1,l} r^{l+j+1} e^{- \frac{r}{a(l+j+1)}}
  \label{estadosfundhierarquiahidrogen}
\end{equation}
com 

\begin{equation*}
\mathcal{N}_{j+1,l} = \left( \left( \frac{a}{2}(l+j+1) \right)^{2l + 2 (j+1) + 1} 
  \Gamma{\left( 2j + 2 (j+1) + 1 \right)} \right)^{-\nicefrac{1}{2}}
\end{equation*}

Comparando \eq{estadosfundhierarquiahidrogen} com as funções de onda radiais do átomo de
hidrogênio, ou seja, com as soluções $u(r)$ da equação de Schrödinger com o potencial 
$V_r(r)$ dado em \eq{potradial} e levando em consideração a forma como essas funções dependem de
$l$ e $j$, temos:

\begin{equation}
  \psi_{0,l}^{j+1}(r) = \psi_{0,l-k}^{j+1+k}(r) = u_{l+j+1,l+j}(r) 
  \label{equivaleestfundradial}
\end{equation}
onde $l,k=0,1,2,\ldots$ e $k \leq l$.

Utilizando agora a expressão \eq{estadosSIP} e escolhendo $j=0$ em
\eq{estadosfundhierarquiahidrogen}, podemos construir, a partir de $\psi_{0,l+j}^{1}(r)$ qualquer
autofunção $\psi_{j,l}^{1}(r)$ de $H_1$:

\begin{equation}
  \psi_{j,l}^1(r) = \left( \prod_{q=1}^{j} \left( E_{2j - q}^{1+q} - E_0^{1+q} \right)^{- \nicefrac{1}{2}} 
  A^{\dagger}_{q,l} \right) \psi_{0,l+j}^1(r)
  \label{estadosSIPhidrogen}
\end{equation}
onde $A^{\dagger}_{q,l} = \left( W_{q,l}(r) - \frac{d}{dr} \right)$ com $W_{q,l}(r)$ dado em \eq{superpothidrogen}.

Escolhendo diferentes valores de $l$, as funções \eq{estadosSIPhidrogen} serão as 
diferentes funções radiais do átomo de hidrogênio. Por exemplo, a função $\psi_{1,l}^{1}(r)$,
obtida por meio de \eq{estadosSIPhidrogen} a partir de $\psi_{0,l+1}^{1}(r)$, corresponde a 
$u_{2+l,l}(r)$, ou seja, $u_{2,0}(r)$ para $l=0$, $u_{3,1}(r)$ para $l=1$, $u_{4,2}(r)$ para
$l=2$ e assim por diante. De modo geral encontraremos a correspondência:

\begin{equation}
  \psi^1_{j,l}(r) = u_{l+j+1,l}(r)
  \label{correspondfuncradiaishidrogen}
\end{equation}

O hamiltoniano $H_1$ difere do hamiltoniano $H_r$ da equação radial apenas por um
fator extra $\frac{1}{a^2(l+1)^2}$, então, para obter os níveis de energia do átomo de
hidrogênio basta subtrair esse fator de \eq{energiashierarquiahidrogen} para obter:

\begin{equation}
  E_{l+j+1} = - \frac{1}{a^2 (l+j+1)^2}
  \label{energiashidrogen}
\end{equation}
que é a energia do nível $(l+j+1)$. 

Seguindo a convenção, podemos definir $n \equiv (l + j + 1)$ e $\kappa \equiv a^{-1}$ 
em \eq{correspondfuncradiaishidrogen} e \eq{energiashidrogen} e obter 
a bem conhecida expressão para o $n$-ésimo nível de energia do átomo de 
hidrogênio, $E_n = - \frac{\kappa^{2}}{n^2}$, correspondendo às autofunções 
$u_{n,l}(r)$, com $l \leq n-1$.  
\end{example}

                                    %

\chapter{Métodos de Aproximação}
\label{chapter:aprox}

\section{Método Variacional} \label{section:aproxvar}

\subsection{O Método Variacional} \label{subsection:var}

Em Mecânica Quântica o \emph{Método Variacional} é um método que permite encontrar aproximações
para a função de onda e para a energia do estado fundamental e de estados excitados do sistema. 

O ponto de partida para o emprego do método é a escolha de uma \emph{função tentativa} $\phi(x)$ 
para fazer o papel de função de onda do estado fundamental do sistema. Embora essa escolha seja 
arbitrária, é interessante notar que é recomendável escolher funções tentativa cuja forma seja
tão próxima quanto possível da forma que se supõe serem as funções de onda reais 
(desconhecidas). Essa escolha pode ser guiada, por exemplo, por características 
que sabemos \emph{a priori} que as funções de onda de determinado tipo de sistema devem ter. Assim a
escolha de funções tentativa com maior ou menor semelhança com as funções de onda reais conduz a
aproximações, respectivamente, melhores ou piores. A função tentativa deve ainda depender de um
ou mais parâmetros $\alpha$ indeterminados que são chamados \emph{parâmetros variacionais}.

O segundo passo consiste na construção de um objeto chamado \emph{funcional da energia}, que 
é definido de forma semelhante ao valor esperado do hamiltoniano do sistema, sendo esse valor 
esperado calculado como se a função de onda do sistema fosse a função tentativa. Para essa 
finalidade a função tentativa deve estar devidamente normalizada, como seria com a função de 
onda real. O funcional da energia assim construído é um funcional dos parâmetros variacionais
sendo denotado por $E[\alpha]$.

Por fim, o método consiste no emprego do princípio variacional tomando como aproximação superior
para o valor da energia o valor de $E[\alpha]$ minimizado com respeito com respeito aos
parâmetros variacionais $\alpha$. Então, encontrando os valores dos parâmetros $\alpha$ que tornam
$E[\alpha]$ mínimo, esse valor mínimo é a aproximação para a energia que o Método Variacional
fornece. A aproximação para a função de onda correspondente é obtida substituindo esses valores dos
parâmetros $\alpha$ na função tentativa.

Na seção seguinte apresentaremos um possível meio de implementação do Método Variacional que
será particularmente útil na resolução de um dos problemas do próximo capítulo.


\subsection{Parâmetros Variacionais como Coeficientes de uma Série de Funções} 
\label{subsection:implementavar}

Nesta seção apresentamos um possível caminho para a resolução de um problema em Mecânica
Quântica por meio do Método Variacional. Mais adiante o Método Variacional assim apresentado 
será especialmente útil na resolução de um problema MQ SUSI envolvendo um oscilador 
anarmônico. Vamos empregar o método partindo de uma função tentativa na forma:

\begin{equation}
\label{functentativa}
\phi (x) = \sum_{j=1}^{m} \alpha_j f_j(x)
\end{equation}
onde $j=1,2,\ldots,m.$ e os coeficientes $\alpha_j \in \mathbbm{C}$ são os parâmetros 
variacionais. Aqui as funções $f_j(x)$ devem ser convenientemente escolhidas podendo levar 
a resultados melhores ou piores dependendo da escolha feita e das características do problema.

A partir da equação de Schrödinger com a função tentativa no lugar da função de onda, definimos 
o funcional da energia como:

\begin{equation}
\label{funcionalenergia}
E[\alpha_1, \alpha_2, \ldots, \alpha_m] = \frac{\braket{\phi|H|\phi}}{\braket{\phi|\phi}}
\end{equation}
onde a presença do denominador do lado direito corresponde a normalizar a função tentativa 
$\phi(x)$.

O Método Variacional diz que encontrando os valores dos parâmetros 
$\alpha_j$, \mbox{$j = 1, \ldots , m.$} que minimizam 
o funcional \eq{funcionalenergia}, esse valor mínimo é uma aproximação para a energia do
estado fundamental no 
sistema descrito pelo hamiltoniano $H$. Além disso, substituindo os
valores de $\alpha_j$ assim encontrados, a função \eq{functentativa} é uma aproximação para a
função de onda desse estado fundamental. A forma como apresentaremos o Método Variacional aqui 
permite determinar não somente uma aproximação para o
estado fundamental, mas também para os $m$ primeiros níveis (lembrando que $m$ é o número de
parâmetros). Vejamos como isso é feito.

Multiplicando os dois lados de \eq{funcionalenergia} por $\braket{\phi|\phi}$, substituindo 
\eq{functentativa} e derivando em relação a $\alpha_k$, temos:

\begin{align*}
E[\alpha_1, \alpha_2, \ldots, \alpha_m] \frac{\partial}{\partial \alpha_k} \sum_{i,j=1}^{m}
\alpha_i^{*} \alpha_j \braket{f_i|f_j} = \frac{\partial}{\partial \alpha_k} \sum_{i,j=1}^{m}
\alpha_i^{*} \alpha_j \braket{f_i|H|f_j}
\end{align*}
onde foi consirerado que $\frac{\partial E}{\partial \alpha_k} = 0$, uma vez que os parâmetros 
$\alpha_k$ correspondem a mínimos de $E$. Além disso devemos notar que as somas duplas
surgem pois as funções $f_j(x)$ não necessariamente formam uma base ortogonal.

Com isso, omitindo a dependência de $E$ com relação aos parâmetros $\alpha_j$ e definindo 
$S_{ij} \equiv \braket{f_i|f_j}$ e $H_{ij} \equiv \braket{f_i|H|f_j}$,
temos:

\begin{align*}
  & E \sum_{i,j=1}^{m} \alpha_i^{*} \delta_{jk} \braket{f_i|f_j} 
  = \sum_{i,j=1}^{m} \alpha_i^{*} \delta_{jk} \braket{f_i|H|f_j} \\
& \Rightarrow \quad
E \sum_{j=1}^{m} \alpha_j^{*} S_{jk} 
= \sum_{j=1}^{m} \alpha_j^{*} H_{jk} \\
& \Rightarrow \quad
\sum_{j=1}^{m} \left( E S_{kj} - H_{kj} \right)
\alpha_j = 0
\end{align*}
que corresponde a um sistema de $m$ equações (uma para cada valor de $k$) e $m$ incógnitas
$\alpha_j$, $j =1,2,\ldots,m.$. Podemos representar esse sistema por um produto de matrizes
simplesmente como:

\begin{equation}
\label{sistmatricial}
M \alpha = 0
\end{equation}
onde $\alpha$ é uma matriz coluna $m \times 1$ cujos elementos são os parâmetros 
$\alpha_j$ e $M$ é uma matriz $m \times m$ com elementos dados por:

\begin{equation}
\label{M}
M_{kj} = \left( E S_{kj} - H_{kj} \right)
\end{equation}

Para que esse sistema de equações tenha solução não trivial devemos exigir que o determinante de
$M$ seja zero:

\begin{equation}
\label{eqE}
\det{M} = 0 
\end{equation}

A equação \eq{eqE} é uma equação de grau $m$ em $E$ e as $m$ soluções dessa equação são, em
ordem crescente, os valores aproximados das energias $E_{j-1}$, $j=1,2,\ldots,m.$ dos $m$ 
primeiros níveis do sistema. 

Uma vez que tenhamos encontrado os valores de $E$ por meio de \eq{eqE}, podemos substituir, por
exemplo, o $n$-ésimo valor $E_n$ novamente no sistema \eq{sistmatricial} e, juntamente com a 
condição de normalização:

\begin{equation}
\label{condnorm}
1 = \sum_{i,j=1}^{m} \alpha_i^{*} \alpha_j S_{ij}
\end{equation}
podemos então determinar os valores dos parâmetros $\alpha^n_j$ (onde o índice $n$ destaca a 
correspondência com o $n$-ésimo nível). Substituindo esses parâmetros $\alpha^n_j$
assim determinados em \eq{functentativa} encontramos uma aproximação para a função de onda
$\psi_n(x)$ associada à energia $E_n$.

Vale destacar que o Método Variacional conforme descrito acima é interessante quando estamos
procurando soluções da equação de Schrödinger que apresentem alguma evidência de terem uma 
forma parecida com \eq{functentativa}. Outras formas de funções tentativa com diferentes tipos
de dependência nos parâmetros são perfeitamente possíveis, porém podem acabar tornando a
resolução complicada. Aumentar o número de parâmetros variacionais também é um meio de melhorar
a precisão do método com a desvantagem de tornar os cálculos mais trabalhosos aumentando o 
esforço computacional.

\vspace{0.5cm}

\begin{example}[Oscilador Harmônico Simples]
  \label{example:OHSvar}
Um exemplo interessante do funcionamento do Método Variacional conforme descrito aqui é 
utilizá-lo para resolver o probelma do OHS. Para o OHS podemos
utilizar o método com a forma de função tentativa dada em
\eq{functentativa} e, escolhendo \mbox{$f_j(x) = x^{j-1} e^{-\frac{1}{2}x^2}$}, encontrar inclusive as 
soluções exatas. Nesse caso os parâmetros $\alpha^n_j$ determinados serão, a menos da 
normalização, os coeficientes dos polinômios de Hermite $\mathcal{H}_n(x^2)$.
\end{example}

\vspace{0.5cm}

\begin{example}[Oscilador Anarmônico do Tipo $x^4$]
  \label{example:OAPvar}
A mesma escolha de $f_j(x)$ feita para o OHS pode ser adotada para um oscilador anar\-mônico com 
$V(x) = \frac{1}{4}x^4$. Nesse caso, utilizando $1$ parâmetro, isto é, fazendo \mbox{$m=1$} em 
\eq{functentativa}, encontramos um valor para a energia do estado fundamental \mbox{$E_0 = 0,6875$}, 
que difere em cerca de $2,9 \%$ do valor numérico dado em \cite{Cooper}, $0,667986$. Com 
$3$ parâmetros, porém, encontramos um valor $E_0 = 0,680159$,
diferindo já por apenas $1,8 \%$. Já com $5$ parâmetros, encontramos $E_0 = 0,668530$, de modo que 
essa diferença cai para $0,08 \%$.

Resolvendo esse mesmo problema para funções tentativa da forma:

\begin{equation*}
\phi(x) = \mathcal{N}_{\beta} \, e^{-\beta x^2}
\end{equation*}
e
\begin{equation*}
  \phi(x) = \mathcal{N}_{\gamma,\rho} \, \exp{\left[ -\frac{1}{2} \left( \frac{x^2}{\rho}
  \right)^{\gamma} \right]}
\end{equation*}
sendo $\beta$, $\gamma$ e $\rho$ os parâmetros variacionais,
encontramos, conforme \cite{Cooper}, diferenças de, respectivamente, $2,0 \%$ e $0,2 \%$, que
embora correspondam a valores mais precisos com um menor número de parâmetros, acabam levando
(no caso de $2$ parâmetros) a um sistema bem mais complicado de se resolver do que
\eq{sistmatricial}.
\end{example}

\vspace{2.0cm}

\section{Teoria de Perturbações Logarítmica} \label{section:aproxlpt}

\subsection{A Teoria de Perturbações Logarítmica}\label{subsection:lpt}

A Teoria de Perturbações Logarítmica é um método de aproximação que determina o estado
fundamental de um sistema em Mecânica Quântica. Embora determinar apenas o estado fundamental
pareça uma desvantagem inicialmente, ao contrário da teoria de perturbações usual da Mecânica
Quântica, que para determinar elementos de certas ordens de correção necessita do conhecimento
de uma base completa de autoestados do caso não perturbado, a teoria aqui apresentada é um
procedimento recursivo que não impõe essa exigência. Além disso, fazendo uso da hierarquia de
hamiltonianos, é possível, conforme descrito na seção \ref{subsection:hierarquia} do capítulo
\ref{chapter:mqsusi}, converter o problema de determinar vários estados excitados em vários 
problemas de determinar estados fundamentais e, desse modo, podemos remover a aparente 
limitação do método.

Vamos considerar a equação de Schrödinger para o estado fundamental de um sistema descrito 
pelo potencial $V(x)$.

\begin{equation}
  p^2 \psi_0(x) + V(x)\psi_0(x) = E_0 \psi_0(x)
  \label{schrestfundlpt}
\end{equation}

Suponhamos agora que o potencial $V(x)$ dependa de algum parâmetro $\delta$ de modo que 
$V(x)$ possa ser escrito como uma expansão em série de potências de $\delta$, ou seja,
suponhamos que:

\begin{equation}
  V(x) = \sum_{n=0}^{\infty} V_n(x) \delta^n
  \label{expandeV}
\end{equation}

Vamos considerar ainda que a energia do estado fundamental $E_0$ também possa ser 
escrita na forma de uma série assim como o potencial $V(x)$. Essa série será então
\footnote{Vamos usar a letra $B$ ao invés de $E$ para representar os coeficiente da série de 
modo a evitar confusão, por exemplo, entre $E_0$, que é a energia do estado fundamental, e 
$B_0$, que é o coeficiente de ordem zero da série (ou 
\emph{correção de ordem zero à energia})}:

\begin{equation}
  E_0 = \sum_{n=0}^{\infty} B_n \delta^n
  \label{expandeE}
\end{equation}

Da mesma forma que fizemos na seção \ref{subsection:hierarquia} do capítulo
\ref{chapter:mqsusi}, podemos considerar um hamiltoniano $\tilde{H}$ com energia 
do estado fundamental $\tilde{E}_0 = 0$ obtido a partir do hamiltoniano original do 
nosso problema por meio da subtração da energia $E_0$ do estado fundamental, ou seja, 
tal que:

\begin{equation*}
  \tilde{V}(x) = V(x) - E_0
\end{equation*}
onde $\tilde{V}(x)$ é o potencial associado ao hamiltoniano $\tilde{H}$. Com isso, supondo que o
hamiltoniano $\tilde{H}$ seja fatorizável, de modo a permitir a definição de um superpotencial
$W(x)$ \footnote{O superpotencial $W(x)$ não precisa ser (e em geral não é) conhecido 
\emph{a priori}. Mesmo que não seja possível encontrar um $W(x)$ que torne $\tilde{H}$ 
fatorizável, vamos supor que $W(x)$ existe.}
e considerando que seja possível encontrar um estado fundamental normalizável a partir de 
$W(x)$, então, conforme \eq{estfundMQsusi}, a função de onda desse estado
fundamental será:

\begin{equation}
\psi_0(x) = \mathcal{N} \exp{\left( - \int ^{x} W(y) dy \right)}
  \label{estfundMQsusilpt}
\end{equation}
e substituindo \eq{estfundMQsusilpt} na equação de Schrödinger \eq{schrestfundlpt} e
rearranjando devidamente os termos, chegamos a:

\begin{equation}
  V(x) - E_0 = W(x)^2 - W'(x)
  \label{riccatilpt}
\end{equation}
que é simplesmente a equação de Riccati \eq{riccati}. 

A equação de Riccati \eq{riccatilpt} que surge como consequência de substituir 
\eq{estfundMQsusilpt} na equação de Schrödinger pode ser entendida como uma equação de 
Schrödinger transformada que surge da escolha de trabalhar com a quantidade 
$\ln{\psi_0}$ em lugar de $\psi_0$. Essa nova quantidade $\ln{\psi_0}$ é um 
\emph{logaritmo} e está relacionada com o superpotencial $W(x)$ por meio de
\eq{estfundMQsusilpt}. Esse é o motivo do nome 
``teoria de perturbações \emph{logarítmica}''. Alguns autores, entretanto, chamam esse 
método de aproximação  simplesmente de ``\emph{expansão $\delta$}'', o que é uma 
nomenclatura talvez um pouco genérica demais.

Por fim, consideramos que o superpotencial $W(x)$ também tem uma expansão em série do tipo:

\begin{equation}
  W(x) = \sum_{n=0}^{\infty} W_n(x) \delta^n
  \label{expandeW}
\end{equation}
e, ainda, que cada $W_n(x)$ satisfaz a condição $W_n(0) = 0$.

Substituindo \eq{expandeV}, \eq{expandeE} e \eq{expandeW} na equação \eq{riccatilpt}, 
temos:

\begin{equation}
  \sum_{n=0}^{\infty} \left[ V_n(x) - B_n \right] \delta^n 
    = \sum_{n,m=0}^{\infty} W_n(x)W_m(x)\delta^{n+m} 
      - \sum_{n=0}^{\infty} W' _n (x) \delta^n
  \label{expandericcatilpt}
\end{equation}

Colecionando um a um os termos correspondentes a potências iguais de $\delta$, a equação
\eq{expandericcatilpt} fornece:

\begin{align*}
& V_0(x) - B_0 = W_0(x)^2 - W' _0(x)                                     &  n=0    \\
& V_1(x) - B_1 = 2 W_0(x) W_1(x) - W' _1(x)                              &  n=1    \\
& V_2(x) - B_2 = 2 W_0(x) W_2(x) - W' _2(x) + W_1(x)^2                   &  n=2    \\
& V_3(x) - B_3 = 2 W_0(x) W_3(x) - W' _3(x) + 2 W_1(x) W_2(x)            &  n=3    \\
& V_4(x) - B_4 = 2 W_0(x) W_4(x) - W' _4(x) + W_2(x)^2 + 2 W_1(x) W_3(x) &  n=4    \\
&  \vdots                                                                &  \vdots  
\end{align*}
onde os diferentes valores de $n$ são as diversas \emph{ordens de perturbação}. Podemos
sumarizar as equações para as diferentes ordens de perturbação como:

\begin{align}
& V_0(x) - B_0 = W_0(x)^2 - W' _0(x)        \; , & 
\qquad \qquad \qquad \qquad \qquad  n=0 \qquad \quad \label{ordem0lpt} \\
& V_1(x) - B_1 = 2 W_0(x) W_1(x) - W' _1(x) \; , & 
\qquad \qquad \qquad \qquad \qquad  n=1 \qquad \quad \label{ordem1lpt}
\end{align}
e
\begin{equation}
  V_n(x) - B_n = 2 W_0(x) W_n(x) - W' _n(x) 
           + \sum_{k=1}^{n-1} W_k(x) W_{n-k}(x) \, , \quad n=2,3,4,\ldots
           \label{ordemnlpt}
\end{equation}

As equações para as diferentes ordens de perturbação podem ser resolvidas recursivamente para
determinar por meio de \eq{expandeE} o valor da energia $E_0$ do estado fundamental do sistema. A
seguir veremos como isso é feito.

\subsubsection{Ordem Zero}
A equação de ordem zero \eq{ordem0lpt}, é simplesmente uma equação de Riccati. Resolvê-la para
$W_0(x)$ e $B_0$ corresponde a encontrar o valor da energia $B_0$ do estado fundamental de um
hamiltoniano $H_0$ cujo potencial é $V_0(x)$ e, em seguida, fatorizar um hamiltoniano 
$\tilde{H}_0 = H_0 - B_0$, encontrando o superpotencial $W_0(x)$. Esse superpotencial $W_0(x)$ pode
ser usado na definição da função de onda do estado fundamental de $H_0$ (ou de $\tilde{H}_0$),
que, de acordo com \eq{estfundMQsusi} será:

\begin{equation}
  \varphi_0(x) = \mathcal{N} e^{ - \int ^{x} W_0(y) dy }
  \label{ordem0estfundlpt}
\end{equation}
onde $\mathcal{N}$ é o fator de normalização correspondente.

Aqui $B_0$ e $\varphi_0(x)$ são, respectivamente, as correções de ordem zero para a energia e para 
a função de onda do estado fundamental do sistema. Encarando esses dois objetos como
\emph{correções}, dizemos então que a energia e a função de onda do estado fundamental do
sistema em \emph{ordem zero de aproximação} são, respectivamente:

\begin{equation}
  E_0 = B_0 
  \qquad \text{e} \qquad 
  \psi_0(x) = \varphi_0(x) = \mathcal{N} e^{ - \int ^{x} W_0(y) dy }
  \label{energiaeestfundordem0aprox}
\end{equation}
que, se pensarmos em $\delta$ como a constante de acoplamento de uma ``perturbação'' ao potencial
$V_0(x)$, corresponderia à solução do caso ``não perturbado''.

\subsubsection{Ordem 1}
Multiplicando os dois lados da equação \eq{ordem1lpt} por $-|\varphi_0(x)|^2$, sendo 
$\varphi_0(x)$ dado por \eq{ordem0estfundlpt}, temos:

\begin{equation}
  B_1 |\varphi_0(x)|^2 - V_1(x)|\varphi_0(x)|^2 = \frac{d}{dx}\left(W_1(x) |\varphi_0(x)|^2 \right)
  \label{ordem1lptcomfatordeconv}
\end{equation}

Considerando que $\varphi_0(x)$ é de quadrado integrável e está normalizada, ou seja, sabendo que:

\begin{equation*}
  \lim_{x \to \pm \infty} |\varphi_0(x)|^2 = 0 
  \qquad \text{e} \qquad
  \int_{-\infty}^{+\infty} dx |\varphi_0(x)|^2 = 1 
\end{equation*}
e integrando a equação \eq{ordem1lptcomfatordeconv} sobre todo o eixo encontramos:

\begin{equation}
  B_1 = \braket{\varphi_0|V_1(x)|\varphi_0}
  \label{ordem1lptenergia}
\end{equation}
que é a correção de primeira ordem para a energia do estado fundamental do sistema.

Para determinar o coeficiente $W_1(x)$ da expansão \eq{expandeW} de $W(x)$, uma vez tendo
determinado $B_1$ por meio de \eq{ordem1lptenergia}, fazemos uso da
condição $W_n(0) = 0$ de modo que, integrando a equação \eq{ordem1lptcomfatordeconv},
encontramos:

\begin{equation}
  W_1(x) = |\varphi_0(x)|^{-2} \int_{0}^{x} dy |\varphi_0(y)|^2 \left[ B_1 - V_1(y) \right]  
  \label{ordem1lptW}
\end{equation}
e com isso, a energia e a função de onda do estado fundamental do sistema 
em \emph{primeira ordem de aproximação} serão, respectivamente:

\begin{equation}
    E_0 = B_0 + \delta B_1 = B_0 + \delta \braket{\varphi_0|V_1(x)|\varphi_0} \\
    \label{energiaordem1aprox}
\end{equation}
e
\begin{equation}
    \psi_0(x) = e^{ - \int ^x dy \left[ W_0(y) + \delta W_1(y) \right] }  
             = e^{ - \int ^x dy W_0(y) } 
                                \left[ 1 - \delta \int ^x dy W_1(y) \right]
    \label{estfundordem1aprox}
\end{equation}

\subsubsection{Ordem n}
Para encontrar a energia e a função de onda em $n$-ésima ordem de aproximação, com
$n \geq 2$, devemos seguir, ordem a ordem, um procedimento idêntico ao do caso de 
primeira ordem. A cada ordem para a qual seguimos esse procedimento, devemos encontrar a
correção à energia e o coeficiente da expansão de $W(x)$ correspondente a essa ordem. Assim ao
chegar a uma ordem $n \geq 2$ qualquer, devemos ter em mãos os resultados de todas as ordens
anteriores.

De maneira idêntica àquela em que chegamos à correção de primeira ordem à energia
\eq{ordem1lptenergia}, mas utilizando \eq{ordemnlpt} ao invés de \eq{ordem1lpt}, chegamos a:

\begin{equation}
  B_n = \braket{\varphi_0|\left[ V_n(x) - \sum_{k=1}^{n-1} W_k(x) W_{n-k}(x) \right]|\varphi_0}
  \label{ordemnlptenergia}
\end{equation}
que é a correção de ordem $n$ à energia do estado fundamental do sistema e, da mesma forma para
$W_n(x)$:

\begin{equation}
  W_n(x) =  |\varphi_0(x)|^{-2} \int_{0}^{x} dy |\varphi_0(y)|^2 
  \left[ B_n - V_n(y) +  \sum_{k=1}^{n-1} W_k(y) W_{n-k}(y) \right]  
  \label{ordemnlptW}
\end{equation}

Assim, a energia e a função de onda do estado fundamental do sistema em 
\emph{$n$-ésima ordem de aproximação} serão, respectivamente:

\begin{equation}
    E_0 = B_0 + \delta B_1 + \delta^2 B_2 + \ldots + \delta^n B_n 
    \label{energiaordemnaprox}
\end{equation}
e
\begin{equation}
    \psi_0(x) = e^{ - \int ^x dy \left[ W_0(y) + \delta W_1(y) + 
                     \delta^2 W_2(x) + \ldots + \delta^n W_n(x) \right] }  
    \label{estfundordemnaprox}
\end{equation}


\subsection{Reparametrização de Potenciais}\label{subsection:lptpotconv}

%

Conforme a seção \ref{subsection:lpt}, um dos primeiros passos para aplicar a teoria de
perturbações logarítmica é escrever o potencial $V(x)$ na forma \eq{expandeV}, ou seja, na forma
de uma série de potências em $\delta$, sendo $\delta$ algum parâmetro do qual depende o
potencial. Nem sempre, porém, o potencial depende de algum parâmetro da forma conveniente para a
aplicação do método. Uma possibilidade de escapar disso é trocar o potencial original $V(x)$ por
um novo potencial $V(x;\delta)$ que depende de $\delta$ de modo conveniente e de modo que, para 
um certo $\delta = \delta_1$, tenhamos $V(x;\delta_1) = V(x)$. Sendo assim, o potencial 
$V(x;\delta)$ é chamado de \emph{potencial reparametrizado}.

Além da condição $V(x;\delta_1) = V(x)$, há um outro fator a ser considerado ao se determinar a forma, 
isto é, a dependência em $\delta$ de um potencial reparametrizado. Esse fator é a forma do
potencial reparametrizado quando $\delta = \delta_0$, onde $\delta_0$ é o valor em torno do 
qual as expansões em série de potências estão sendo feitas. No caso da seção 
\ref{subsection:lpt} as expansões foram todas tomadas em torno de $\delta_0 = 0$, mas
poderíamos, é claro, ter considerado outro valor de $\delta_0$. Se o potencial reparametrizado
for tal que $V(x;\delta_0) = V_0(x)$, onde $V_0(x)$ seja um potencial para o qual conhecemos bem
a solução da equação de Schrödinger para o estado fundamental, então a resolução da equação de 
ordem zero \eq{ordem0lpt} pode se tornar muito mais simples.

Na aplicação da Teoria de Perturbações Logarítmica sempre efetuamos os cálculos tratando
$\delta$ como um parâmetro pequeno (tipicamente $\delta \ll 1$). Isso, porém, não é
necessariamente verdade. Conforme a forma da reparametrização, por exemplo, o valor de
$\delta = \delta_1$ para o qual $V(x;\delta_1) = V(x)$ pode levar as séries a divergirem. Um
procedimento frequentemente sugerido \cite{Cooper} \cite{CooperLPT} \cite{BenderLPT1}
\cite{BenderLPT2} \cite{BenderBook} para contornar isso consiste em substituir as séries
divergentes por aproximantes de Padé e tomar esses aproximantes como resultado.

\begin{example}[Oscilador Anarmônico do Tipo $x^4$]
  \label{example:OAPlpt}
Um exemplo de aplicação da Teoria de Perturbações Logarítmica empregando a reparametrização do
potencial é apresentado em \cite{CooperLPT} e depois reapresentado no livro \cite{Cooper} pelo
mesmo autor. Nesse exemplo é resolvido o problema do potencial de oscilador anarmônico
$V(x) = \frac{1}{4} x^4$ que, sendo reparametrizado como:

\begin{equation}
  V(x;\delta) = \left[ \left( \frac{1}{4} \right)^{\nicefrac{1}{3}} \right]^{2 + \delta} \left( x^{2} \right)^{1+\delta}
  \label{x4lptreparametrizado}
\end{equation}
fornece, por meio do método da Teoria de Perturbações Logarítmica em primeira ordem de
aproximação, o seguinte resultado para a energia do estado fundamental:

\begin{equation}
  E_0 = \left( \frac{1}{4} \right)^{\nicefrac{1}{3}} \left[ 1 + \frac{1}{2} \psi(\nicefrac{3}{2}) \right]
  \label{estfundx4lpt}
\end{equation}
onde $\psi(z) \equiv \frac{\Gamma'(z)}{\Gamma(z)}$ é a função digama.

Substituindo o valor da função digama em \eq{estfundx4lpt} para obter a energia do estado 
fundamental, encontramos $E_0 = 0,6415$ que difere em cerca de $4 \%$ 
do resultado numérico dado em \cite{Cooper}. Com isso, comparando esse resultado 
com aquele obtido para o mesmo potencial $\frac{1}{4}x^{4}$ pelo Método Variacional, 
conforme apresentado no exemplo \ref{example:OAPvar} da seção 
\ref{subsection:implementavar}, vemos que, nesse 
caso, a aproximação em primeira ordem na Teoria de Perturbações Logarítmica é pior 
do que aquela encontrada pelo Método Variacional, mesmo quando se aplica este 
método com apenas um único parâmetro. Esperamos, é claro, melhorar esse resultado para maiores
ordens de aproximação.
\end{example}

                                    %

\chapter{Aplicações}
\label{chapter:aplic}

\section{Superpotenciais do Tipo $W(x) = g x^{2n+1}$}

Vamos considerar superpotenciais do tipo:

\begin{equation}
 W(x) = g x^{2n+1}  
  \label{superpotencialpotimpar}
\end{equation}
isto é, na forma de monômios com potências ímpares de $x$. Utilizando a equação de Riccati
\eq{riccati}, sabemos que os potenciais associados a esse tipo de superpotencial são:

\begin{equation}
  V_{\pm}(x) =  W(x)^2 \pm W'(x) = g^2 x^{4n+2} \pm g (2n + 1) x^{2n}
  \label{potenciaisparceirospotimpar}
\end{equation}

O exemplo mais simples de superpotenciais da forma \eq{superpotencialpotimpar} ocorre 
para $n=0$. Nesse caso os potenciais parceiros associados são, conforme
\eq{potenciaisparceirospotimpar}:

\begin{equation}
  V_{\pm}(x) = g^2 x^{2} \pm g
  \label{potenciaisparceirospotimparOHS}
\end{equation}
que são os potenciais de osciladores harmônicos deslocados.

Para o caso de qualquer $n \geq 0$ em superpotenciais como \eq{superpotencialpotimpar}, é 
possível encontrar um estado fundamental normalizável por meio de \eq{estfundMQsusi}. Esse 
estado fundamental normalizado, associado ao sistema definido pelo potencial $V_{-}(x)$, 
é dado por:

\begin{equation}
  \psi^{-}_0(x) = \mathcal{N} e^{- \int ^x dy W(y)} =
  \left( \frac{g (n+1)^{2n+1}}{\Gamma \left( \frac{1}{2(n+1)} \right)^{2(n+1)}} \right)^{\nicefrac{1}{4(n+1)}} 
  e^{- \nicefrac{g (x^2)^{n+1}}{2(n+1)}}
  \label{estfundpotimparnormalizado}
\end{equation}
onde o fator de normalização correspondente foi calculado explicitamente.

Conforme explicado na seção \ref{subsection:quebra} do capítulo \ref{chapter:mqsusi}, esperamos
que para superpotenciais que obedecem à regra $W(x) \lessgtr 0$ para $x \lessgtr 0$ a SUSI se manifeste. 
Este é precisamente o caso dos superpotenciais da forma \eq{superpotencialpotimpar} que são
monômios com potências ímpares de $x$. Ao contrário, para monômios com potências pares de $x$ devemos observar
a quebra da SUSI. Para contornar isso podemos utilizar a função sinal
$\varepsilon(x)$ e estudar superpotenciais da forma $W(x) = g \varepsilon(x) x^{2n}$, que é uma possibilidade 
ainda não explorada na literatura. Faremos isso nas próximas seções para $n=0$ e $n=1$.

\section{Superpotencial $W(x) = g \varepsilon(x)$}

Vamos considerar o superpotencial:

\begin{equation}
\label{superpotEpsilon}
W(x) = g \varepsilon(x)
\end{equation}
onde $g$ é uma constante positiva e $\varepsilon(x) = \theta(x) - \theta(-x)$ é, em termos da
função degrau de Heaviside, a função sinal.

Para esse superpotencial a equação de Riccati \eq{riccati} fornece os seguintes potenciais 
parceiros supersimétricos:

\begin{equation}
\label{riccatiEpsilon}
V_{\mp}(x) = W(x)^2 \mp W'(x) = g^2 \mp 2g \delta(x)
\end{equation}
onde $\delta(x)$ é a função delta de Dirac.

\begin{figure}[ht]
\begin{center}
\includegraphics[width=0.70\textwidth]{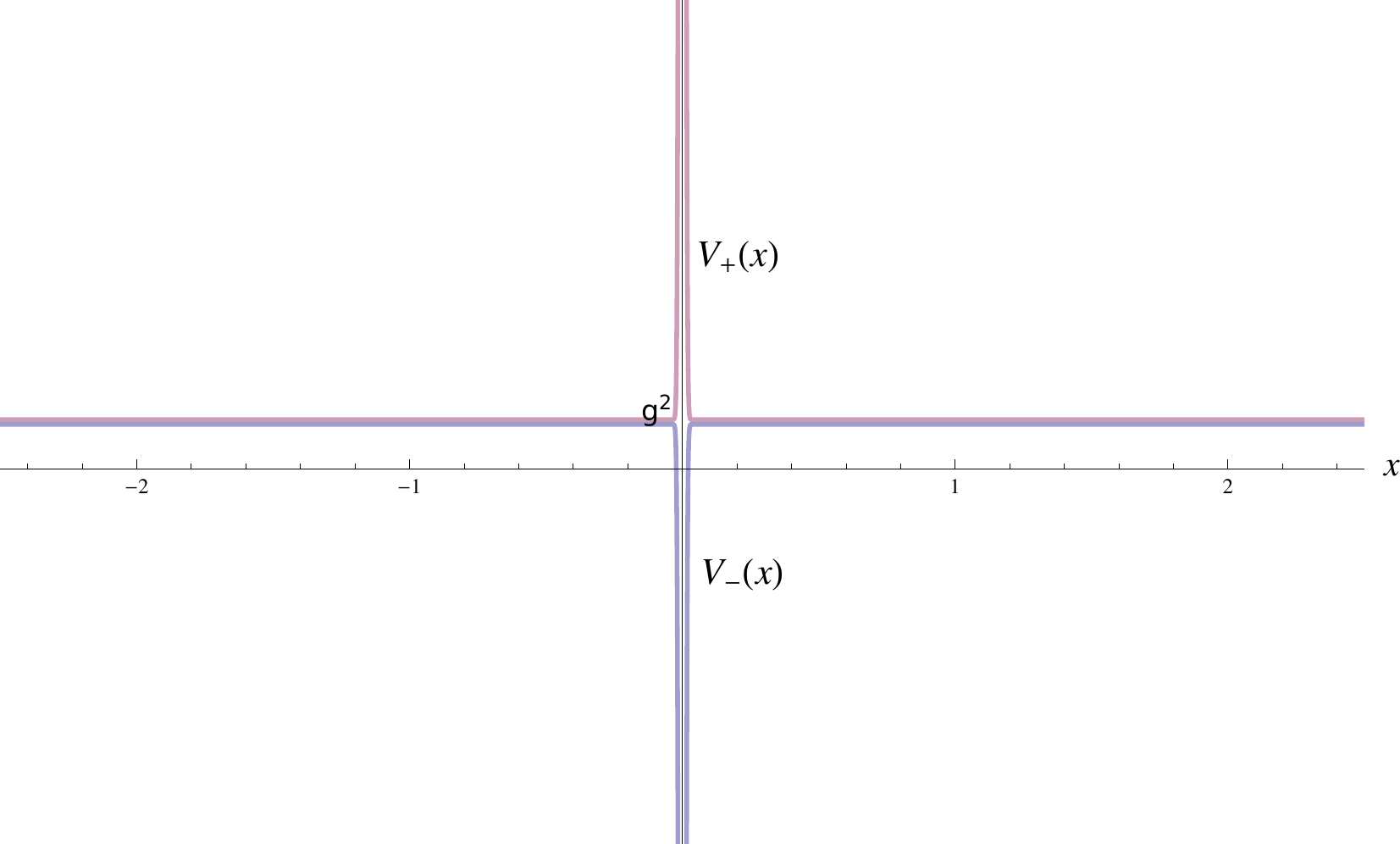}
\caption{\label{fig:potenciaisEpsilon}Potenciais parceiros SUSI associados ao superpotencial $W(x) =
 g \varepsilon(x)$.}
\end{center}
\end{figure}

A forma desses potenciais pode ser vista na figura \ref{fig:potenciaisEpsilon}. Um deles, 
$V_{-}$, tem a forma de um poço delta enquanto o outro, $V_{+}$, é uma barreira delta.

\vbox{
A equação de Schrödinger para os potenciais $V_{\pm}(x)$ é:

\begin{equation}
\label{schrEpsilon1}
\left( p^2 + g^2 \pm 2g \delta(x) \right) \psi^{\pm}(x) = E^{\pm} \psi^{\pm}(x)
\end{equation}
}

Utilizando a representação do operador momento no espaço das posições, $p = -i \frac{d}{dx}$, e
rearranjando os termos de forma conveniente, reescrevemos \eq{schrEpsilon1} como:

\begin{equation}
\label{schrEpsilon1rearranja}
- \psi^{\pm \prime \prime}(x) \pm 2g \delta(x) \psi^{\pm}(x) = \left( E^{\pm} - g^2 \right) \psi^{\pm}(x)
\end{equation}

Esse é o bem conhecido problema do poço delta de Dirac (no caso do potencial ser $V_{-}$) 
e da barreira delta de Dirac (no caso do potencial ser $V_{+}$). Procurando, para o caso
do poço, soluções de \eq{schrEpsilon1rearranja} que sejam estados ligados, ou seja, com 
$\left( E^{-} - g^2 \right) \leq 0$, encontramos: 

\begin{equation}
\begin{aligned}
\label{estadoligEpsilon1}
& E^{-}_0 = 0 \\
& \psi^{-}_0 = \sqrt{g} e^{-g |x|}
\end{aligned}
\end{equation}
que é o estado fundamental com energia zero e também o único estado ligado desse sistema.

Todos os outros estados tem a forma de ondas planas com um espectro contínuo de energia. Tanto
para o caso do poço quanto para o caso da barreira delta, as soluções da equação de Schrödinger
\eq{schrEpsilon1rearranja} que são estados de espalhamento, ou seja, com 
$\left( E^{\pm} - g^2 \right) > 0$ tem a forma de ondas planas sendo dadas por:

\begin{align}
  & \psi^{\pm}_{I}(x) =\mathcal{A}_{\pm}e^{i k x}+\mathcal{B}_{\pm}e^{-i k x} &,\qquad x \leq 0
  \label{solEspalha1} \\
  & \psi^{\pm}_{II}(x)=\mathcal{C}_{\pm}e^{i k x}+\mathcal{D}_{\pm}e^{-i k x} &,\qquad x \geq 0
  \label{solEspalha2}
\end{align}
onde $k = \sqrt{E^{\pm}-g^2}$ e as energias $E^{\pm}$ podem assumir qualquer valor real positivo.

Assim, percebemos que o sistema definido pelo hamiltoniano $H_{-}$ tem um estado fundamental de
energia $E_0=0$ e infinitos estados com energias $E^{-} > 0$ tais que, para cada um desses
valores $E^{-} \neq 0$ existe um valor de energia $E^{+}$ de $H_{+}$ satisfazendo 
$E^{+} = E^{-}$. Essas são as primeiras características da manifestação da SUSI nesse sistema.

Se considerarmos o caso de uma partícula que se move sobre o eixo $x$ da esquerda para a
direita de modo que não possa haver nenhuma partícula sendo refletida do infinito positivo,
então devemos escolher $\mathcal{D}_{\pm} = 0$ em \eq{solEspalha2}. Além disso, impondo a condição de
continuidade da solução da equação de Schrödinger em $x=0$, encontramos que:

\begin{equation}
  \mathcal{C}_{\pm} = \mathcal{A}_{\pm} + \mathcal{B}_{\pm}
  \label{condicaoContinuidade}
\end{equation}
e, integrando a equação de Schrödinger de $-\epsilon$ até $+\epsilon$ e depois tomando o limite
$\epsilon \rightarrow 0$, chegamos a:

\begin{equation}
  \mathcal{C}_{\pm} = \left( 1 + i \frac{2 (\mp g)}{k} \right) \mathcal{A}_{\pm} + 
  \left( -1 + i \frac{2(\mp g)}{k} \right) \mathcal{B}_{\pm}
  \label{condicaoDerivada}
\end{equation}

Solucionando o sistema formado por \eq{condicaoContinuidade} e \eq{condicaoDerivada} para 
$\mathcal{B}_{\pm}$ e $\mathcal{C}_{\pm}$ em função de $\mathcal{A}_{\pm}$ obtemos:

\begin{equation}
  \mathcal{B}_{\pm} =  i \frac{ \frac{(\mp g)}{k}}{ \left( 1 - i \frac{ (\mp g )}{k} \right)}
  \mathcal{A}_{\pm} 
  \qquad \text{e} \qquad 
  \mathcal{C}_{\pm} =  \frac{1}{ \left( 1 - i \frac{(\mp g)}{k} \right)} 
  \mathcal{A}_{\pm} 
  \label{solucaoSistemaEspalha}
\end{equation}

Podemos definir os coeficientes de reflexão e transmissão, que são, respectivamente, as 
probabilidades da partícula incidente ser refletida pelo potencial ou ser transmitida 
através dele, como:

\begin{align}
  & R_{\pm} = \frac{|\mathcal{B}_{\pm}|^2}{|\mathcal{A}_{\pm}|^2} 
            = \frac{\left( \frac{g}{k} \right)^2}{1+ \left( \frac{g}{k} \right)^2} 
            = \frac{g^2}{E}                                          \label{reflete}\\
  & T_{\pm} = \frac{|\mathcal{C}_{\pm}|^2}{|\mathcal{A}_{\pm}|^2} 
            = \frac{1}{1+ \left( \frac{g}{k} \right)^2} 
            = \frac{E - g^2}{E}                                      \label{transmite}
\end{align}

Substituindo \eq{solucaoSistemaEspalha} em \eq{solEspalha1} e \eq{solEspalha2}, temos:

\begin{align}
  & \psi^{\pm}_{I}(x) =\mathcal{A}_{\pm} \left\{  e^{i k x}+ 
  i \frac{ \frac{(\mp g)}{k}}{ \left( 1 - i \frac{ (\mp g )}{k} \right)}   e^{-i k x}  \right\} &,\qquad x \leq 0
  \label{solEspalhaFinal1} \\
  & \psi^{\pm}_{II}(x)= \mathcal{A}_{\pm} \frac{1}{ \left( 1 - i \frac{(\mp g)}{k} \right)} e^{i k x} &,\qquad x \geq 0
  \label{solEspalhaFinal2}
\end{align}

Para verificar que $\psi^{\pm}_{I}(x)$ e $\psi^{\pm}_{II}(x)$ podem ser relacionados,
respectivamente, com $\psi^{\mp}_{I}(x)$ e $\psi^{\mp}_{II}(x)$ por meio de regras como
\eq{conectaestados1} e \eq{conectaestados2}, devemos aplicar sobre essas funções operadores
$A$ e $A^{\dagger}$ e averiguar que isso converte uma função ``+'' em uma ``-'' e vice-versa.
Fazendo isso para, por exemplo, o operador $A$ atuando em $\psi^{-}_{I}(x)$ e desconsiderando
fatores multiplicativos constantes como $\mathcal{A}_{\pm}$, temos:

\begin{align*}
  A \psi^{-}_{I}(x) & \propto \left( g \varepsilon(x) + \frac{d}{dx} \right) \left\{   e^{i k x}+ 
  i \frac{ \frac{g}{k}}{ \left( 1 - i \frac{g}{k} \right)}   e^{-i k x}  \right\} \\ 
                    & \propto  \left\{   e^{i k x}+ 
  i \frac{ \frac{-g}{k}}{ \left( 1 - i \frac{-g}{k} \right)}   e^{-i k x}  \right\} \\ 
  & \propto  \psi^{+}_{I}(x)
\end{align*}
o que equivale a satisfazer \eq{conectaestados1} para $\psi^{-}_{I}$. 

Aplicando $A$ e $A^{\dagger}$ nas demais funções verificamos que as soluções 
\eq{solEspalhaFinal1} e \eq{solEspalhaFinal2} satisfazem relações do tipo 
\eq{conectaestados1} e \eq{conectaestados2}, o que corresponde, juntamente com 
a equiparação de níveis de energia, à manifestação da SUSI nesse sistema.

\section{Superpotencial $W(x) = g \varepsilon(x) x^2$}

\subsection{Buscando Soluções Analíticas da Equação de Schrödinger}
\label{subsection:solanaliticax2}

Vamos considerar agora o superpotencial:

\begin{equation}
\label{superpotEpsilonx2}
W(x) = g \varepsilon(x) x^2
\end{equation}
onde $g$ é uma constante positiva e $\varepsilon(x) = \theta(x) - \theta(-x)$ é, em termos da
função degrau de Heaviside, a função sinal.

Para esse superpotencial a equação de Riccati \eq{riccati} fornece os seguintes potenciais 
parceiros supersimétricos:

\begin{equation}
\label{riccatiEpsilonx2}
V_{\mp}(x) = W(x)^2 \mp W'(x) = g^2 x^4 \mp 2g |x|
\end{equation}

\begin{figure}[ht]
\begin{center}
\includegraphics[width=0.70\textwidth]{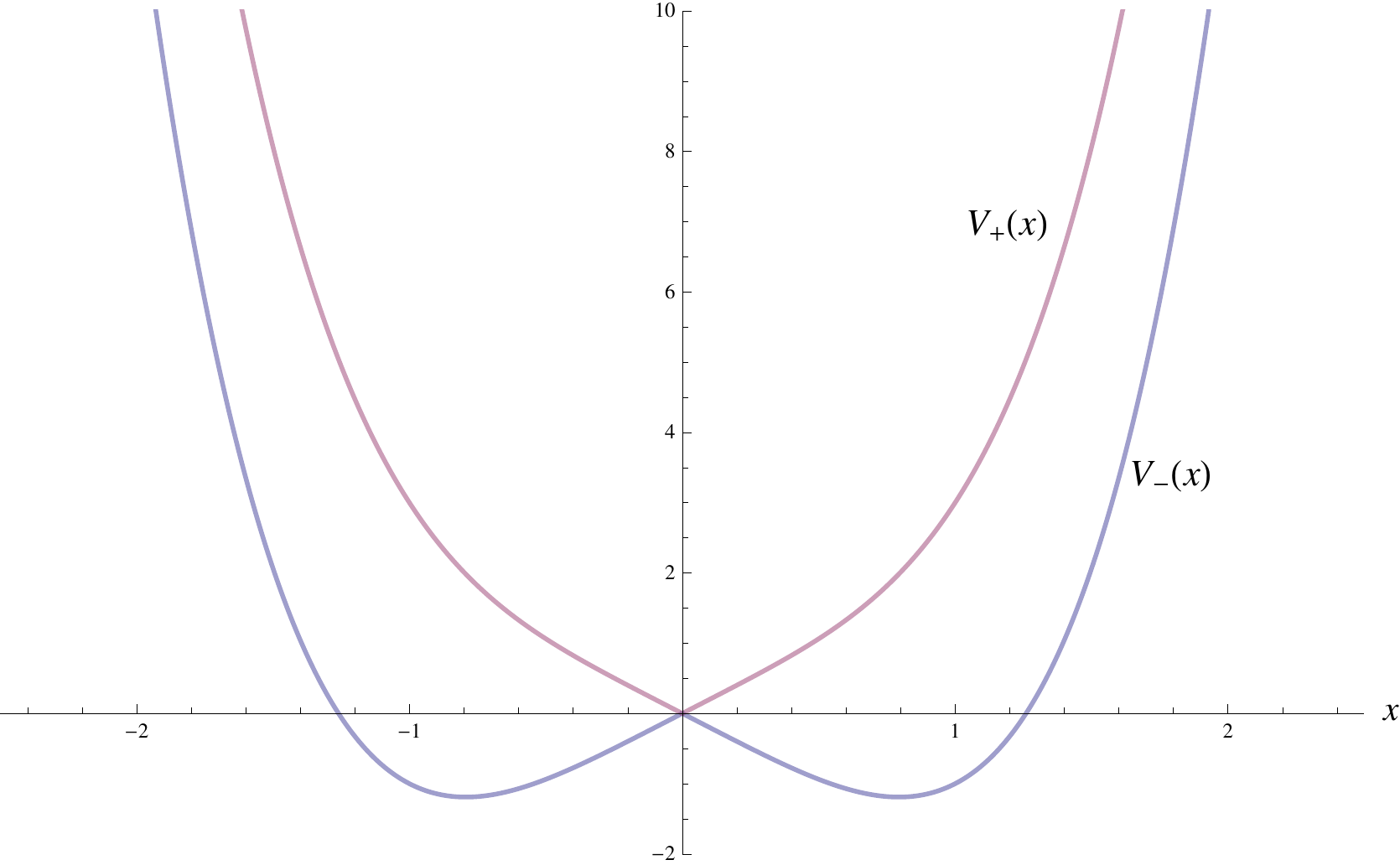}
\caption{\label{fig:potenciaisEpsilonx2}Potenciais parceiros SUSI associados ao superpotencial $W(x) =
 \varepsilon(x) x^2$.}
\end{center}
\end{figure}

A forma desses potenciais pode ser vista na figura \ref{fig:potenciaisEpsilonx2}. Um fato
curioso aqui é que o desenho de $V_{+}$ forma na origem uma continuação suave do desenho de 
$V_{-}$.

A equação de Schrödinger para esses potenciais é:

\begin{equation}
\label{schrEpsilonx2}
\left( p^2 + g^2 x^4 \pm 2g |x| \right) \psi^{\pm}(x) = E^{\pm} \psi^{\pm}(x)
\end{equation}

Do mesmo modo que foi feito em \eq{schrEpsilon1rearranja}, reescrevemos \eq{schrEpsilonx2},
agora para o potencial $V_{-}$, como:

\begin{equation}
\label{schrH1Epsilonx2rearranja}
- \psi''(x) + \left( g^2 x^4 - 2g |x| \right) \psi(x) = E \psi(x)
\end{equation}
onde os índices ``$-$'' foram omitidos para não carregar demais a notação.

Procurando para a equação \eq{schrH1Epsilonx2rearranja}, para o caso $E = 0$, soluções da forma 
\mbox{$\psi(x) = \mathcal{N} e^{\alpha (x)}$}, encontramos que, para 
$\alpha (x) = - \frac{1}{3} g |x|^3$, temos
uma solução exata sem nós, correspondendo ao estado fundamental de energia zero de 
$H_{-}$. Esse é o primeiro passo para observar a manifestação da SUSI nesse sistema. Essa 
solução (já normalizada) é então:

\begin{equation}
  \psi_0 (x) =
  \left( \frac{3}{2} \right)^{\nicefrac{1}{3}} 
  \frac{g^{\nicefrac{1}{6}}}{\Gamma \left( \nicefrac{1}{3} \right)^{\nicefrac{1}{2}}}
   e^{-\nicefrac{g |x|^3}{3}}
  \label{solucaoanaliticaenergiazero}
\end{equation}

Inspirados pelo método analítico de resolução do problema do OHS e pela forma da solução já
encontrada em \eq{solucaoanaliticaenergiazero}, vamos supor que as soluções gerais
sejam da forma\footnote{No caso do OHS, as soluções são consideradas da forma 
$H(x)e^{-\nicefrac{x^2}{2}}$ e a imposição das condições de contorno acaba restringindo 
as funções $H(x)$ a serem os polinômios de Hermite $\mathcal{H}_n(x^2)$.}:

\begin{equation}
\label{assintEpsilonx2}
\psi(x) = F(x) e^{-\nicefrac{g |x|^3}{3}}
\end{equation}

Substituindo \eq{assintEpsilonx2} na equação de Schrödinger \eq{schrH1Epsilonx2rearranja},
temos:

\begin{equation}
\label{schrHeunT}
F''(x) - 2g \varepsilon(x) x^2 F'(x) + E F(x) = 0
\end{equation}

No caso do OHS esse passo levaria à equação de Hermite. No nosso caso levou à equação
\eq{schrHeunT}, que é, para uma escolha particular de parâmetros, a 
\emph{equação triconfluente de Heun}. 

A equação triconfluente de Heun é em geral dada por:

\begin{equation}
\label{HeunT}
y''(z) + \left( - \gamma - 3 z^2 \right) y'(z) + \left( \alpha + \beta z - 3 z \right) y(z) = 0
\end{equation}
que, para $\gamma = 0$, $\alpha = \left( \frac{3}{2g} \right)^{\nicefrac{2}{3}} E$, 
$\beta = 3$ e $z = \left( \frac{2g}{3} \right)^{\nicefrac{1}{3}} x$, se reduz à \eq{schrHeunT}.

Vamos tentar solucionar a equação \eq{schrHeunT} por meio do método de Frobenius. Começamos
supondo que as soluções $F(x)$ podem ser escritas na forma de uma série de potências como:

\begin{equation}
  F(x) = \sum_{j=0}^{\infty} a_j x^j
  \label{solucaoHeunTserie}
\end{equation}

Substituindo $F(x)$ na forma \eq{solucaoHeunTserie} na equação diferencial \eq{schrHeunT}, temos:

\begin{equation*}
  \sum_{j=0}^{\infty} j (j-1) a_j x^{j-2} - 2 g \varepsilon(x) \sum_{j=0}^{\infty} j a_j
  x^{j+1} + E \sum_{j=0}^{\infty} a_j x^j = 0
\end{equation*}
e, redefinindo os índices de soma e rearranjando os termos convenientemente nessa expressão,
chegamos a:

\begin{equation*}
  2 a_2 + E a_0 + \sum_{j=1}^{\infty} \left[ (j+2)(j+1) a_{j+2} 
  - 2 g \varepsilon(x)(j-1)a_{j-1} + E a_j \right] = 0
\end{equation*}
de modo que, dados $a_0$ e $a_1$, a equação acima é satisfeita para coeficientes $a_j$,
$j \geq 2$, dados por:

\begin{align}
  &  a_2 = - \frac{E}{2}a_0 & , \qquad  j = 2 \label{coeficientea2} \\
&  a_j = \frac{2 g \varepsilon(x)(j-3) a_{j-3} - E a_{j-2}}{j(j-1)} & , \qquad  j \geq 3
\label{coeficienteaj}
\end{align}

Assim, a equação diferencial \eq{schrHeunT} é satisfeita por funções $F(x)$ na forma dada em
\eq{solucaoHeunTserie} com coeficientes $a_0$ e $a_1$ determinados por condições de contorno,
$a_2$ dado por \eq{coeficientea2} e $a_j$, $j \geq 3$ dados por \eq{coeficienteaj}. 

Se estivéssemos resolvendo a equação de Hermite para o problema do OHS, teríamos chegado, no 
lugar de \eq{coeficientea2} e \eq{coeficienteaj}, na relação:

\begin{equation*}
  a_{j+2} = \frac{2j+1 - E}{(j+1)(j+2)} a_j
\end{equation*}
e, para construir soluções que fossem funções de onda de quadrado integrável, exigiríamos, naquele 
caso, que a série $H(x)$ (o análogo de $F(x)$) fosse truncada para um certo valor 
$j_{\text{máx}} = n$ tal que tivéssemos $a_{n+2} = 0$ e $a_n$ fosse o último coeficiente não nulo da série. 
Isso podia ser alcançado para $E = 2n+1$, o que determinava os possíveis valores da energia e 
as autofunções correspondentes.

Voltando ao nosso caso, notamos que a relação \eq{coeficienteaj}, que determina 
os coeficientes da série $F(x)$, é uma relação de recorrência de 3 índices e não permite escolher
uma expressão para $E$ que trunque a série em determinado ponto. Essa é a dificuldade que
encontramos ao tentar resolver o problema associado ao potencial $V_{-}(x)$ dado em
\eq{riccatiEpsilonx2} e, para tentar contornar isso, partimos então para tentativas de
encontrar soluções aproximadas.

\subsection{Buscando Soluções Aproximadas pelo Método Variacional}

Vamos empregar o Método Variacional conforme descrito na seção \ref{subsection:implementavar} do
capítulo \ref{chapter:aprox}. Aqui vamos utilizar uma função tentativa da forma
\eq{functentativa} com $f_j(x)$ dado por:

\begin{equation}
\label{tentativax3}
f_j(x) = x^{j-1} e^{-\nicefrac{|x|^3}{3}}
\end{equation}
sendo que estamos, a partir de agora, considerando unidades tais que $g = 1$. Mais tarde 
recolocaremos $g$ nos devidos lugares.

Essa escolha de função tentativa é semelhante àquela que leva às soluções exatas do OHS (ver
exemplo \ref{example:OHSvar} na seção \ref{subsection:implementavar} do capítulo 
\ref{chapter:aprox}). Naquele caso os parâmetros
variacionais encontrados são, a menos da normalização, os coeficientes dos polinômios de Hermite
$\mathcal{H}_n(x^2)$. Ao escolher uma função tentativa com $f_j(x)$ conforme \eq{tentativax3} a
melhor situação possível seria se as soluções da equação triconfluente de Heun \eq{schrHeunT}
que fazem com que \eq{assintEpsilonx2} seja de quadrado integrável fossem algum polinômio 
de grau $m$, $\mathcal{F}_m(x)$. Se fosse assim bastaria considerar um número de parâmetros 
variacionais maior ou igual a $m$ e o Método Variacional forneceria, com a função tentativa 
proposta, as soluções exatas do problema. Nesse caso os parâmetros $\alpha_j$ seriam então, a 
menos da normalização, os coeficientes desses polinômios $\mathcal{F}_m(x)$. 

O que ocorre em geral porém, é que as soluções exatas da equação \eq{schrHeunT} são, conforme
encontrado por meio do método de Frobenius, uma série com infinitos termos e não um
polinômio. Nesse caso o que o Método Variacional fará então será aproximar essa série por um
polinômio de grau $m$. Se utilizarmos um número muito grande de parâmetros variacionais, devemos
encontrar, a menos da normalização, valores de $\alpha_j$ que se aproximam dos valores dos 
coeficientes da série.

A partir da equação de Riccati \eq{riccatiEpsilonx2} com $g=1$ temos:

\begin{equation}
\label{riccatiEpsilonx2reescrita}
V_{\pm}(x) = x^4 \pm 2 |x|
\end{equation}

Uma vez que escolhemos $f_j(x)$ real , os elementos de matriz $S_{kl}$ e $H_{kl}$ serão 
simétricos pela troca dos índices e, conforme \eq{M}, os elementos da matriz $M$ para a 
nossa escolha de função tentativa, serão dados por:

\begin{equation}
\label{Msim}
M_{kl} = \left(E S_{kl} - H_{kl} \right)
\end{equation}

Para os potenciais parceiros \eq{riccatiEpsilonx2reescrita}, os elementos de matriz 
$S_{kl}$ e $H_{kl}$ são dados por:

\begin{align}
  \label{S}  S_{kl} &= \braket{f_k|f_l} = \int_{-\infty}^{+\infty} dx \, e^{-\frac{2}{3}|x|^3} x^{k+l-2} \\
  \label{H}  \left(H_{\pm} \right)_{kl} &= \braket{f_k|H_{\pm}|f_l} = \int_{-\infty}^{+\infty} dx \, e^{-\frac{2}{3}|x|^3}
\left[ -(l-1)(l-2)x^{k+l-4} + 2 (l \pm 1) \varepsilon(x) x^{k+l-1} \right]
\end{align}

No caso de $(k+l)$ ser um número ímpar, as integrais \eq{S} e \eq{H} acima tem integrandos que são
funções ímpares e então, como a integração é sobre um intervalo simétrico, encontramos que 
$S_{kl} = \left( H_{\pm} \right)_{kl} = 0$. Caso contrário, se $(k+l)$ for um número par,
encontramos:

\begin{align}
\label{Spar}  S_{kl} &= \left( \frac{3}{2} \right)^{\frac{k+l-4}{3}} \Gamma \left( \frac{k+l-1}{3} \right) \\
\label{Hpar}  \left(H_{\pm} \right)_{kl} &= - 2 \left( \frac{3}{2} \right)^{\frac{k+l-3}{3}}
\left[ \frac{(l-1)(l-2) - (l \pm 1)(k+l-3)}{(k+l-3)}  \right] \Gamma \left( \frac{k+l}{3} \right)
\end{align}
e com isso determinamos, conforme \eq{Msim} a forma da matriz $M$:

\begin{equation}
\label{Mmatriz}
M_{\pm} = 
\begin{pmatrix}
(M_{\pm})_{11}&0             &(M_{\pm})_{13}&0             &&\ldots&&(M_{\pm})_{1m}\\
0             &(M_{\pm})_{22}&0             &(M_{\pm})_{24}&&\ldots&&(M_{\pm})_{2m}\\
(M_{\pm})_{31}&0             &(M_{\pm})_{33}&0             &&\ldots&&(M_{\pm})_{3m}\\
              &              &              &              &&      &&              \\
\vdots        &\vdots        &\vdots        &\vdots        &&\ddots&&\vdots        \\
              &              &              &              &&      &&              \\
(M_{\pm})_{m1}&(M_{\pm})_{m2}&(M_{\pm})_{m3}&(M_{\pm})_{m4}&&\ldots&&(M_{\pm})_{mm}\\
\end{pmatrix}
\end{equation}

Nessa matriz todos os elementos nas posições $(k,l)$ tais que $(k+l)$ é ímpar são nulos enquanto
aqueles para os quais $(k+l)$ é par são dados por \eq{Msim} com $S_{kl}$ e $H_{kl}$ dados por
\eq{Spar} e \eq{Hpar}. 

Para encontrar os valores das energias devemos resolver a equação \eq{eqE} com a matriz $M$ dada
em \eq{Mmatriz} acima. As tabelas \ref{table:energiasH1Epsilonx2} e \ref{table:energiasH2Epsilonx2} 
mostram alguns resultados encontrados para diferentes números de parâmetros variacionais. Nessas duas
tabelas $m$ é o número de parâmetros variacionais utilizado e os valores de energia foram 
calculados para $g=1$. Para diferentes valores de $g$, porém, obteríamos simplesmente os mesmos 
valores da energia multiplicados por $g^{\nicefrac{2}{3}}$.

\begin{table}[h!]
\caption{\label{table:energiasH1Epsilonx2} Valores de energia associados a $H_{-}$ calculados com diferentes números de
parâmetros variacionais.}
\centering
\small
\begin{tabular}{ccccccccc}
\toprule
$m$  &   $E^{-}_0$   &   $E^{-}_{1}$   &   $E^{-}_{2}$   &   $E^{-}_{3}$   & $E^{-}_{4}$  &   $E^{-}_{5}$   &   $E^{-}_{6}$   &   $E^{-}_{7}$  \\
\midrule
 1  &  0,00000 &          &          &          &          &          &          &         \\
 2  &  0,00000 &  2,04441 &          &          &          &          &          &         \\
 3  &  0,00000 &  2,04441 &  5,76541 &          &          &          &          &         \\
 4  &  0,00000 &  1,97852 &  5,76541 & 10,00191 &          &          &          &         \\
 5  &  0,00000 &  1,97852 &  5,54135 & 10,00191 & 14,94174 &          &          &         \\
 6  &  0,00000 &  1,97115 &  5,54135 &  9,49446 & 14,94174 & 20,37028 &          &         \\
 7  &  0,00000 &  1,97115 &  5,51302 &  9,49446 & 14,06558 & 20,37028 & 26,29953 &         \\
 8  &  0,00000 &  1,96991 &  5,51302 &  9,41370 & 14,06558 & 19,02962 & 26,29953 & 32,64399\\
 9  &  0,00000 &  1,96991 &  5,50842 &  9,41370 & 13,90148 & 19,02962 & 24,43194 & 32,64399 \\
10  &  0,00000 &  1,96963 &  5,50842 &  9,39868 & 13,90148 & 18,73498 & 24,43194 & 30,18755 \\
\bottomrule
\end{tabular}
\end{table}

\begin{table}[h!]
\caption{\label{table:energiasH2Epsilonx2} Valores de energia associados a $H_{+}$ calculados com diferentes números de
parâmetros variacionais.}
\centering
\small
\begin{tabular}{ccccccccc}
\toprule
$m$  &  $\,\,\,\,\,\,\,\,\,\,\,\,\,\,\,\,\,\,\,\,$  & $E^{+}_0$   &   $E^{+}_{1}$   &   $E^{+}_{2}$   &   $E^{+}_{3}$   & $E^{+}_{4}$  &   $E^{+}_{5}$   &   $E^{+}_{6}$  \\ 
\midrule
 1  & $\,\,\,\,\,\,\,\,\,\,\,\,\,\,\,\,\,\,\,\,$ & 2,31447 &          &          &          &          &          &          \\
 2  & $\,\,\,\,\,\,\,\,\,\,\,\,\,\,\,\,\,\,\,\,$ & 2,31447 &  6,13324 &          &          &          &          &          \\
 3  & $\,\,\,\,\,\,\,\,\,\,\,\,\,\,\,\,\,\,\,\,$ & 2,04493 &  6,13324 & 10,54940 &          &          &          &          \\
 4  & $\,\,\,\,\,\,\,\,\,\,\,\,\,\,\,\,\,\,\,\,$ & 2,04493 &  5,63655 & 10,54940 & 15,63469 &          &          &          \\
 5  & $\,\,\,\,\,\,\,\,\,\,\,\,\,\,\,\,\,\,\,\,$ & 1,99066 &  5,63655 &  9,66470 & 15,63469 & 21,21933 &          &          \\
 6  & $\,\,\,\,\,\,\,\,\,\,\,\,\,\,\,\,\,\,\,\,$ & 1,99066 &  5,53888 &  9,66470 & 14,30956 & 21,21933 & 27,28556 &          \\
 7  & $\,\,\,\,\,\,\,\,\,\,\,\,\,\,\,\,\,\,\,\,$ & 1,97666 &  5,53888 &  9,46567 & 14,30956 & 19,36916 & 27,28556 & 33,76558 \\
 8  & $\,\,\,\,\,\,\,\,\,\,\,\,\,\,\,\,\,\,\,\,$ & 1,97666 &  5,51611 &  9,46567 & 13,98107 & 19,36916 & 24,86727 & 33,76558 \\
 9  & $\,\,\,\,\,\,\,\,\,\,\,\,\,\,\,\,\,\,\,\,$ & 1,97235 &  5,51611 &  9,41524 & 13,98107 & 18,85787 & 24,86727 & 30,72924 \\
10  & $\,\,\,\,\,\,\,\,\,\,\,\,\,\,\,\,\,\,\,\,$ & 1,97235 &  5,51007 &  9,41524 & 13,89369 & 18,85787 & 24,13659 & 30,72924 \\
\bottomrule
\end{tabular}
\end{table}

As tabelas \ref{table:energiasH1Epsilonx2} e \ref{table:energiasH2Epsilonx2} indicam a 
manifestação da supersimetria do sistema no que se refere à equiparação dos níveis de 
energia $E_{n}^{-}$ e $E_{n-1}^{+}$, $n > 0$, de $H_{-}$ e $H_{+}$. Como esperado, a 
energia do estado fundamental de $H_{-}$ 
é zero e não tem um equivalente em $H_{+}$. Além disso, para $n>0$, aumentando o número 
de parâmetros variacionais, encontramos, principalmente nos primeiros níveis, energias 
$E^{-}_n$ cada vez mais próximas de $E^{+}_{n-1}$. Isso quer dizer que, quanto melhor for a
aproximação que fizermos, mais próximos estaremos de satisfazer \eq{degE}. Além disso, como para
o estado fundamental de $H_{-}$ a função tentativa de um parâmetro tem exatamente a forma da
solução exata, o valor $E^{-}_0 = 0$ encontrado também é o valor exato e \eq{naodegE0}
é naturalmente satisfeita.

\begin{figure}[h!]
\begin{center}
\includegraphics[width=0.70\textwidth]{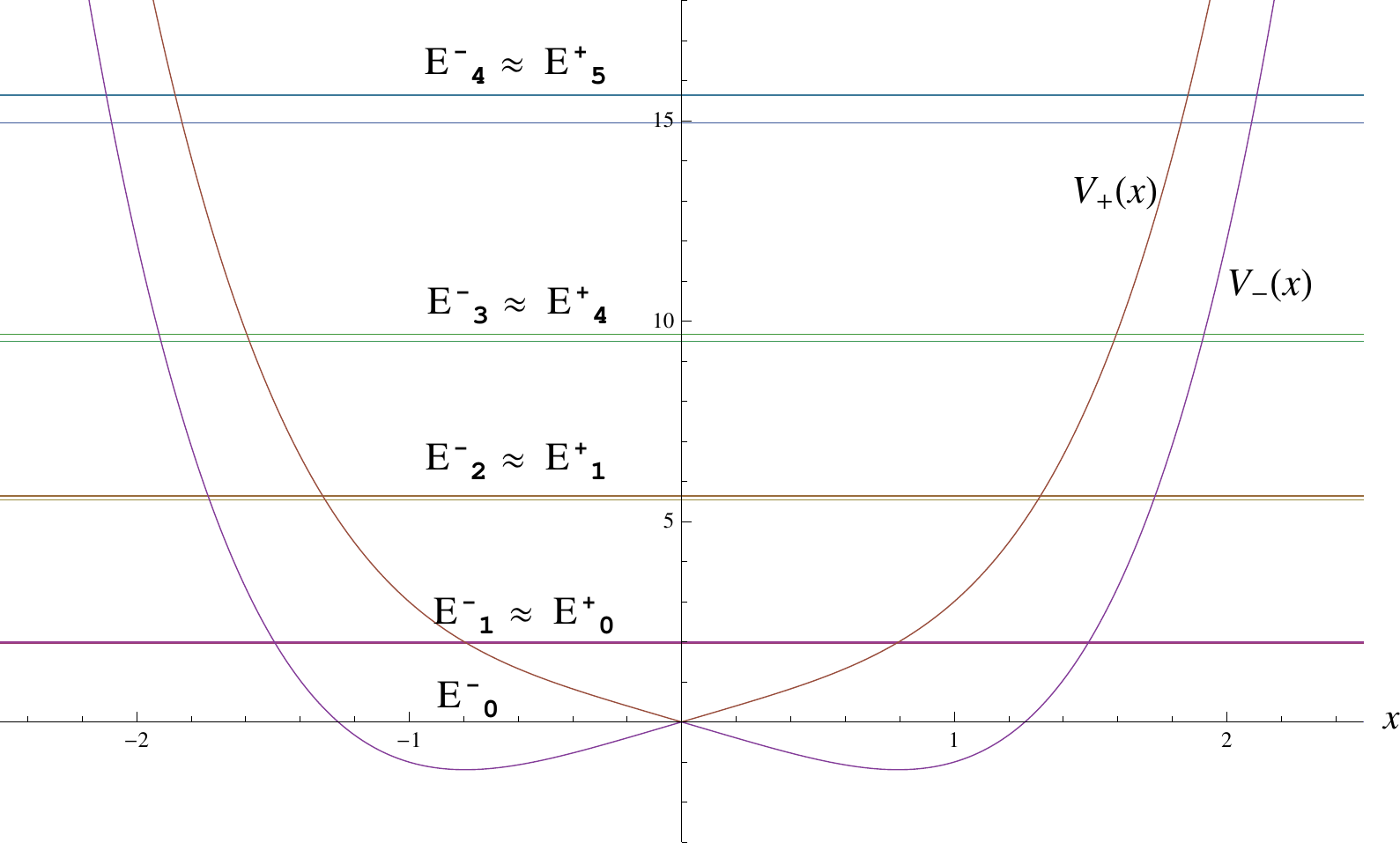}
\caption{\label{fig:autoenergiasEpsilonx2}Esquema dos 5 primeiros níveis de energia de $H_{-}$ e 4 primeiros de $H_{+}$ utilizando 6 parâmetros variacionais.}
\end{center}
\end{figure}

Na figura \ref{fig:autoenergiasEpsilonx2} estão esquematizados os primeiros níveis de energia de
$H_{-}$ e de $H_{+}$. Devemos lembrar os valores obtidos devem ser melhores quanto maior for o 
número de parâmetros utilizado e quanto mais baixo for o nível considerado. Assim, esperamos
encontrar para o nível $n=4$ uma aproximação muito mais pobre do que aquela feita para o nível
$n=1$ ou $n=0$, por exemplo.

\begin{figure}[h!]
\begin{center}
\vbox{
\subfigure[Autofunções de $H_{-}$]{\includegraphics[width=0.58\columnwidth]{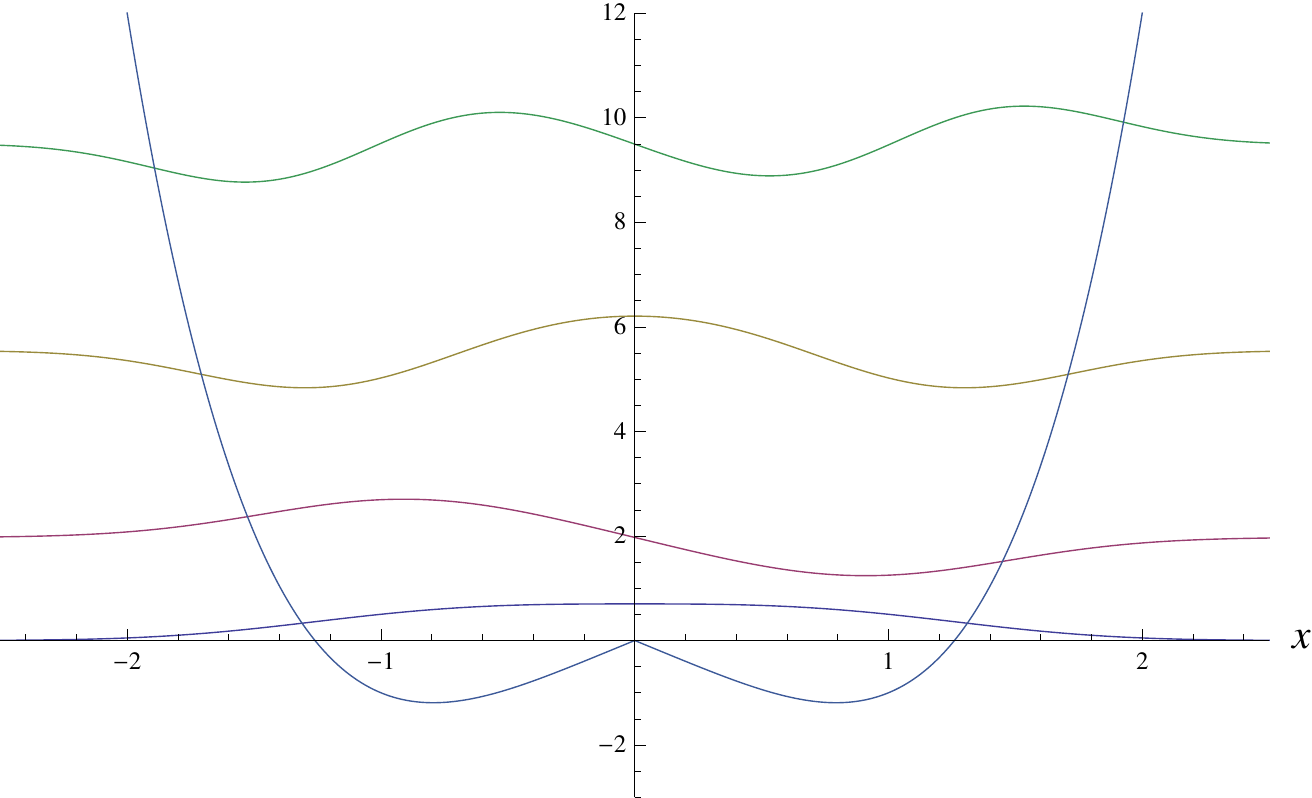}}
\\
\subfigure[Autofunções de $H_{+}$]{\includegraphics[width=0.58\columnwidth]{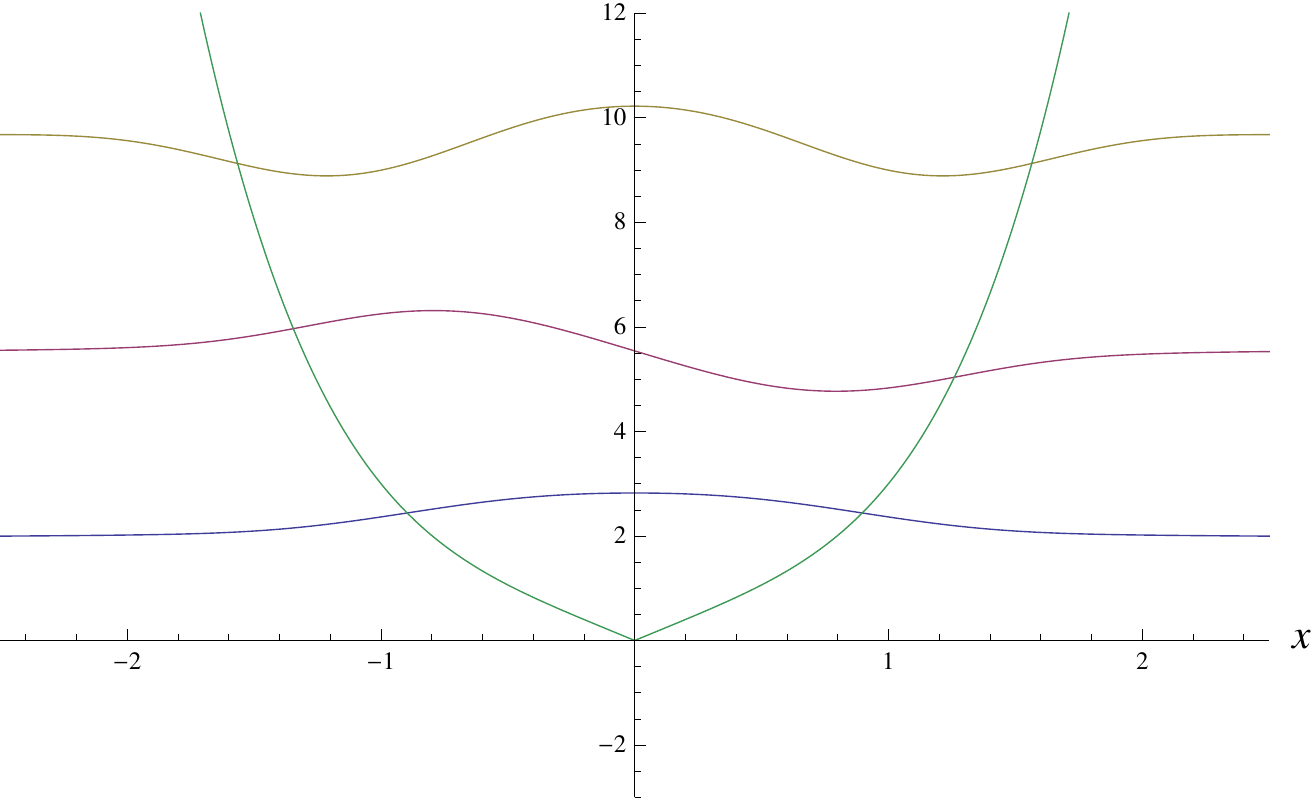}}
}
\end{center}
\caption{\label{fig:autofuncoesEpsilonx2}Esboço das autofunções dos primeiros níveis de $H_{-}$
e $H_{+}$ utlizando 6 parâmetros variacionais.}
\end{figure}

Os gráficos da figura \ref{fig:autofuncoesEpsilonx2} mostram as aproximações para as funções de onda
dos primeiros níveis, respectivamente, de $H_{-}$ e $H_{+}$. Essas aproximações foram obtidas com 6 
parâmetros variacionais.

Como seria esperado, vemos que as funções de onda obtidas apresentam paridades bem definidas,
intercalando soluções pares e ímpares e partindo de soluções pares para os estados fundamentais.

A seguir, avaliamos a diferença $\Delta$ entre o lado esquerdo e o lado direito da equação de 
Schrödinger \eq{schrEpsilonx2}. Essa diferença $\Delta$ é obtida simplesmente substituindo na equação de 
Schrödinger as aproximações para função de onda e energia encontradas por meio do método
variacional e tomando a diferença entre o lado direito e o lado esquerdo, isto é:

\begin{equation}
  \Delta = H \psi_n (x) - E_n \psi_n (x)
  \label{Delta}
\end{equation}
onde $\psi_n (x)$ e $E_n$ são, respectivamente, as aproximações para a função de onda e para a
energia do nível $n$ do sistema descrito pelo hamiltoniano $H$.

\begin{figure}[h!]
\begin{center}
\vbox{
\subfigure[Diferença $\Delta$ em função de $x$ para o primeiro estado excitado de $H_{-}$]{\includegraphics[width=0.48\columnwidth]{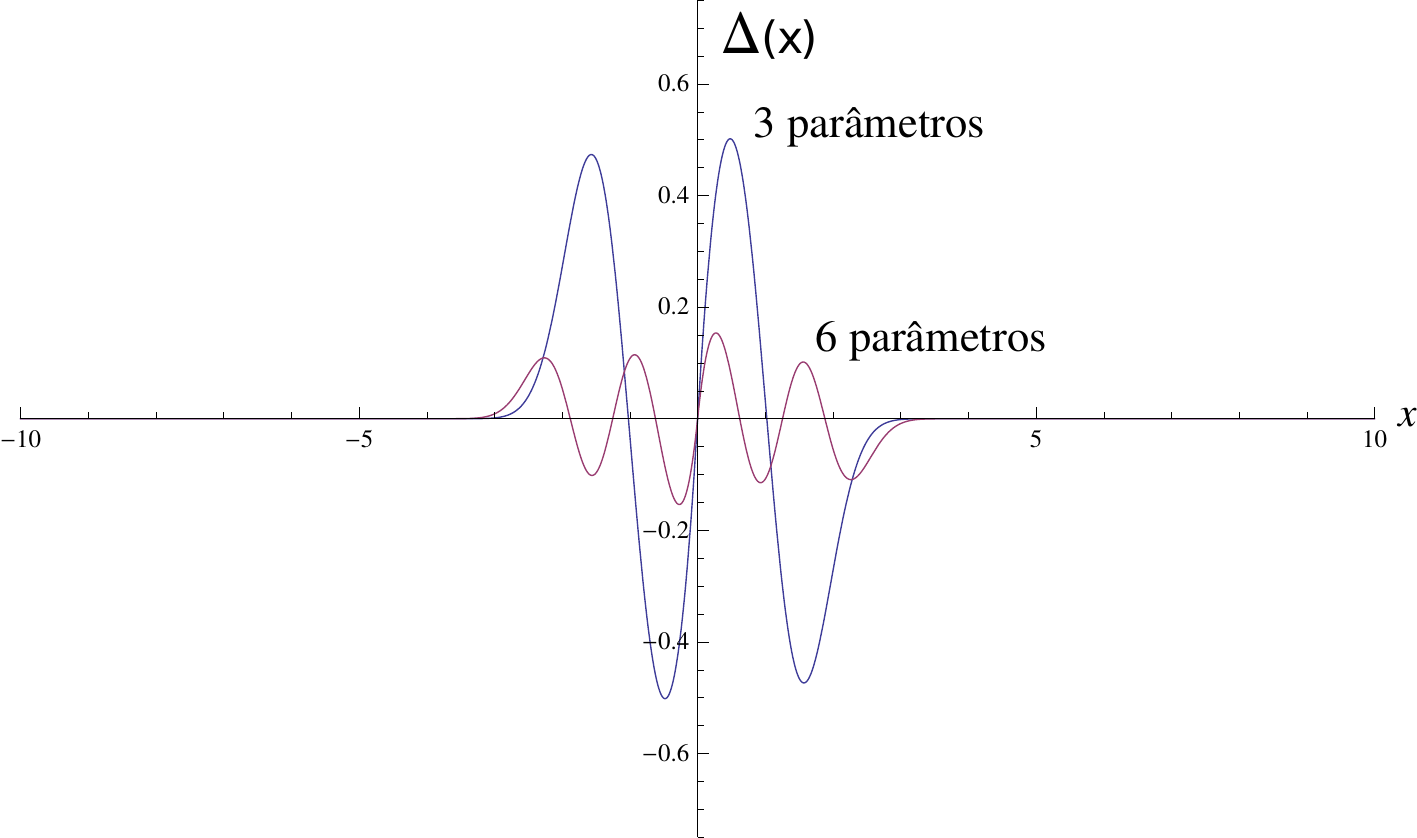}}
\\
\subfigure[Diferença $\Delta$ em função de $x$ para o estado fundamental de $H_{+}$]{\includegraphics[width=0.48\columnwidth]{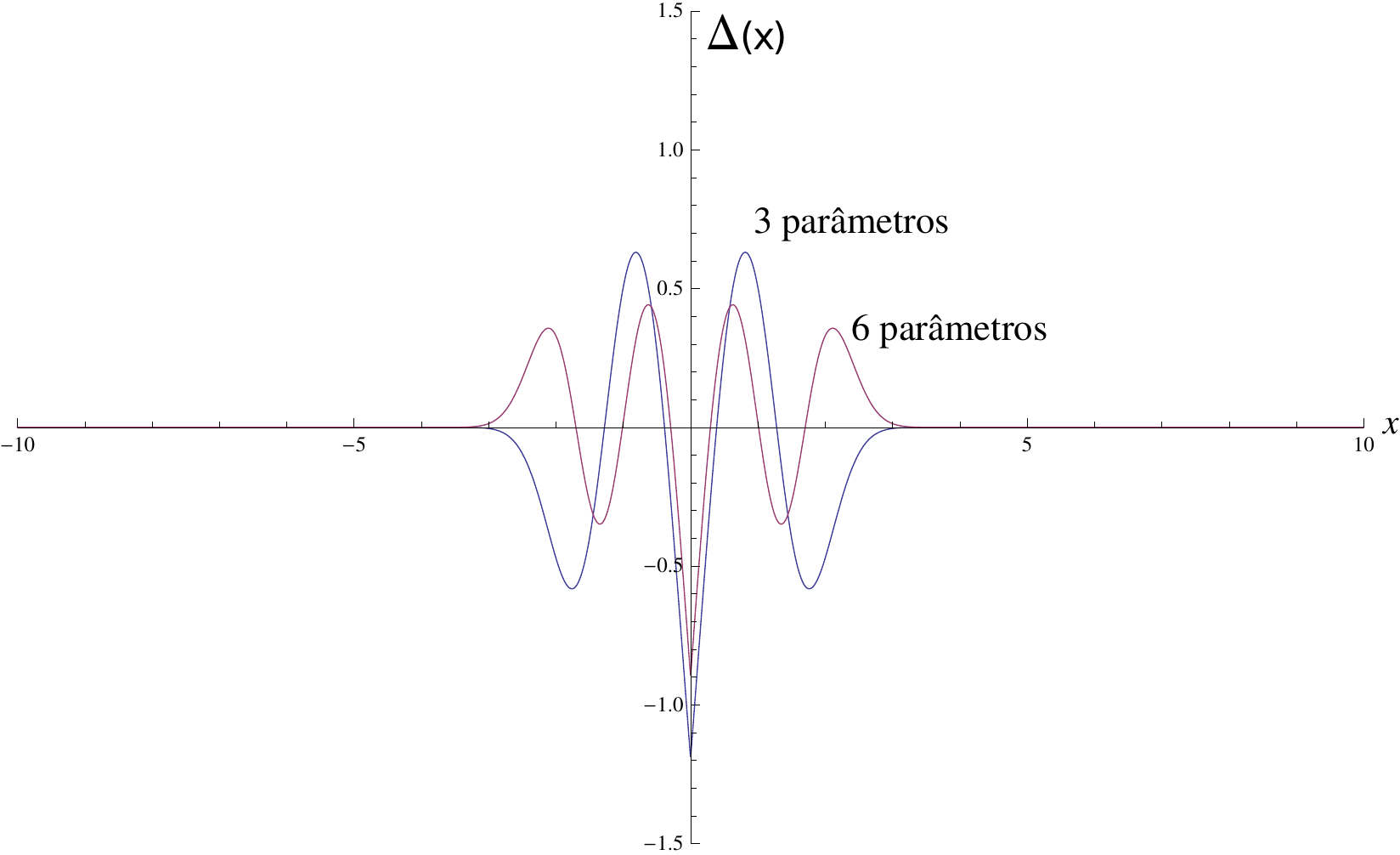}}
}
\end{center}
\caption{Gráficos das diferenças entre o lado esquerdo e o lado
direito da equação de Schrödinger \eq{schrEpsilonx2}.}
\label{fig:diferencaVarEpsilonx2}
\end{figure}

A diferença $\Delta$ foi calculada substituindo as aproximações para função de onda e 
energia do primeiro estado excitado de $H_{-}$ e depois do estado fundamental de $H_{+}$ e fazendo 
isso com 3 e depois com 6 parâmetros variacionais, representamos as diferenças obtidas nos
gráficos da figura \ref{fig:diferencaVarEpsilonx2}.

Conforme escrito acima, um aumento no número de parâmetros variacionais leva a aproximações
melhores, o que pode ser visto nos gráficos da figura \ref{fig:diferencaVarEpsilonx2} pela
diferença entre os máximos das curvas correspondentes a diferentes números de parâmetros.
Notamos ainda, que a aproximação melhora drasticamente para valores de $x$ mais distantes da
origem. Isso poderia estar relacionado à maior influência dos termos dos potenciais que contém
$|x|$ sobre o termo que contém $x^4$ em valores de $x$ próximos da origem.

Os resultados para os valores de energia obtidos aqui por meio do Método Variacional podem
ainda ser comparados com resultados numéricos. Os resultados numéricos necessários para essa
comparação podem ser obtidos, por exemplo, por meio do Método do Abano do Rabo do Cão (do
inglês, \emph{Wag the Dog Method}), que é sugerido em \cite{Griffiths} como um possível meio de 
determinar as autoenergias do OHS. Esse método consiste simplesmente em resolver numericamente a
equação de Schrödinger para um valor de energia escolhido arbitrariamente e observar o
comportamento assintótico da solução encontrada. Para isso convém ter alguma ideia de quanto
vale a energia procurada. Então, escolhendo um certo valor de energia $E_a$
(supostamente menor que o valor real da energia) e 
observando que a função que satizfaz a equação de Schrödinger diverge para $+ \infty$ 
quando $x \rightarrow \pm \infty$ e depois escolhendo um outro valor de energia $E_b$ 
(supostamente maior que o valor real da energia) para o qual
a solução diverge para $- \infty$ quando $x \rightarrow \pm \infty$, esperamos que a energia $E$ 
associada à solução normalizável, isto é, à solução que tende a \emph{zero} para 
$x \rightarrow \pm \infty$, esteja entre $E_a$ e $E_b$. Redefinindo o valor de
$E_a$ ou $E_b$ como a média entre $E_a$ e $E_b$ e repetindo essa análise gráfica até atingir um
intervalo suficientemente pequeno, determinamos o valor da energia $E$ como sendo o valor médio
entre $E_a$ e $E_b$.

Aqui a ideia que temos de quanto devem valer as energias associadas aos hamiltonianos 
$H_{-}$ e $H_{+}$ resulta dos valores aproximados encontrados por meio do Método Variacional e
listados nas tabelas \ref{table:energiasH1Epsilonx2} e \ref{table:energiasH2Epsilonx2}. Assim,
tomamos esses valores como referência e aplicamos o Método do Abano do Rabo do Cão. As tabelas
\ref{table:WTD1} e \ref{table:WTD2} apresentam uma comparação entre os valores numéricos e os resultados
obtidos com $10$ parâmetros variacionais para os primeiros níveis de energia de $H_{-}$ e
$H_{+}$. Além disso, como o valor $E^{-}_0 = 0$ é exato, tendo sido calculado analiticamente,
não nos preocupamos em exibí-lo na tabela \ref{table:WTD1}.

\begin{table}[h!]
\caption{\label{table:WTD1} Comparação entre os valores de energia associados a $H_{-}$ calculados por meio do Método Variacional com $10$ parâmetros variacionais e numericamente.}
\centering
\small
\begin{tabular}{ccccccccc}
\toprule
                  & $E^{-}_1$& $E^{-}_2$& $E^{-}_3$& $E^{-}_4$& $E^{-}_5$& $E^{-}_6$& $E^{-}_7$\\
\midrule
Variacional       &  1,96963 &  5,50842 &  9,39868 & 13,90148 & 18,73498 & 24,43194 & 30,18755 \\           
Valor Numérico    &  1,96951 &  5,50718 &  9,39427 & 13,85837 & 18,64598 & 23,80719 & 29,23255 \\
Desvio(\%)        &  0,00609 &  0,02252 &  0,04694 &  0,31108 &  0,47731 &  2,62421 &  3,26691 \\
\bottomrule
\end{tabular}
\end{table}

\begin{table}[h!]
\caption{\label{table:WTD2} Comparação entre os valores de energia associados a $H_{+}$ calculados por meio do Método Variacional com $10$ parâmetros variacionais e numericamente.}
\centering
\small
\begin{tabular}{ccccccccc}
\toprule
                  & $E^{+}_0$& $E^{+}_1$& $E^{+}_2$& $E^{+}_3$& $E^{+}_4$& $E^{+}_5$& $E^{+}_6$\\
\midrule
Variacional       &  1,97235 &  5,51007 &  9,41524 & 13,89369 & 18,85787 & 24,13659 & 30,72924 \\          
Valor Numérico    &  1,96951 &  5,50718 &  9,39427 & 13,85837 & 18,64598 & 23,80719 & 29,23255  \\
Desvio(\%)        &  0,14420 &  0,05248 &  0,22322 &  0,25486 &  1,13638 &  1,38362 &  5,11994 \\
\bottomrule
\end{tabular}
\end{table}

O desvio entre os valores obtidos por meio do Método Variacional e os valores numéricos tende a
aumentar para níveis de energia mais altos. Isso mostra que, conforme esperávamos, temos
aproximações melhores para os primeiros níveis de energia. Para os $6$ primeiros níveis de
energia de $H_{-}$ e para os $4$ primeiros níveis de $H_{+}$ vemos que os valores encontrados
por meio do Método Variacional diferem por menos de $1 \%$ dos valores numéricos
correspondentes. Além disso, comparando os resultados numéricos dados nas tabelas 
\ref{table:WTD1} e \ref{table:WTD2}, conseguimos observar a concordância desses valores com a
equiparação de níveis de energia em até $5$ casas decimais, o que está relacionado à
manifestação da SUSI nesse sistema.

\subsection{Buscando Soluções Aproximadas pela Teoria de Perturbações \mbox{Logarítmica}}

Vamos agora empregar a Teoria de Perturbações Logarítmica conforme descrito na seção
\ref{section:aproxlpt} do capítulo \ref{chapter:aprox}.

Na Teoria de Perturbações Logarítmica a energia $E^{\pm}_0$ do estado fundamental de $H_{\pm}$ 
e o superpotencial $W_{\pm}(x)$ \footnote{Aqui estamos usando índices ``$\pm$'' para os
superpotenciais $W_{\pm}(x)$ que devem ser entendidos como os índices dos
superpotenciais em uma hierarquia formada pelos hamiltonianos $H_{\pm}$ (ver seção 
\ref{subsection:hierarquia} do capítulo \ref{chapter:mqsusi}), isto é, $W_{\pm}(x)$ 
são tais que satisfazem uma equação de Riccati $V_{\pm} = W_{\pm}(x)^2 - W' _{\pm}(x)$. } 
em $n$-ésima ordem de aproximação são, conforme \eq{energiaordemnaprox} e \eq{ordemnlptW}, 
escritos na forma de expansões em série de potências de $\delta$ como:

\begin{align}
  & E^{\pm}_0 = \sum_{m=0}^{n} B^{\pm}_m \delta^m \label{energiaemordemn}\\
  & W_{\pm}(x) = \sum_{m=0}^{n} W^{\pm}_m(x) \delta^m \label{Wemordemn}
\end{align}

Os potenciais $V_{\pm}(x)$ com $g=1$ (ver equação \eq{riccatiEpsilonx2reescrita}), podem ser
reparametrizados em termos de $\delta$ como:

\begin{equation}
  V_{\pm}(x;\delta) = \left( x^2 \right)^{1+\delta} 
  \pm \left( 4 x^2 \right)^{\nicefrac{\delta}{2}}
  \label{potenciaisreparametrizados}
\end{equation}
sendo que, para $\delta = 0$, temos o caso de dois osciladores harmônicos deslocados 
parceiros SUSI e, para $\delta = 1$, temos novamente os potenciais parceiros gerados pelo
superpotencial $W(x) = \varepsilon(x) x^2$, que é o problema que estamos interessados em
resolver. Para outros valores de $\delta$ podemos obter potenciais de problemas 
diferentes, embora a convergência das séries associadas a esses outros problemas deva 
ser verificada com cuidado.


Expandindo os potenciais reparametrizados $V_{\pm}(x;\delta)$ dados por 
\eq{potenciaisreparametrizados} em série de Taylor até ordem $n$ em torno de 
$\delta_0 = 0$, temos \footnote{O surgimento de fatores contendo logaritmos em expansões como
\eq{expandepotenciaisreparametrizados} seria talvez uma outra possível fonte
de inspiração para o nome ``teoria de perturbações \emph{logarítmica}'', mais restritiva, é
claro, do que aquela mencionada na seção \ref{subsection:lpt} do capítulo \ref{chapter:aprox}
logo após a equação \eq{riccatilpt}.}:

\begin{equation}
  \begin{aligned}
  V_{\pm}(x;\delta) &= \sum_{m=0}^{n} \frac{ \delta^m}{m!} 
  \left\{ x^2 \left[ \ln{(x^2)} \right]^m \pm \left[ \ln{2} + 
                                    \frac{1}{2}\ln{(x^2)} \right]^m \right\} \\
             &= \sum_{m=0}^{n} \frac{ \delta^m}{m!} 
\left\{ x^2 \left[ \ln{(x^2)} \right]^m \pm \sum_{j=0}^{m} \frac{m!}{2^j j! (m-j)!} 
                \left[ \ln{2} \right]^{m-j} \left[ \ln{(x^2)} \right]^j \right\}
  \label{expandepotenciaisreparametrizados}
\end{aligned}
\end{equation}

Conforme \eq{ordem0lpt}, a equação da ordem zero de perturbação é:

\begin{equation}
  \left( x^2 \pm 1 \right) - B^{\pm}_0 = W^{\pm}_0(x)^2 - W^{\pm \prime}_0(x) 
  \label{ordem0expansao}
\end{equation}
sendo satisfeita por:

\begin{equation}
  W^{\pm}_0(x) = x \qquad \text{e} \qquad B^{\pm}_0 = 1 \pm 1
  \label{ordem0correcoes}
\end{equation}
e sendo $\varphi^{\pm}_0(x)$, conforme \eq{ordem0estfundlpt}, dado por:

\begin{equation}
  \varphi^{\pm}_0(x) = \mathcal{N} e^{- \nicefrac{x^2}{2}}
  \label{ordem0estfundexpansao}
\end{equation}

A equação da primeira ordem de perturbação, conforme \eq{ordem1lpt}, é por sua vez:

\begin{equation}
  \left( x^2 \ln{(x^2)} \pm \frac{1}{2} \ln{(x^2)} \pm \ln{2} \right) - B^{\pm}_1 =
  2 W^{\pm}_0(x) W^{\pm}_1(x) - W^{\pm \prime}_1(x)
  \label{ordem1expansao}
\end{equation}
e, tendo em vista \eq{ordem1lptenergia} e \eq{ordem1lptW}, a correção de primeira ordem à
energia $B^{\pm}_1$ e o coeficiente $W^{\pm}_1(x)$ de $W_{\pm}(x)$ que satisfazem 
\eq{ordem1expansao} são, respectivamente:

\begin{equation}
  B^{\pm}_1 = \frac{\braket{\varphi^{\pm}_ 0|V^{\pm}_1(x)|\varphi^{\pm}_0}}{\braket{\varphi^{\pm}_0|\varphi^{\pm}_0}}
  = \frac{\int_{-\infty}^{+\infty} dx e^{-x^2} \left[  x^2 \ln{(x^2)} \pm \frac{1}{2} \ln{(x^2)} \pm \ln{2}  \right]}{\int_{-\infty}^{+\infty} dx e^{-x^2}}
  \label{ordem1correcaoenergia}
\end{equation}
(onde o denominador extra serve para garantir a normalização de $\varphi^{\pm}_0(x)$) e
\begin{equation}
  \begin{aligned}
  W^{\pm}_1(x) &=  |\varphi^{\pm}_0(x)|^{-2} \int_{0}^{x} dy |\varphi^{\pm}_0(y)|^2 \left[ B^{\pm}_1 - V^{\pm}_1(y) \right]  \\
  &= e^{x^2} \int_{0}^{x} dy e^{-y^2} \left[ B^{\pm}_1 - y^2 \ln{(y^2)} \mp \frac{1}{2} \ln{(y^2)} \mp \ln{2}  \right]
  \label{ordem1correcaoW}
\end{aligned}
\end{equation}

Assim, calculando $B^{\pm}_1$ por meio de \eq{ordem1correcaoenergia} e substituindo em
\eq{ordem1correcaoW} para encontrar $W^{\pm}_1(x)$, chegamos a:

\begin{equation}
  B^{\pm}_1 = \frac{1}{2} \psi \left( \nicefrac{3}{2} \right) \pm 
  \frac{1}{2} \psi \left( \nicefrac{1}{2} \right) \pm \ln{2}
  \label{ordem1correcaoenergiacalculada}
\end{equation}
e
\begin{equation}
  W^{\pm}_1(x) = e^{x^2} \int_{0}^{x} dy e^{-y^2} \left[ \frac{1}{2} \psi \left( \nicefrac{3}{2} \right) \pm 
  \frac{1}{2} \psi \left( \nicefrac{1}{2} \right)
 - y^2 \ln{(y^2)} \mp \frac{1}{2} \ln{(y^2)}  \right]
  \label{ordem1correcaoWpseudocalculado}
\end{equation}
onde $\psi(z) \equiv \frac{\Gamma'(z)}{\Gamma(z)}$ é a função digama.

Fazendo $n=1$, substituindo \eq{ordem1correcaoenergiacalculada} em \eq{energiaemordemn} e
tomando $\delta = 1$, que corresponde a retornar ao problema original, as energias dos estados
fundamentais de $H_{\pm}$ em primeira ordem de aproximação são dadas por:

\begin{equation}
  E^{\pm}_0 = (1 \pm 1) + \left(  \frac{1}{2} \psi \left( \nicefrac{3}{2} \right) \pm 
  \frac{1}{2} \psi \left( \nicefrac{1}{2} \right) \pm \ln{2} \right)
  \label{energiaemprimeiraordemaproxcalculada}
\end{equation}
o que resulta em $E^{-}_0 = 0,30685$ e $E^{+}_0 = 1,72964$.

Os melhores valores obtidos por meio do Método Variacional, conforme as tabelas 
\ref{table:energiasH1Epsilonx2} e \ref{table:energiasH2Epsilonx2}, foram  
$E^{-}_0 = 0,00000$ e $E^{+}_0 = 1,97235$. Lembramos que, no caso de
$E^{-}_0$, o valor zero corresponde ao valor exato, tendo sido também 
calculado analiticamente na seção \ref{subsection:solanaliticax2}.

Ao comparar os resultados do Método Variacional com os da Teoria de Perturbações Logarítmica até
primeira ordem, notamos que esses resultados ainda diferem significativamente. Esperamos porém
que essa diferença se torne menor ao aplicar a Teoria de Perturbações Logarítmica para obter
resultados em maiores ordens de aproximação. 

Cabe também notar que a escolha da forma dos
potenciais reparametrizados pode influir no resultado obtido até determinada ordem. Nesse caso
poderia haver uma escolha na forma da reparametrização que levasse a um resultado mais próximo
daquele encontrado pelo Método Variacional já na primeira ordem.
                                    %

\chapter{Considerações Finais}

Nesse trabalho apresentamos uma introdução à noção de SUSI e à MQ SUSI em $1$ dimensão
espacial. Para isso utilizamos o exemplo do OHS para introduzir o conceito de fatorização de
hamiltonianos. Utilizando ainda o exemplo de osciladores harmônicos, o oscilador bosônico e o
oscilador fermiônico, introduzimos como um primeiro caso de manifestação de SUSI na Mecânica
Quântica o sistema do oscilador supersimétrico. Generalizando a idéia do oscilador
supersimétrico, desenvolvemos o formalismo da MQ SUSI chegando às relações de comutação e
anti-comutação que constituem a chamada super-álgebra. Introduzimos os conceitos de 
parceiros supersimétricos, quebra da supersimetria, hierarquia de hamiltonianos e
invariância de forma.

Em um ponto intermediário desse trabalho nos dedicamos a expor dois métodos de aproximação 
úteis no tratamento de problemas em Mecânica Quântica. Primeiramente abordamos o bem 
conhecido Método Variacional e discutimos uma forma particular do seu emprego. Essa forma 
particular foi útil no capítulo seguinte ao tratar o problema do superpotencial 
$W(x) = g \varepsilon(x) x^{2}$, sendo responsável pelo melhor resultado aqui encontrado 
para esse problema. O segundo método apresentado foi o da Teoria de Perturbações Logarítmica,
que é um método perturbativo intimamente relacionado com os conceitos da MQ SUSI. Vimos que esse 
método permite calcular recursivamente aproximações para a energia e para a função de onda do estado 
fundamental de um sistema e também para o superpotencial que fatoriza o hamiltoniano
correspondente e, por meio da hierarquia de hamiltonianos, pode permitir encontrar também os estados
excitados do sistema estudado. A Teoria de Perturbações Logarítmica foi também empregada no 
tratamento do problema do superpotencial $W(x) = g \varepsilon(x)x^{2}$ com a finalidade de comparação 
com o resultado encontrado por meio do Método Variacional.

Como aplicação dos conhecimentos expostos ao longo do trabalho propusemos uma nova classe de
superpotenciais para os quais esperamos observar a manifestação de SUSI; os superpotenciais da
forma $W(x) = g \varepsilon(x) x^{2n}$, com $n = 0, 1, 2,\ldots$. Estudamos o caso mais simples
do superpotencial $W(x) = g \varepsilon(x)$ que origina como potenciais parceiros,
respectivamente, os bem conhecidos potenciais do poço e da barreira Delta de Dirac. Conforme 
esperado, observamos que a SUSI se manifesta nesse sistema com um estado fundamental de energia
zero (o estado ligado associado ao potencial do poço Delta) e um contínuo de estados
de energia positiva para os dois potenciais parceiros (os estados de espalhamento do poço e da
barreira).

Por fim, estudamos o problema do superpotencial $W(x) = g \varepsilon(x) x^{2}$. Esse
superpotencial origina como potenciais parceiros dois osciladores anarmônicos. Para esse
problema encontramos analiticamente um estado fundamental de energia zero para um dos potenciais
parceiros. Demonstramos as dificuldades de se encontrar soluções analíticas para os estados
excitados uma vez que a solução via Método de Frobenius da equação diferencial do problema 
conduzia a relações de recorrência de $3$ termos entre os coeficientes da série, o que 
impossibilitava encontrar uma fórmula fechada para os níveis de energia quantizados do sistema.
Com isso, empregamos o Método Variacional conforme exposto no capítulo anterior, levando a
soluções aproximadas que concordaram com a manifestação da SUSI nesse sistema. Essa
conclusão, porém, é limitada pela precisão do método que, por sua vez, depende, entre outros 
fatores, do número de parâmetros variacionais utilizado.

Com o intuito de empregar um caminho alternativo na resolução do problema, o que poderia
eventualmente levar a um resultado melhor ou pelo menos reforçar o resultado do Método 
Variacional, empregamos a Teoria de Perturbações Logarítmica. O cálculo correspondente foi feito
até primeira ordem de aproximação adotando uma reparametrização conveniente para os potenciais
parceiros. O cálculo de segunda ordem seria muito semelhante àquele que deve ser feito para 
obter o resultado dado em \cite{CooperLPT} para um potencial do tipo $x^4$. Entretanto, embora 
esse resultado exista na literatura, os detalhes do cálculo que leva ao mesmo não são expostos 
e, como estamos tratando um problema similar, entender os detalhes 
do cálculo seria essencial para fornecer resultados em maiores ordens de aproximação. Assim, até
primeira ordem, a Teoria de Perturbações Logarítmica levou a resultados distantes daqueles
encontrados pelo Método Variacional, mantendo esse último como aquele que forneceu o melhor
resultado.

Como perspectivas no estudo da MQ SUSI podemos destacar a sua extensão para
mais de uma dimensão espacial, incluindo, por exemplo, espaços não-comutativos. Dois exemplos 
interessantes na
literatura que exploram essas possibilidades são \cite{DasNonCommut} e \cite{Correa}. Ambos
mostram, entre outras coisas, a aplicação da MQ SUSI em $2$ dimensões espaciais a problemas 
de partículas se movendo
em um plano sob a ação de um potencial vetor $A^{\mu}(x)$ (problema de Landau e efeito Aharanov-Bohm). Outra
possibilidade seria tentar estudar a MQ SUSI no contexto da Mecânica Quântica Relativística, o
que parece ser um campo ainda inexplorado.

                                    %





\end{document}